\documentclass[traditabstract]{aa}

\usepackage{graphicx}
\usepackage{natbib}
\usepackage{amssymb}
\usepackage{txfonts}

\begin{document}

\title{A double component in GRB 090618: a proto-black hole and a genuinely long GRB}
\titlerunning{GRB 090618}
\authorrunning{L. Izzo et al.}
\author{L. Izzo\inst{1,2}, R. Ruffini\inst{1,2,5}, A. V. Penacchioni\inst{1,5}, C. L. Bianco\inst{1,2}, L. Caito\inst{1,2}, S. K. Chakrabarti\inst{3,4}, Jorge A. Rueda\inst{1,2}, A. Nandi\inst{4}, B. Patricelli\inst{1,2} }
\institute{Dip. di Fisica and ICRA, Sapienza Universit\`a di Roma, Piazzale
Aldo Moro 5, I-00185 Rome, Italy, \and ICRANet, Piazza della Repubblica 10, I-65122 Pescara, Italy,
\and S. N. Bose National Center for Basic Sciences, Salt Lake, Kolkata - 700098, India, \and Indian Center for Space Physics, Garia, Kolkata - 700084, India, \and Universite de Nice Sophia Antipolis, CEDEX 2, Grand Chateau Parc Valrose, Nice, France.}

\abstract{
\emph{Context:} The joint X-ray and gamma-ray observations of GRB 090618 by a large number of satellites offer an unprecedented possibility of testing crucial aspects of theoretical models. In particular, it allows us to test (a) in the process of gravitational collapse, the formation of an optically thick $e^+e^-$ -baryon
plasma self-accelerating to Lorentz factors in the range 200 $< \Gamma <$ 3000; (b) its transparency condition with the emission of a component of $10^{53-54}$ baryons in the TeV region and (c) the collision of these baryons with the circumburst medium (CBM) clouds, characterized by dimensions of $10^{15-16}$ cm. In addition, these observations offer the possibility of testing a new understanding of the thermal and power-law components in the early phase of this GRB.\\
\emph{Aims:} We test the fireshell model of GRBs in one of the closest ($z = 0.54$) and most energetic ($E_{iso} = 2.90 \times 10^{53}$ ergs) GRBs, namely GRB 090618. It was observed at ideal conditions by several satellites, namely Fermi, Swift, Konus-WIND, AGILE, RT-2 and Suzaku, as well as from on-ground optical observatories.\\
\emph{Methods:} We analyze the emission from GRB 090618 using several spectral models, with special
attention to the thermal and power-law components. We determine the fundamental parameters of a 
canonical GRB within the context of the fireshell model, including the identification of the total energy of the $e^+e^-$ plasma, $E_{tot}^{e^+e^-}$, the Proper GRB (P-GRB), the baryon load, the density and structure of the CBM.\\
\emph{Results:} We find evidences of the existence of two different episodes in GRB 090618. The first episode lasts $50$ s and is characterized by a spectrum consisting of thermal component, which evolves between $kT = 54$ keV and $kT = 12$ keV, and a power law with an average index $\gamma = 1.75 \pm 0.04$. The second episode, which lasts for $\sim$ 100s, behaves as a canonical long GRB with a Lorentz gamma factor at transparency of $\Gamma = 495$, a temperature at transparency of $29.22$ keV and with characteristic size of the surrounding clouds of $R_{cl} \sim 10^{15-16}$ cm and masses of $\sim 10^{22-24}$ g.\\
\emph{Conclusions:} We support the recently proposed two-component nature of GRB 090618, namely, 
\textit{Episode 1} and \textit{Episode 2}, by using specific theoretical analysis. We further illustrate that the episode 1 cannot be considered to be either a GRB or a part of  a GRB event, but it appears to be related to the progenitor of the collapsing bare core leading to the formation of the black hole which we call a ``proto-black hole''. Thus, for the first time, we are witnessing the process of formation of a black hole from the phases just preceding the gravitational collapse all the way up to the GRB emission.}


\date{}

\maketitle

\section{Introduction}\label{sec:1}

After the discovery of the GRBs by the Vela satellites \citep{Klebesadel1973,Strong1974a,Strong1974b, Strong1975}, the first systematic analysis on a large sample of GRBs was possible thanks to the observations of the BATSE instrument on board the Compton Gamma-Ray Observer (CGRO) satellite \citep{Meegan1992}.
The 4BATSE catalog \citep{Meegan1997,Paciesas1999,Kaneko2006} consists of 2704 confirmed GRBs, and it is widely used by the science community as a reference for spectral and timing analysis on GRBs. 
One of the outcomes of this early analysis of GRBs led to the classification of GRBs as a function of their observed time duration.
$T_{90}$ was defined as the time interval over which the 90$\%$ of the total BATSE background-subtracted counts are observed.
The distribution of the $T_{90}$ duration was bi-modal: the GRBs with $T_{90}$ less than $2$s were classified as ``short'' while the ones with $T_{90}$ larger than $2$s were classified as ``long'' \citep{Klebesadel1992,Dezalay1992,Koveliotou1993,Tavani1998}.

After the success of BATSE, a large number of space missions dedicated to the GRB observations were launched.
Particularly significant was the discovery of an additional prolonged soft X-ray emission by Beppo-SAX \citep{Costa1997}, following the usual hard X-ray emission observed by BATSE.
The Beppo-SAX observed emission was named as the ``afterglow'', while the BATSE one as the ``prompt'' radiation.
The afterglow allowed to pinpoint more accurately the GRB position in the sky and permitted the identification of their optical counterpart by space and ground based telescopes.
The measurement of the cosmological redshift for GRBs became possible and their cosmological nature was firmly established \citep{vanParadjis1997}.

The Beppo-SAX and related results led to rule out literally hundreds of theoretical models of GRBs, \citep[see for a review][]{Ruffini2001K}. 
Among the handful of surviving models, there was the one by \citet{Damour} based on the mass-energy formula of Black Holes. This model can naturally explain the energetics up to 10$^{54-55}$ erg, as requested by the cosmological nature of GRBs, through the creation of an $e^+e^-$-plasma by vacuum polarization processes in the Kerr-Newman geometry \citep[for a recent review see][]{PhysRep}.
This model was proposed a few months after the presentation of the discovery of GRBs 
by Strong \citep{Strong1975} at the AAAS meeting in San Francisco.

It soon became clear that, as suggested by Goodman and Paczynski \citep{Goodman1986,Paczynski1986}, the presence of a Lorentz gamma factor larger than 100 could overcome the problem of opacity of the $e^+e^-$-plasma and justify the $\gamma$-ray emission of GRBs at cosmological distances \citep[see e.g.][]{Piran2005}.
That the dynamics of an $e^+e^-$-plasma with a baryon load with mass $M_B$ would naturally lead to Lorentz gamma factor in the range (10$^{2}$- 10$^{3}$) was demonstrated by \citet{ShemiPiran1990, PiranShemiNarayan,1993ApJ...415..181M}.
The general solution for a baryon load $B = M_B c^2/E_{tot}^{e^+e^-}$ between $0$ and $10^{-2}$ was obtained in \citet{RSWX}. The interaction between the accelerated baryons with the CBM, 
indicated by \citet{MeszarosRees1993}, was advocated to explain the nature of 
the afterglow \citep[see e.g.][and references therein]{Piran1999}.

The unprecedented existence of such large Lorentz gamma factors led to formulate the Relativistic Space-Time Transformations paradigm for GRBs \citep{Ruffini2001L107}.
Such a paradigm made it a necessity to have a global, instead of a piecewise, description of a GRB phenomenon \citep{Ruffini2001L107}.
This global description led to the conclusion that the emission by the accelerated baryons interacting with the CBM indeed occurs already in the prompt emission phase in a fully radiative regime.  
A new interpretation of the burst structure paradigm was then introduced \citep{Ruffini2001}: the existence of a characteristic emission at the transparency of an $e^+e^-$-plasma, the Proper-GRB, followed by an extended-afterglow emission.
The relative intensity of these two components is a function of the baryon load.
It was proposed that the case of $B < 10^{-5}$ corresponds to the short GRBs, 
while the case of $B > 3 \times 10^{-4}$ corresponds to the long GRBs.
 
This different parametrization of the prompt -- afterglow versus the one of the P-GRB -- extended-afterglow could have originated years of academic discussions. However a clear cut observational evidence came from the Swift satellite, in favour of the second parametrization. The Norris-Bonnell sources, characterized by an initial short spike-like emission in the hard X-rays followed by a softer extended emission, had been indicated in the literature as short bursts. There is a clear evidence that they belong to a new class of ``disguised'' short GRBs, \citep{Bernardini2007, Caito2009, Caito2010, deBarros2011}, where the initial spike is identified as the P-GRB while the prolonged soft emission occurring from the extended-afterglow emission in a CBM typically of the galatic halo. These sources have a baryon load $10^{-4} < B < 7 \times 10^{-4}$: they are just long GRBs exploding in a particularly low density CBM of the order of $10^{-3}$ particles/cm$^3$.
This class of sources has given the first evidence of GRBs originating from binary mergers, strongly supported also from direct optical observations \citep{Bloom2006,Fong2010}.
    
It is interesting that, independent of the development of new missions, the BATSE data continue to attract 
full scientific interests, even after the end of the mission in the 2000. 
Important inferences, based on the BATSE data, on the spectra of the early emission of the GRB have been made by \citet{Ryde2004} and \citet{Rydeetal2006}. They have convincingly demonstrated that the spectral feature composed by a blackbody and a power-law plays an important role in selected episodes in the early part of the GRB emission. They have also shown, in some cases, a power-law variation of the thermal component as a function of time, following a broken power-law behavior, see Fig. \ref{fig:no17}.

The arrival of the Fermi and other satellites allowed further progresses in the understanding of the GRB phenomenon in a much wider energy range. Thanks to the Gamma-Ray Burst Monitor (GBM) \citep{Meegan2009} and the Large Area Telescope (LAT) \citep{Atwood2009}, additional data are obtained in the 8 keV - 40 MeV and 100 MeV - 300 GeV energy range.
It has allowed, among others, this first evidence of a GRB originating from the collapse of a core in the late evolution of a massive star, what we have called the Proto Black Hole \citep{Ruffini2011,Penacchioni2012}.

In the specific case of GRB 090618, it has been possible to obtain a complete temporal coverage of the emission in gamma and X-rays, due to the joint observations by Swift, Fermi, AGILE, RT-2/Coronas-PHOTON, Konus-WIND and Suzaku-WAM telescopes.
A full coverage in the optical bands, up to 100 days from the burst trigger, has been obtained. This has allowed the determination of the redshift, $z = 0.54$, of the source from spectroscopical identification of absorption lines \citep{GCN9518} and a recent claim of a possible supernova emission $\sim$ 10 days after the GRB trigger.
This GRB lasts for $\sim$ 150 s in hard X-rays, and it is characterized by four prominent pulses.
In the soft X-rays there are observations up to 30 days from the burst trigger.

We have pointed out in \citet{COSPAR} that two different episodes are present in GRB 090618. We have also showed that while the second episode may fit a canonical GRB, the first episode is not expected to be either a part of a GRB or an independent GRB \citep{TEXAS}. 

In the present paper we enter in the merit of the nature of these two episodes. In particular: 
\begin{itemize}
\item in Section 2, we describe the observations and data reduction and analysis. We obtain the Fermi GBM (8 keV - 1 MeV and 260 keV - 40 MeV) flux light curves, shown in Fig. \ref{fig:1}, following the standard data reduction procedure, and make a detailed spectral analysis of the main emission features, using a Band and a power-law with an exponential high energy cut-off spectral models.
\item in Section 3, after a discussion about the most quoted GRB model, the fireball, we recall the main features of the fireshell scenario, focusing on the reaching of transparency at the end of the initial optically thick phase, with the emission of the Proper-GRB (P-GRB). In Fig. \ref{fig:no4g} we give the theoretical evolution of the Lorentz $\Gamma$ factor as a function of the radius, for selected values of the baryon load, corresponding to fixed values of the total energy $E_{tot}^{e^+e^-}$. The identification of the P-GRB is crucial in determining the main fireshell parameters, which describe the canonical GRB emission. The P-GRB emission is indeed  characterized by the temperature, the radius and the Lorentz $\Gamma$ factor at the transparency, which are related with the $E_{tot}^{e^+e^-}$ energy and the baryon load, see Fig. \ref{fig:no4}.  We then recall the theoretical treatment, the simulation of the light curve and spectrum of the extended-afterglow and, in particular, the determination of the equations of motion, the role of the EQuiTemporal Surfaces (EQTS) \citep{Bianco2004, Bianco2005b}, as well as the ansatz of the spectral energy distribution in the fireshell comoving frame, \citep[see][and references therein]{Patricelli2011}.

The temporal variability of a GRB light curve has been interpreted in some current models as due to internal shock \citep{ReesMeszaros1994}. In the fireshell model instead such temporal variability is produced by the interaction of the ultra-relativistic baryons colliding with the inhomogeneities of the CircumBurst Medium (CBM). This allows to perform a tomography of the CBM medium around the location of the black hole formation, see Fig. \ref{fig:rad}, gaining important information on its structure. These collisions are described by three parameters: the $n_{CBM}$ average density, the filling factor $\mathcal{R}$, the clumpiness on scales of 10$^{15-16}$ cm  and average density contrast $10^{-1} \lesssim \langle\delta n/n\rangle \lesssim 10$.
We then refer also to the explanation of the observed hard-to-soft behavior due to the drop of the Lorentz $\Gamma$ factor and the curvature effect of the EQTS. We then recall the determination of the instantaneous spectra and the simulations of the observed multi-band light curves in the chosen time interval, taking into account all the thousands of convolutions of comoving spectra over each EQTS leading to the observed spectrum. We also emphasize how these simulations have to be performed together and optimized.
\item in Section 4, we perform a spectral analysis of GRB 090618. 
We have divided the total GRB emission in 6 time intervals, see Table \ref{tab:no1}, each one identifying a significant feature in the emission process, see also \citet{Rao2011}. 
We have considered two different spectral models in the data fitting procedure: a Band model \citep{Band1993} and one by a blackbody plus a power-law component, following e.g. \citet{Ryde2004}.
We find that the first $50$s of emission are well-fitted by both models, equally the following $9$s, from $50$ to $59$s.
The remaining part, from $59$ to $151$ s, is fitted satisfactorily only by the Band model, see Table \ref{tab:no1}.
\item in Section 5, we proceed to the analysis of GRB 090618 in the fireshell scenario. 
In Section 5.1, we attempt our first interpretation of GRB 090618 assuming it to be a single GRB.
We recall that the blackbody is an expected feature in the theory of P-GRB.
From the spectral analysis of the first $50$ s, we find a spectral distribution consistent with a blackbody plus a power-law component. We have first attempted a fit of the source identifying these first $50$s as the P-GRB, see Fig. \ref{fig:big1}.
We confirm the conclusion reached in \citet{COSPAR} that this interpretation is not sustainable for three different reasons, based on: 1) the energetics of the source, 2) the time duration and 3) the theoretical expected temperature for the P-GRB.
We then proceed, in the sub-section 5.2, to an interpretation of GRB 090618 as a multi component system, following the procedure outlined in \citet{TEXAS}, in which we outline the possibility of the second episode between $50$ and $151$s
to be an independent GRB.

We identify the P-GRB of this second episode, as the first $4$s of emission.
We find that the spectrum in this initial emission can be fitted by a blackbody plus a power-law component, see Fig. \ref{fig:pgrb}.
Since this extra power-law component can be due to the early onset of the extended-afterglow, we take it into account to perform a fireshell simulation which is shown in Fig. \ref{fig:pgrb}, with an energy $E_{tot}^{e^+e^-} = 2.49 \times 10^{53}$ erg and a baryon load $B = 1.98 \pm 0.15 \times 10^{-3}$.
In Figs. \ref{fig:rad},\ref{fig:firesh},\ref{fig:comp}, we report the results of our simulations, summarized in Table \ref{tab:ris}.
We notice, in particular, the presence of a strong time lag in this GRB.
A detailed analysis, see \citet{Rao2011}, about the time lags in the mean energy ranges of $35$ keV, $68$ keV and $125$ keV, reports a quite large lag , $ \sim$ 7 s, in the first $50$s of the emission  which is unusual for GRBs, while in the following emission, from $51$ to $151$ s, the observed lags are quite normal, $\sim 1$ s. 
\item in Section 6,  we perform a spectral analysis of the first $50$s, where we find a strong spectral variation with time, as reported in Table \ref{tab:no} and in Figs. \ref{fig:no6},\ref{fig:no17}, with a chacteristic power-law time variation similar to the ones identified by \citet{Ryde2009} in a sample of 49 BATSE GRBs.
\item in Section 7, we estimate the variability of the radius emitter, Fig. \ref{fig:no18}, and proceed to an estimate of the early expansion velocity. We interpret this data as originating in the expansion process occurring previous to the collapse of the core of a massive star to a black hole, see e.g. \citet{Arnett2011}: this early $50$s of the emission are then defined as the proto-black hole phenomenon.
\item In Section 8, we proceed to the conclusions.

\end{itemize}

\section{Observations}\label{sec:2}

On 18th of June 2009, the Burst Alert Telescope (BAT) on board the Swift satellite \citep{Gehrels2009} triggered on GRB 090618 \citep{GCN9512}. 
After 120 s the X-Ray Telescope (XRT) \citep{Burrows2005} and the UltraViolet Optical Telescope (UVOT) \citep{Roming2005} on board the same satellite, started the observations of the afterglow of GRB 090618.
UVOT found a very bright optical counterpart, with a white filter magnitude of 14.27 $\pm$ 0.01 \citep{GCN9527} not corrected for the extinction, at the coordinates RA(J2000) = 19:35:58.69 = 293.99456, DEC(J2000) = +78:21:24.3  = 78.35676.   
The BAT light curve shows a multi peak structure, whose total estimated duration is of $\sim$ 320 s, with the T$_{90}$ duration in the (15-350) keV range was of 113 s \citep{GCN9530}. 
The first 50 s of the light curve presents a smooth decay trend, followed by a spiky emission, with three prominent peaks at 62, 80 and 112 seconds after the trigger time, respectively, and each have the typical appearance of the FRED pulse (see e.g. \citet{Fishman1994}), see Fig. \ref{fig:1}. 
The time integrated spectrum, (t$_0$ - 4.4, t$_0$ + 213.6) s in the (15-150)keV range, was found to be in agreement with a power-law spectral model with an exponential cutoff, whose photon index was $\gamma$ = 1.42 $\pm$ 0.08 and a cut-off energy $E_{peak}$ = 134 $\pm$ 19 keV \citep{GCN9534}.  
The XRT observations started 125 s after the BAT trigger time and lasted $\sim$ 25.6 ks \citep{GCN9528} and reported an initially bright uncatalogued source, identified as the afterglow of GRB 090618.
Its early decay was very steep, ending at 310 s after the trigger time, when it starts a shallower phase, the plateau.
Then the light curve breaks to a more steep last phase.

GRB 090618 was observed also by the Gamma-ray Burst Monitor (GBM) on board the Fermi satellite \citep{Meegan2009}.
From a first analysis, the time-integrated spectrum, ($t_0$, $t_0$ + 140) s in the (8-1000)keV range, was fitted by a Band \citep{Band1993} spectral model, with a peak energy $E_{peak}$ = 155.5  keV, $\alpha$ = $-1.26$ and $\beta$ = $-2.50$ \citep{GCN9535}, but with strong spectral variations within the considered time interval.

It is appropriate to compare and contrast the considerations of the time-integrated spectral analysis, often adopted in the current literature of GRBs, with the information from the time-resolved spectral analysis, as presented e.g. in this article.
For a traditional astrophysical source, steady during the observation time, the time-integrated and time-resolved spectral analysis usually coincide.
In the case of GRBs, although the duration is only a few seconds, each instantaneous observation corresponds to a very different physical process and the two approaches  have an extremely different physical and astrophysical content. 

The redshift of the source is $z =  0.54$ and it was determined thanks to the identification of the MgII, Mg I and FeII absorption lines, using the KAST spectrograph mounted at the 3-m Shane telescope at the Lick observatory \citep{GCN9518}.
Given the redshift, and the distance of the source, we computed the emitted isotropic energy in the 8 - 10000 keV energy range, using the Schaefer formula \citep{Schaefer2007}: using the fluence in the (8-1000 keV) as observed by Fermi-GBM, S$_{obs}$ = 2.7 $\times$ 10$^{-4}$ \citep{GCN9535}, and the $\Lambda$CDM cosmological standard model $H_0$ = 70 km/s/Mpc, $\Omega_m$ = 0.27, $\Omega_{\Lambda}$ = 0.73, we obtain for the isotropic energy emitted the value of E$_{iso}$ = 2.90 $\times$ 10$^{53}$ erg.

\begin{center}
\begin{figure}
\includegraphics[width=12cm, height=8cm, angle=270]{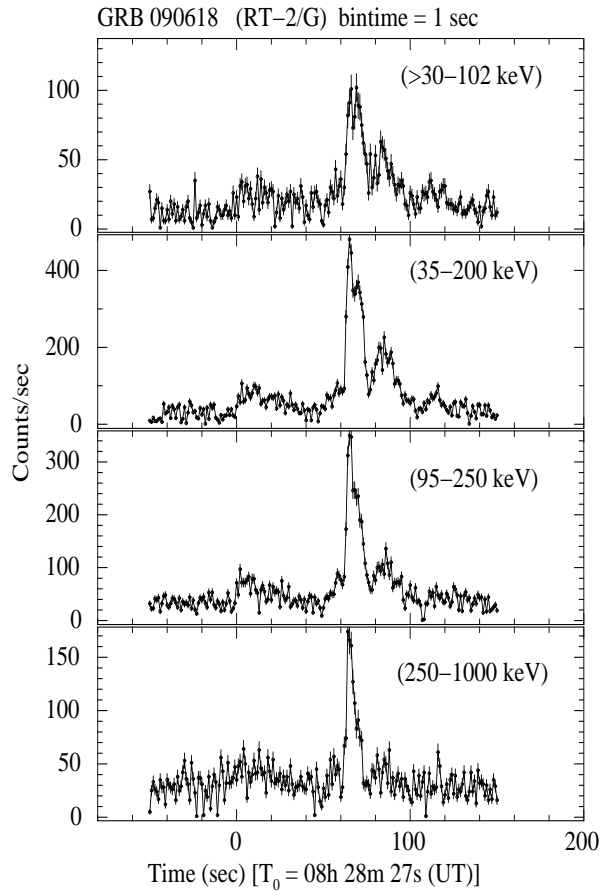}
\caption{RT2 light curves of GRB 090618. }
\label{fig:Chak}
\end{figure}
\end{center}

This GRB was observed also by Konus-WIND \citep{GCN9553}, Suzaku-WAM \citep{GCN9568} and by the AGILE satellite \citep{GCN9524}, which detected emission in the (18-60) keV and in the MCAL instrument, operating at energies greater than 350 keV, but it did not observe high energy photons above 30 MeV.
GRB 090618 was the first GRB observed by the Indian payloads RT-2 on board Russian Satellite CORONAS-PHOTON \citep{Kotov2008,Nandi2009,Rao2011}.
Two detectors, namely, RT-2/S and RT-2/G consist of NaI(Tl)/CsI(Na) scintillators in phoswich assembly viewed by a photomultiplier tube (PMT). RT-2/S has a viewing angle of $4^\circ \times 4^\circ$ and covers an energy range of 15 keV to 1 MeV, whereas RT-2/G has an Al filter which sets the lower energy to $\sim 20$keV.
The Mission was launched from Plesetsk Cosmodrom, Russia on  January 30, 2009. 
During the event RT-2 payload was in the SHADOW mode (away from the Sun) during 08:16:10.207 UT and ended at 08:37:35.465 UT and the GRB 090618 was detected at $77^\circ$ off-axis angle.
During this period, the spectra was accumulated in every 100s while the eight channel count rates for each detector are accumulated every second. The entire episode was observed for a duration of more than 200 seconds. 
A closer examination of the data in the accumulated Channels 1:15-102 keV, 2:95-250 keV and 3:250-1000 keV indicates that the most significant counts is in Channel 2 with a clear evidence of the followings: (a) The emission in the first 50 s is prominent and broader in the lower channels, see Fig. \ref{fig:Chak}, (b) After the first 50 s, there is an evidence of a precursor of about 6 seconds duration before the main pulse (c) a break up into two peaks of the main pulse at intermediate energies (35-200 keV) while at higher energies (250-1000 keV) only the first peak of the main pulse survives, see \citet{Rao2011} and also Fig. \ref{fig:1} in this manuscript. 

Thanks to the complete data coverage of the optical afterglow of GRB 090618, the possible presence of a supernova underlying the emission of the GRB 090618 optical afterglow \citep{Cano2010} was reported.
The evidence of a supernova emission came from the presence of several bumps in the light curve and by the change in $R_c$ - $i$ color index over time: in the early phases, the blue color is dominant, typical of the GRB afterglow, but then the color index increases, suggesting a presence of a core-collapse SN. 
At late times, the contribution from the host galaxy was dominant.

\subsection{Data Analysis}\label{sec:2.2}

We consider the BAT and XRT data of the Swift satellite, together with the Fermi-GBM and RT2 data of the Coronas-PHOTON satellite.
The data reduction was done using the Heasoft v6.10 packages\footnote{http://heasarc.gsfc.nasa.gov/lheasoft/} for BAT and XRT, and the Fermi-Science tools for GBM.

We obtained the BAT light curve and spectra using the standard \texttt{headas} procedure. 
After the data download from the \texttt{gsfc} website\footnote{ftp://legacy.gsfc.nasa.gov/swift/data/obs/}, we made a detector quality map and corrected the event data for the known errors of the detector and the hot pixels. 
We subtracted the background from the data, corrected for the improved position, using the tool \texttt{batmaskwtevt} and obtained the 1-s binned light curves and spectra in the main BAT energy band 15 - 150 keV and its subranges, using the tool \texttt{batbinevt}. 
After the systematic corrections to the spectrum, we created the response matrices and obtained the final spectra.

For the XRT data, we obtained a total dataset using the standard pipeline, while for a time-resolved analysis we considered the on-line recipe, which is well described in literature, see \citet{Evans, Evans2}.
The GBM data\footnote{ftp://legacy.gsfc.nasa.gov/fermi/data/gbm/}, in particular the fourth NaI detector in the (8 - 440 keV) and the b0 BGO detector (260 keV - 40 MeV), were analyzed using the \texttt{gtbindef} tool to obtain a GTI file for the energy distribution and the \texttt{gtbin} for the light curves and final spectra.
In order to obtain an energy flux lightcurve, we made a time resolved spectral analysis dividing the count lightcurve in six time intervals, each of them corresponding to a particular pulse, as described in the work of \citet{Rao2011}.
All the time resolved spectra were fitted using the XSPEC data analysis software \citep{XSPEC} version 12.6.0q, included in the Heasoft data package, and considering for each spectrum a classical Band spectral model \citep{Band1993} and a power-law model with an exponential energy cut-off, folded through the detector response matrix. 
After the subtraction of the background, we fit the spectrum by minimizing the $\chi^2$ between the spectral models described above and the observed data, obtaining the best-fit spectral parameters and the respective model normalization.
In Table \ref{tab:no1} we give the results of our spectral analysis. 
The time reported in the first column corresponds to the time after the GBM trigger time t$_{trig}$ = 267006508 s, where the $\beta$ parameter was not constrained, we used its averaged value, as delineated in \citet{Guetta2011} $\beta$ = -2.3 $\pm$ 0.10. We have considered the chi-square statistic for testing our data fitting procedure. The reduced chi-square $\tilde{\chi}^2 = \chi^2/N$, where $N$ is the number of degrees of freedom (dof) which is $N = 82$ for the NaI dataset and $N = 121$ for the BGO one.

\begin{center}
\begin{figure}
\begin{tabular}{|c|}
\hline
\includegraphics[height=6cm,width=8cm,angle=0]{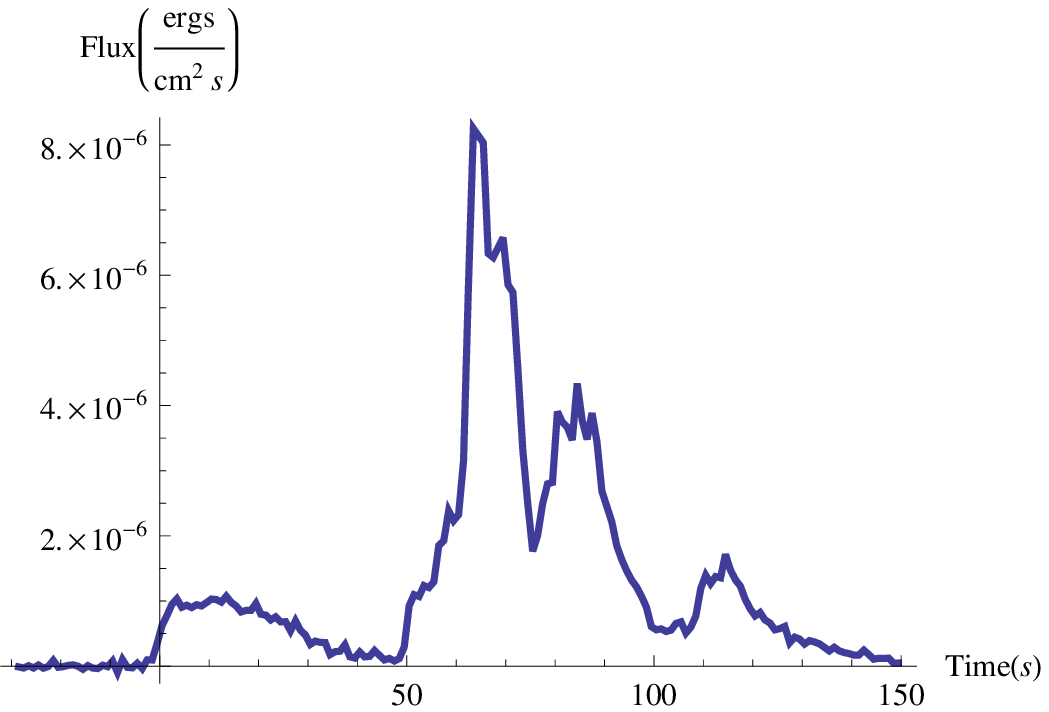}\\
\hline
\includegraphics[height=6cm,width=8cm,angle=0]{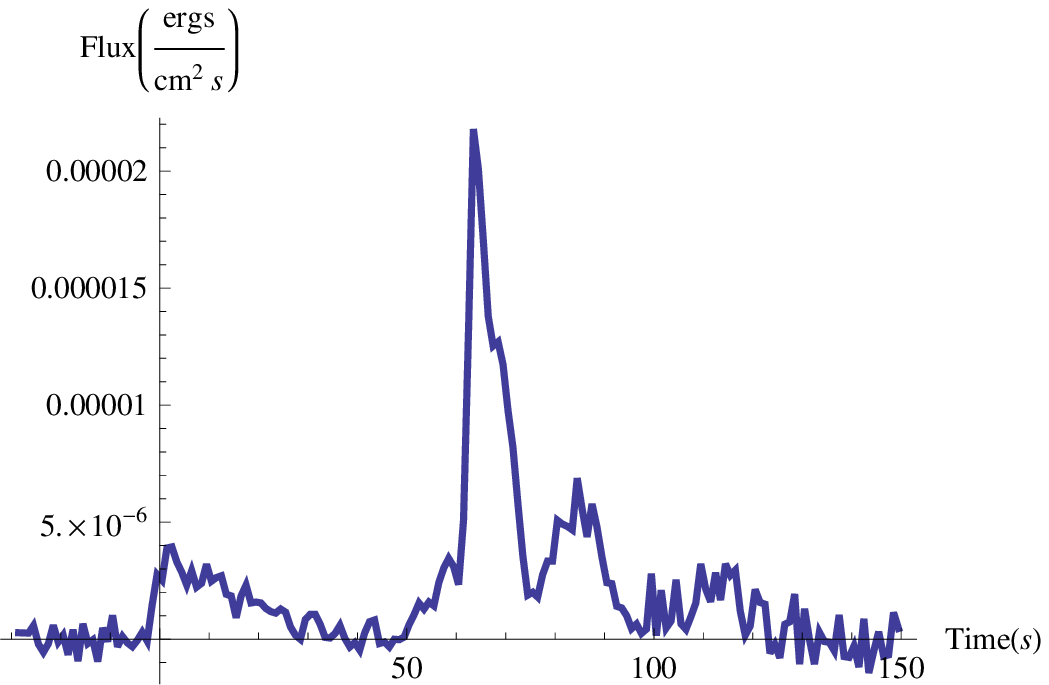}\tabularnewline
\hline
\end{tabular}
\caption{Fermi-GBM flux light curve of GRB 090618 referring to the NaI (8-440 keV, \emph{upper panel}) and BGO (260 keV - 40 MeV, \emph{lower panel}) detectors.}
\label{fig:1}
\end{figure}
\end{center}

For the last pulse of the second episode, the Band model is not very precise ($\tilde{\chi}^2$ = 2.24), but a slightly better approximation is given by the power-law with an exponential cut-off, whose fit results are shown for the same intervals in the last two columns.
From these values, we build the flux light curves for both the detectors, which are shown in fig. \ref{fig:1}.

\begin{table*}
\centering
\caption{Time-resolved spectral analysis of GRB 090618. We have considered six time intervals, each one corresponding to a particular emission feature in the light curve. We fit the GBM (8 keV - 10 MeV) observed emission with a Band model \citep{Band1993} and a power-law function with an exponential cut-off. In the columns 2-4 are listed the Band low energy index $\alpha$, the high-energy $\beta$ and the break energy $E_0^{BAND}$, with the reduced chi-square value in the 6$^{th}$ column. In the last three columns are listed the power-law index $\gamma$, the cut-off energy $E_0^{cut}$ and the reduced chi-square value respectively, as obtained from the spectral fit with the cut-off power-law spectral function. } 
\label{tab:no1} 
\begin{tabular}{l c c c c c c c}
\hline\hline
Time Interval  & $\alpha$ & $\beta$ & $E_0^{BAND}$ (keV) &  $\tilde{\chi}^2_{BAND}$ & $\gamma$ & $E_0^{cut}$ (keV) &  $\tilde{\chi}^2_{cut}$\\ 
\hline 
0 - 50 & -0.77$^{+0.38}_{-0.28}$ & -2.33$^{+0.33}_{-0.28}$ & 128.12$^{+109.4}_{-56.2}$  & 1.11 & 0.91$^{+0.18}_{-0.21}$ & 180.9$^{+93.1}_{-54.2}$ & 1.13\\
50 - 57 & -0.93$^{+0.48}_{-0.37}$ & -2.30 $\pm$ 0.10 & 104.98$^{+142.3}_{-51.7}$ & 1.22 & 1.11$^{+0.25}_{-0.30}$ & 168.3$^{+158.6}_{-70.2}$ & 1.22 \\
57 - 68 & -0.93$^{+0.09}_{-0.08}$ & -2.43$^{+0.21}_{-0.67}$ & 264.0$^{+75.8}_{-54.4}$ & 1.85 & 1.01$^{+0.06}_{-0.06}$ & 340.5$^{+56.0}_{-45.4}$ & 1.93 \\
68 - 76 & -1.05$^{+0.08}_{-0.07}$ & -2.49$^{+0.21}_{-0.49}$ & 243.9$^{+57.1}_{-53.0}$  & 1.88  & 1.12$^{+0.04}_{-0.04}$ & 311.0$^{+38.6}_{-32.9}$ & 1.90 \\
76 - 103 & -1.06$^{+0.08}_{-0.08}$ & -2.65$^{+0.19}_{-0.34}$ & 125.7$^{+23.27}_{-19.26}$ & 1.23 & 1.15$^{+0.06}_{-0.06}$ & 157.7$^{+22.2}_{-18.6}$ & 1.39 \\          
103 - 150 & -1.50$^{+0.20}_{-0.18}$ & -2.30 $\pm$ 0.10 & 101.1$^{+58.3}_{-30.5}$ & 1.07  & 1.50$^{+0.18}_{-0.20}$ & 102.8$^{+56.8}_{-30.4}$ & 1.06\\
\hline
\end{tabular}
\end{table*}

We turn now to the XRT which started to observe GRB 090618 $\sim$ 120 s after the BAT trigger. Its early data show a continued activity of the prompt emission, fading away $\sim$ 200 s after the BAT trigger time. Then the light curve is well approximated with a power-law decay. 
In view of the lack of soft X-ray data before the onset of the XRT, we cannot exclude a previous pulse in the X-ray light curve emission of GRB 090618.
The following shallow and late decay phases, well-known in literature \citep{Sarietal1999, Nousek2006}, will not be analyzed in this paper since we focus in the first 200 s of the GRB emission.

\section{A brief review of the fireshell and alternative models}\label{sec:3}

\subsection{The GRB prompt emission in the fireball scenario}\label{sec:3.0}

A variety of models have been developed to theoretically explain the observational properties of GRBs. 
One of the most quoted is the fireball model (see for a review \citet{Piran2005}).
The model was first proposed by \citet{CavalloRees}, \citet{Goodman1986} and \citet{Paczynski1986}, who have shown that the sudden release of a large quantity of energy in a compact region can lead to an optically thick photon-lepton plasma and to the production of $e^+e^-$ pairs.
The total annihilation of the $e^+e^-$ plasma was assumed, leading to a vast release of energy pushing on the CBM: the ``fireball''.

An alternative approach, originating in the gravitational collapse to a black hole, is the fireshell model (see for a review \citet{PhysRep} and \citep{Ruffini2011}).
There the GRBs originate from an optically thick electron--positron plasma in thermal equilibrium, having a total energy of $E_{tot}^{e^\pm}$.
Such plasma is initially confined between the radius of a black hole $r_h$ and the dyadosphere radius 
\begin{equation}\label{eq:rh}
r_{ds}= r_h \left[2 \alpha \frac{E_{tot}^{e^+e^-}}{m_e c^2}\left(\frac{\hbar /m_e c}{r_h}\right)^3\right]^{1/4},
\end{equation}
where, $\alpha$ is the usual fine structure constant, $\hbar$ and $c$ the Planck constant and the speed of light, and $m_e$ the mass of the electron.
The lower limit of $E_{tot}^{e^\pm}$ coincides with $E_{iso}$. 
The condition of thermal equilibrium assumed in this model as shown by \citet{2007PhRvL..99l5003A}, differentiates this approach from the alternative ones \citep[e.g. the one by][]{CavalloRees}, see Section 3.2.

In the fireball model, the prompt emission, including the sharp luminosity variations \citep{Ramirez2000} are due to the prolonged and variable activity of the ``inner engine'' \citep{ReesMeszaros1994, Piran2005}.
The conversion of the fireball energy to radiation originates by shocks, either internal (when faster moving matter takes over a slower moving shell, see \citet{ReesMeszaros1994}) or external (when the moving matter is slowed down by the external medium surrounding the burst, see \citet{ReesMeszaros1992}).
Much attention has been given to the Synchrotron emission from relativistic electrons, possibly accompanied by SSC emission to explain the observed GRB spectrum. 
These processes were found to be consistent with the observational data of many GRBs \citep{Tavani1996, Frontera2000}. 
However, several limitations have been evidenced in relation with the low energy spectral slopes of time-integrated spectra \citep{Crider1997, Preece2002, Ghirlanda2002, Ghirlanda2003, Daigne2009} and time-resolved
spectra \citep{Ghirlanda2003}.
Additional limitations on SSC have also been pointed out by \citet{Kumar2008} and \citet{Piran2009}.

The latest phases of the afterglow are described in the fireball model by assuming an equation of motion given by the  Blandford-McKee self-similar power-law solution \citep{BlandfordMcKee}. 
The maximum Lorentz factor of the fireball is estimated from the temporal occurrence of the peak of the optical emission, which is identified with the peak of the forward external shock emission \citep{Molinari2007, Rykoff2009} in the thin shell
approximation \citep{SariPiran1999}.
There have been developed partly alternative and/or complementary scenarios to the fireball model, e.g. the ones based on: quasi-thermal Comptonization \citep{Ghisellini1999}, Compton drag emission \citep{Zdziarski1991, Shemi1994}, Synchrotron emission from a decaying magnetic field \citep{Peer2006b}, jitter radiation \citep{Medvedev2000}, Compton scattering
of synchrotron self absorbed photons \citep{Panaitescu2000, Stern2004}, photospheric emission \citep{Eichler2000, MeszarosRees2000, Meszaros2002, Daigne2002, Giannios2006, Ryde2009, Lazzati2010}. 
In particular, \citet{Ryde2009} pointed out that the photospheric emission overcomes some of the difficulties of pure
non-thermal emission models.

\subsection{The fireshell scenario}\label{sec:3.1}

\begin{figure}
\includegraphics[width=\hsize]{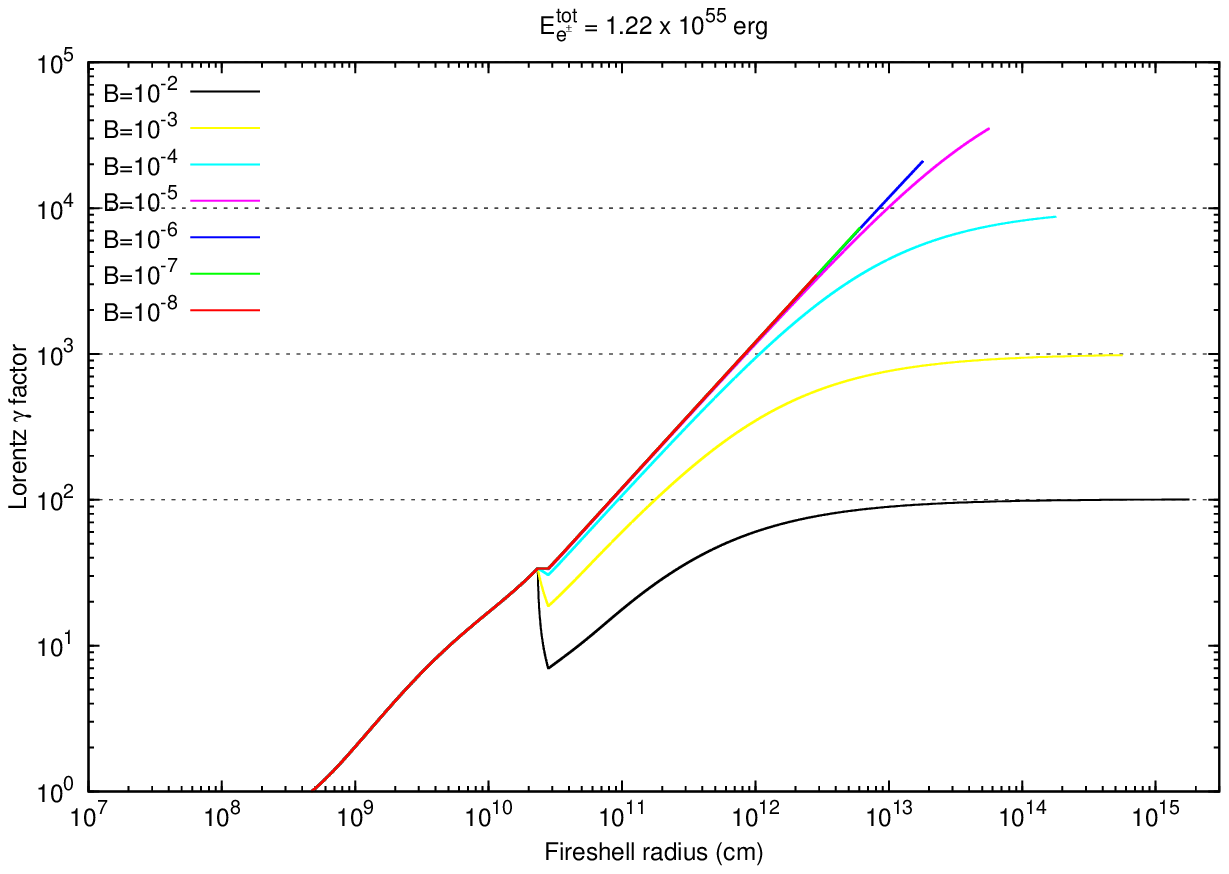}
\includegraphics[width=\hsize]{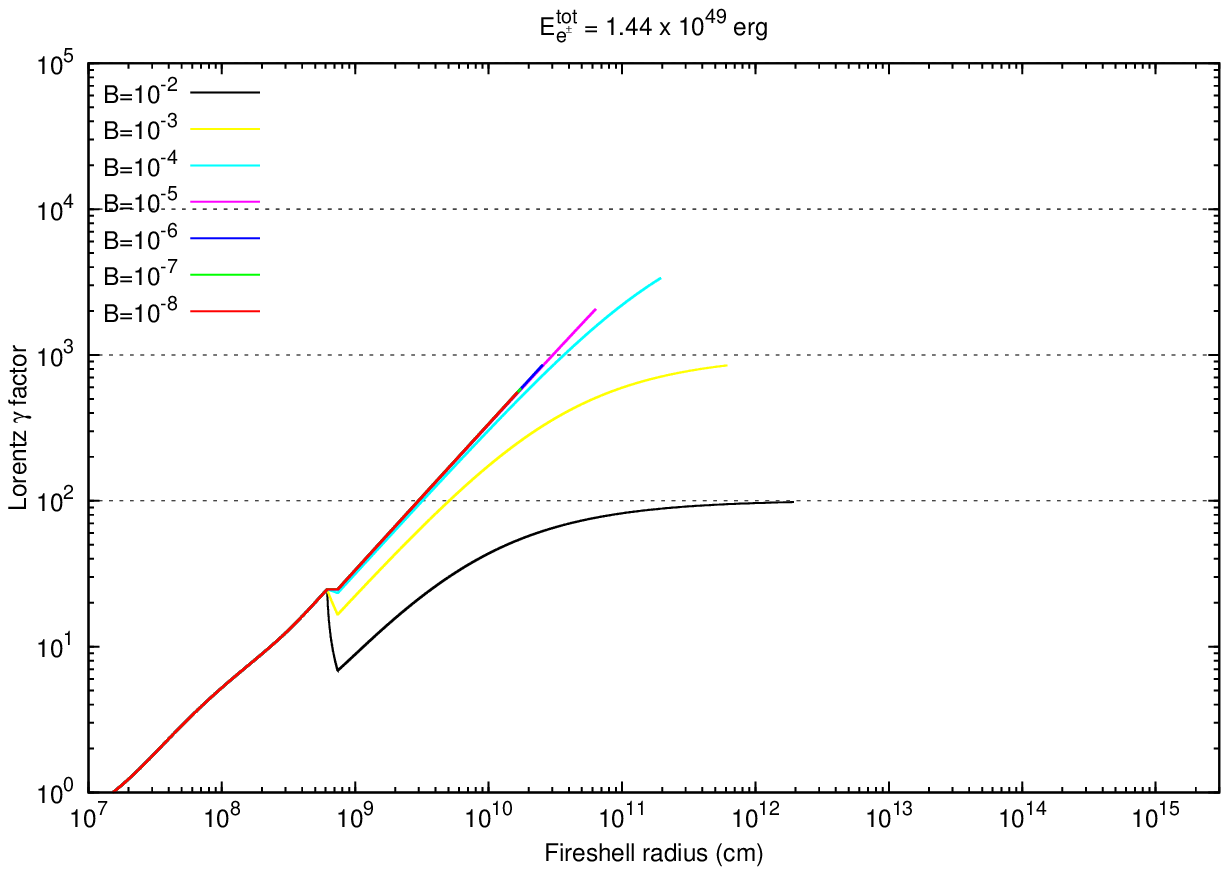}
\caption{The evolution of the Lorentz $\Gamma$ factor until the transparency emission, for a GRB of a fixed $E_{tot}^{e^+e^-}$ = 1.22 $\times$ 10$^{55}$ (upper panel),and $E_{tot}^{e^+e^-}$ = 1.44 $\times$ 10$^{49}$, for different values of the baryon load $B$. This computation refers to a mass of the black hole of 10 M$_{\sun}$ and a $\tau$ = $\int_R dr (n_{e^{\pm}} + n_{e^-}^b) \sigma_T = 0.67$, where $\sigma_T$ is the Thomson cross-section and the integration is over the thickness of the fireshell \citep{RSWX2}.}
\label{fig:no4g}
\end{figure}

In the fireshell model, the rate equation for the $e^+e^-$ pairs and its dynamics have been given by \citet{RSWX} (the pair-electromagnetic pulse or PEM pulse for short). 
This plasma engulfs the baryonic material left over in the process of gravitational collapse having mass $M_B$, still keeping thermal equilibrium between electrons, positrons and baryons. 
The baryon load is measured by the dimensionless parameter $B=M_B c^2/E_{tot}^{e^+e^-}$. 
It was shown \citep{RSWX2} that no relativistic expansion of the plasma can be found for $B > 10^{-2}$. The fireshell is still optically thick and self-accelerates to ultrarelativistic velocities \citep[the pair-electromagnetic-baryonic pulse or PEMB pulse for short,][]{RSWX2}. 
Then the fireshell becomes transparent and the \mbox{Proper - GRB} (P-GRB) is emitted \citep{Ruffini2001}. 
The final Lorentz gamma factor at transparency can vary in a vast range between $10^2$ and $10^3$ as a function of $E_{tot}^{e^+ e^-}$ and $B$, see Fig. \ref{fig:no4g}.
For the final determination it is necessary to integrate explicitly the rate equation of the $e^+ e^-$ annihilation process and evaluate, for a given black hole mass and a given $e^+e^-$ plasma radius, the reaching of the transparency condition \cite{RSWX}, see Fig. \ref{fig:no4}. 

The fireshell scenario does not require any prolonged activity of the inner engine.
After transparency, the remaining accelerated baryonic matter still expands ballistically and starts to slow down by the collisions with the CBM, having average density $n_{cbm}$. In the standard fireball scenario \citep{Meszaros2006}, the spiky light curve is assumed to be caused by internal shocks.
In the fireshell model the entire extended-afterglow emission is assumed to originate from an expanding thin shell enforcing energy and momentum conservation in the collision with the CBM. 
The condition of a fully radiative regime is assumed \citep{Ruffini2001}.
This, in turn, allows to estimate the characteristic inhomogeneities of the CBM, as well as its average value.

It is appropriate to recall a further difference between our treatment and the ones in the current literature. 
The complete analytic solution of the equations of motion of the baryonic shell has been developed \citep{Bianco2004,Bianco2005a}, while in the current literature usually the Blandford - McKee \citep{BlandfordMcKee} self-similar solution has been uncritically adopted \citep[e.g.][]{1993ApJ...415..181M,Sari1997,Sari1998,Waxman1997,ReesMeszaros1998,Granot1997,Panaitescu1998,1999ApJ...511..852G,Van2000,Meszaros2002}. 
The analogies and differences between the two approaches have been explicitly pointed out in \citet{Bianco2005b}.

From this general approach, a canonical GRB bolometric light curve composed of two different parts is defined: 
the P-GRB and the extended-afterglow. 
The relative energetics of these two components, the observed temporal separation between the corresponding peaks, is a function of the above three parameters $E_{tot}^{e^+e^-}$, $B$, and the average value of the $n_{cbm}$; the first two parameters are inherent to the accelerator characterizing the GRB, i.e., the optically thick phase, while the third one is inherent to the GRB surrounding environment which gives rise to the extended-afterglow. 
If one goes to the observational properties of this model of a relativistic expanding shell, a crucial concept has been the introduction of the EQTS. In this topic, also, our model differs from the ones in the literature for deriving the analytic expression of the EQTS from the analytic solutions of the equations of motion \citep{Bianco2005b}.

In this paper, we assume $E_{tot}^{e^+e^-} = E_{iso}$.
This assumption is based on the very accurate information we have on the luminosity and the spectral properties of the source.
In other GRBs, we have assumed $E_{tot}^{e^+e^-} > E_{iso}$ to take into account the observational limitations, due to detector thresholds, distance effects and lack of data.

\subsection{The emission of the P-GRB}\label{sec:3.2}

\begin{figure}
\centering
(a)\\
\includegraphics[width=0.80\hsize,clip]{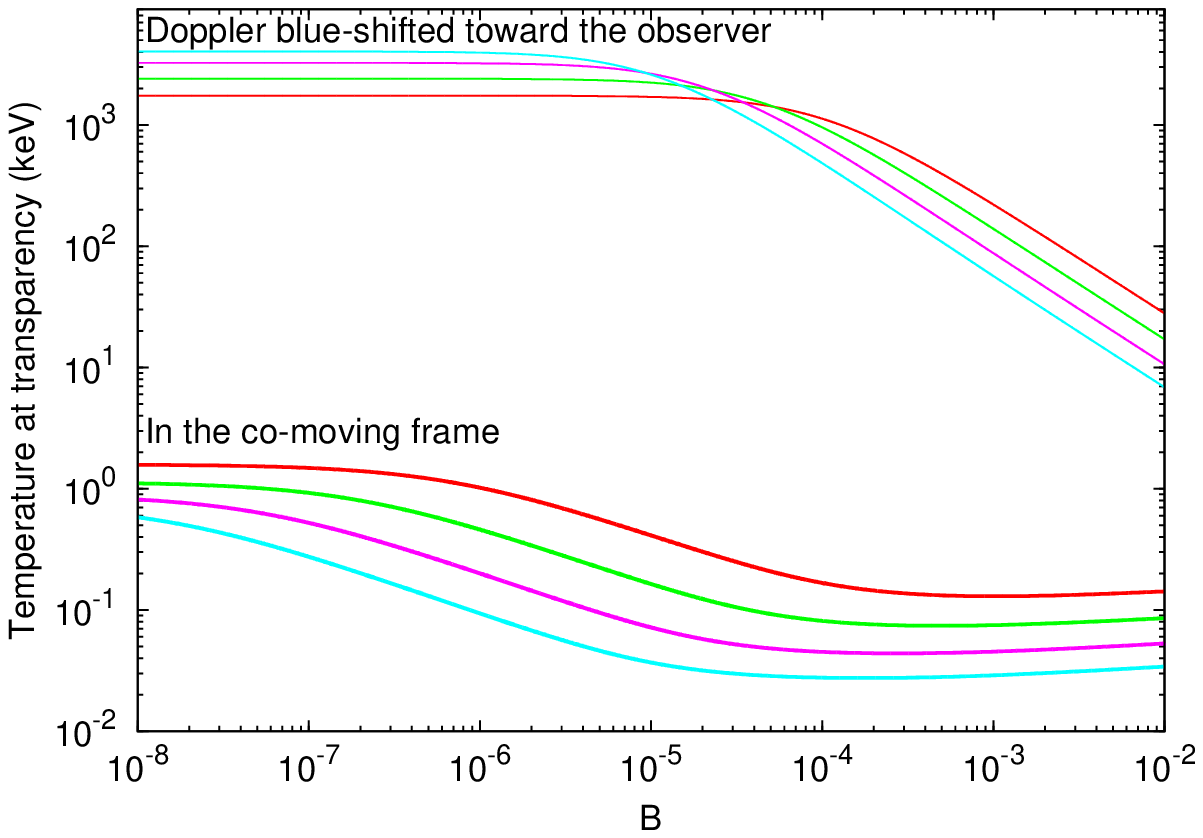}\\
(b)\\
\includegraphics[width=0.80\hsize,clip]{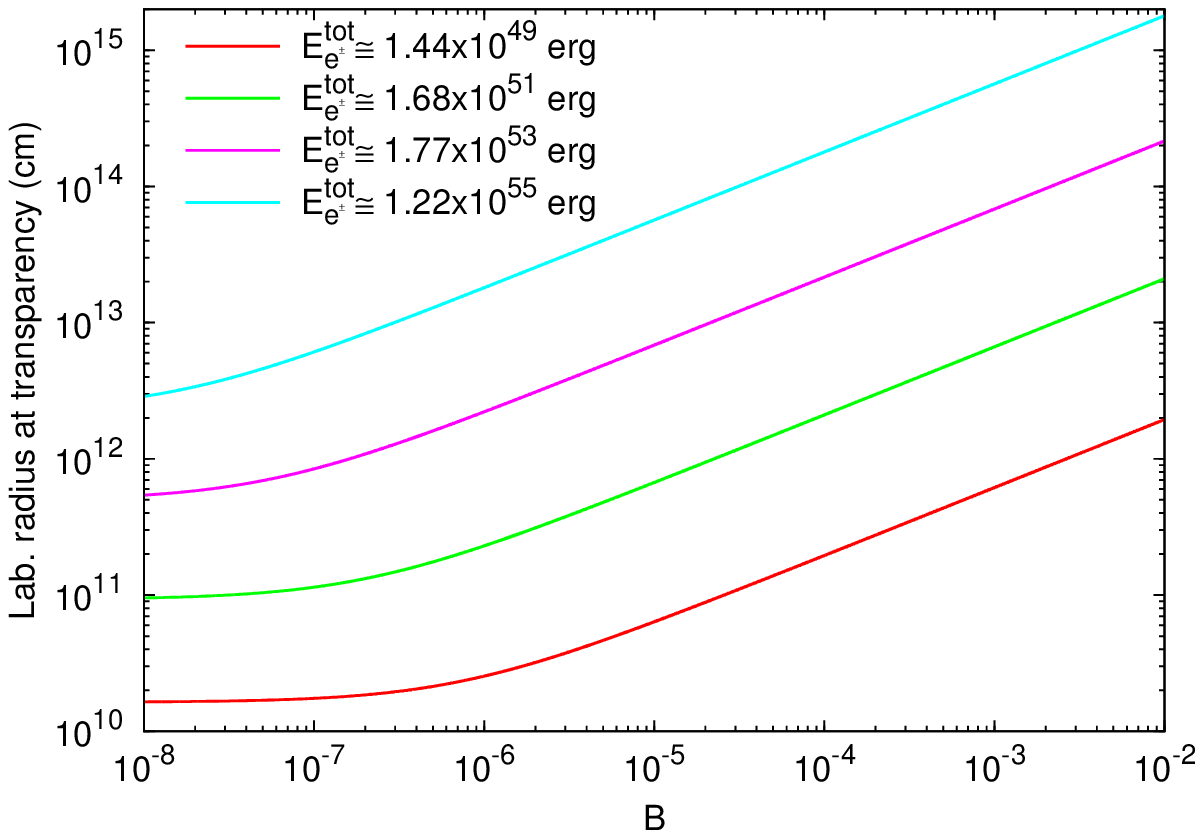}\\
(c)\\
\includegraphics[width=0.80\hsize,clip]{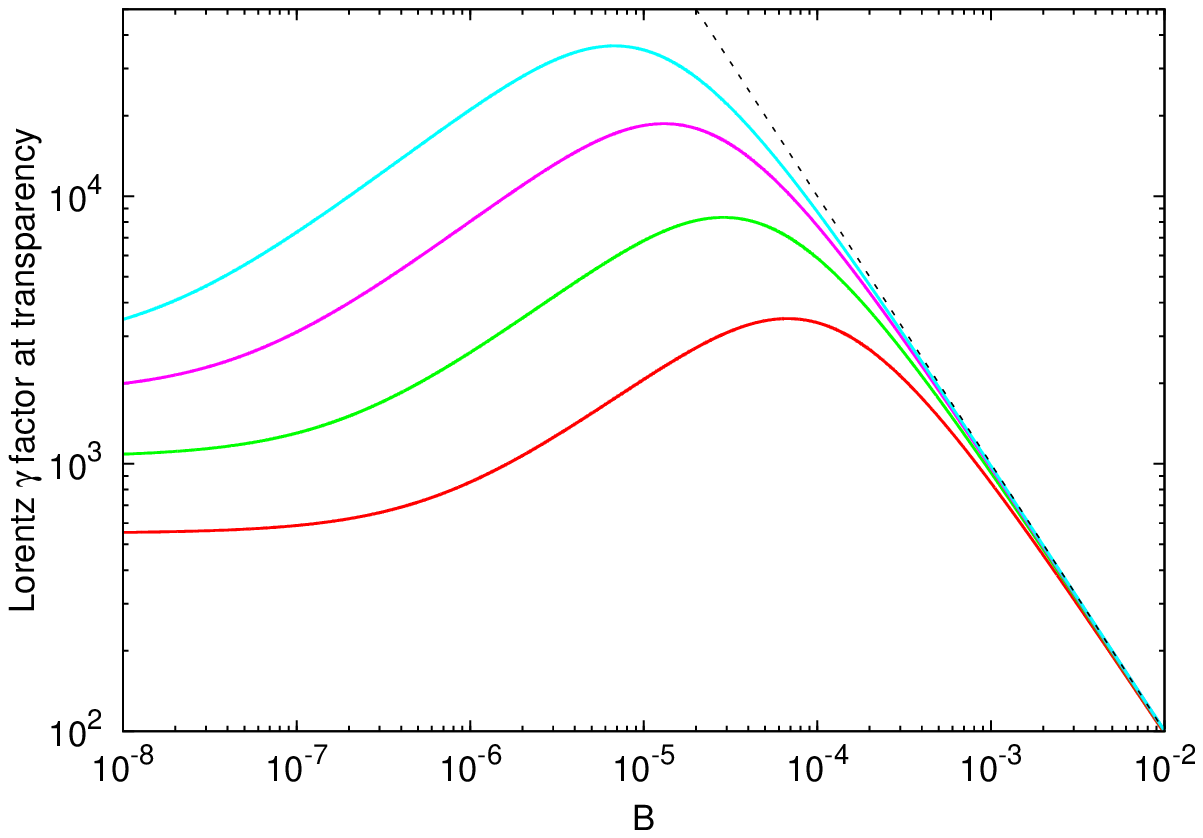}\\
(d)\\
\includegraphics[width=0.80\hsize,clip]{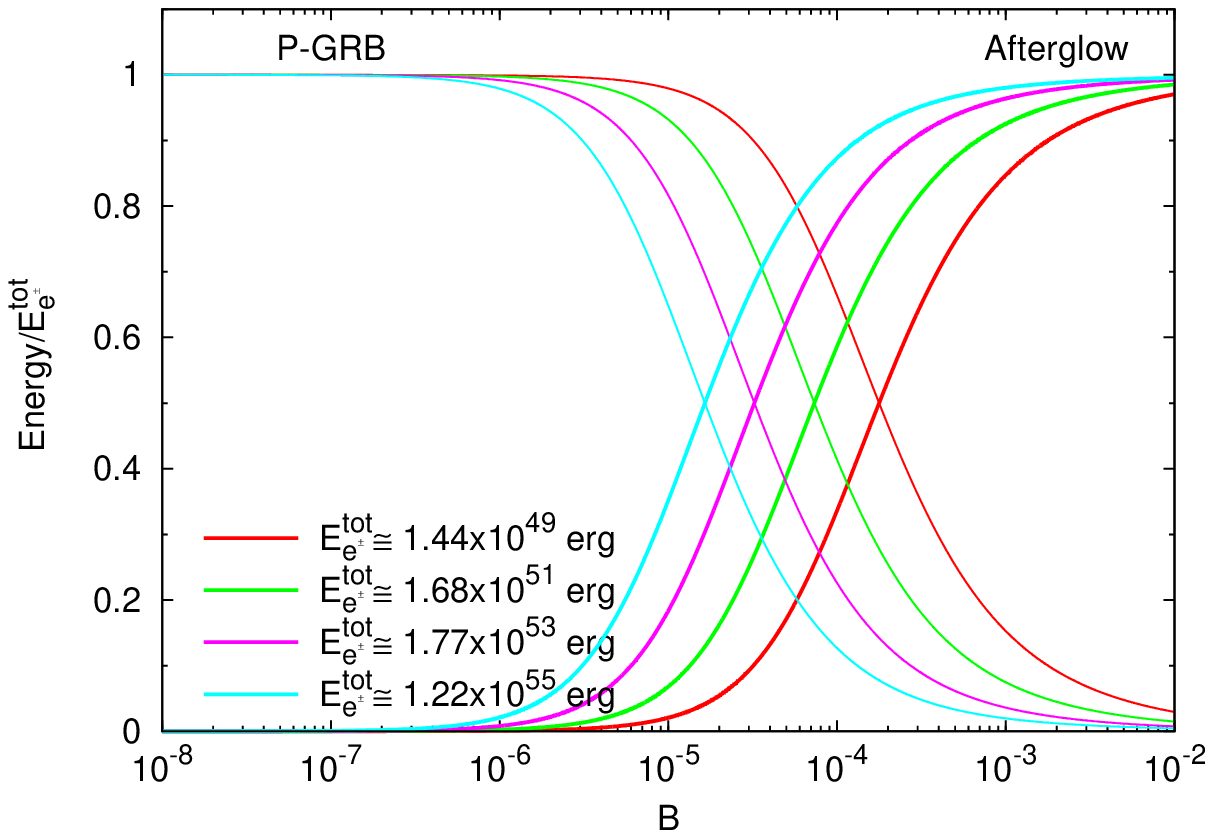}\\
\caption{The fireshell temperature in the comoving and observer frame and the laboratory radius at the transparency emission (panels (a) and (b)), the Lorentz $\Gamma$ factor at the transparency (panel (c)) and the energy radiated in the P-GRB and in the afterglow in units of $E_{tot}^{e^+e^-}$ (panel (d)) as a function of the baryon load $B$, for 4 different values of $E_{tot}^{e^+e^-}$.}
\label{fig:no4}
\end{figure}

The lower limit of $E_{tot}^{e^+e^-}$ is given by the observed isotropic energy emitted in the GRB, $E_{iso}$. 
The identification of the energy of the afterglow and of the P-GRB determines the baryon load $B$ and, from these, it is possible to determine, see Fig. \ref{fig:no4}: the value of the Lorentz $\Gamma$ factor at transparency, the observed temperature as well as the temperature in the comoving frame and the laboratory radius at transparency.
We can determine indeed from the spectral analysis of the P-GRB candidate, the temperature $kT_{obs}$ and the energy emitted in the transparency $E_{PGRB}$.
The relation between these parameters can not be expressed by an analytical formulation: they can be only obtained by a numerical integration of the entire fireshell equations of motion.
In practice we need to perform a trial and error procedure to find the set of values which fit the observations.

As we are going to see in the case of GRB 090618, the direct measure of the temperature of the thermal component at the transparency offers a very important new information in the determination of the GRB parameters.
In the emission of the P-GRB two different phases are present: 
one corresponding to the emission of the photons when the transparency is reached, 
and the second is the early interaction of the ultra-relativistic protons and electrons with the CBM.
A spectral energy distribution with a thermal component and a non-thermal one should be expected to occur.

\subsection{The extended-afterglow}\label{sec:3.3}

The majority of works in the current literature has addressed the analysis of the afterglow emission as due to various combinations of Synchrotron and Inverse Compton processes, see e.g. \citet{Piran2005}.
It appears, however, that this description is not fully satisfactory \citep[see e.g.][]{Ghirlanda2003,KumarMcMahon2008,Piran2009}.

We have adopted in the fireshell model a pragmatic approach by making the full use of the knowledge of the equations of motion, of the EQTS formulations \citep{Bianco2005a} as well as of the correct relativistic transformations between the comoving frame of the fireshell and the observer frame.
These equations, that relate the four time variables, are necessary for the interpretation of the GRB data.
They are: a) the comoving time, b) the laboratory time, c) the arrival time, and d) the arrival time at the detector corrected by the cosmological effects.
This is the content of the Relative Space-Time Transformations paradigm, essential for the interpretation of GRBs data \citep{Ruffini2001L107}.
Such a paradigm made it a necessity to have a global, instead of a piecewise, description of a GRB phenomenon \citep{Ruffini2001L107}.
This global description led to a new interpretation of the burst structure paradigm \citep{Ruffini2001}. 
As recalled in the Introduction, a new conclusion, arising from the burst structure paradigm, has been that the emission by the accelerated baryons interacting with the CBM is indeed occurring already in the prompt emission phase, just after the P-GRB emission.
This is the extended-afterglow emission, which presents in its ``light curve'' a rising part, a peak, and a decaying tail.
Following this paradigm, the prompt emission phase is therefore composed by both the P-GRB emission and the peak of the extended-afterglow.

To evaluate the extended-afterglow spectral properties, we have adopted an ansatz on the spectral properties of the emission in the collisions between the baryons and the CBM in the comoving frame. We have then evaluate all the observational properties in the observer frame by integrating on the EQTS.
The initial ansatz of thermal spectrum \citep{Ruffini2001}, has been recently modified to
\begin{equation}\label{eq:no2}
\frac{d N_{\gamma}}{d V d \epsilon} = \left(\frac{8 \pi}{h^3 c^3}\right)\left(\frac{\epsilon}{k_B T}\right)^{\alpha}\frac{\epsilon^2}{exp\left(\frac{\epsilon}{k_B T}\right)-1},
\end{equation}
where $\alpha$ is a phenomenological parameter defined in the comoving frame of the fireshell \citep{Patricelli2011}, determined by the optimization of the simulation of the observed data.
It is well known that in the ultrarelativistic collision of protons and electrons with the CBM, collective processes of ultrarelativistic plasma physics are expected, not yet fully explored and understood (e.g. Weibel instability, see \citet{MedvedevLoeb}).
Promising results along this line have been already obtained by \citet{Spitkovsky2008} and \citet{MedvedevSpitkovsky2009}, and may lead to the understanding of the physycal origin of the $\alpha$ parameter in Eq. \ref{eq:no2}.

In order to take into due account the filamentary, clumpy and porous structure of the CBM, we have introduced the additional parameter $\mathcal{R}$, which describes the fireshell surface filling factor.
It is defined as the ratio between the effective emitting area of the fireshell $A_{eff}$ and its total visible area $A_{vis}$ \citep{Ruffini2002, Ruffini2005}.
 
One of the main features of the GRB afterglow has been the observation of hard to soft spectral variation, which is generally absent in the first spike-like emission, which we have identified as the P-GRB, \citet{Bernardini2007,Caito2009,Caito2010,deBarros2011}.
An explanation of the hard-to-soft spectral variation has been advanced on the ground 
of two different contributions: the curvature effect and the intrinsic spectral evolution.
In particular, in the work of \citet{Peng2011} the authors use the model developed in \citet{Qin2002} for the spectral lag analysis, taking into account an intrinsic Band model for the GRBs and a Gaussian profile for the GRB pulses, in order to take into account the angular effects, and they find that both causes provide a very good explanation for the observed time lags.
Within the fireshell model we can indeed explain a hard-to-soft spectral variation very naturally, in the extended-afterglow emission.
Since the Lorentz $\Gamma$ factor decreases with time, the observed effective temperature of the fireshell will drop as the emission goes on, so the peak of the emission will occur at lower energies. 
This effect is amplified by the presence of the curvature effect, which has origin in the EQTS concept.
Both these observed features are considered as the responsible for the time lag observed in GRBs.

\subsection{The simulation of a GRB light curve and spectra of the extended-afterglow}\label{sec:3.4}

The simulation of a GRB light curve and the respective spectrum requires also the determination of the filling factor $\mathcal{R}$ and of the CBM density $n_{CBM}$.
These extra parameters are extrinsic and they are just functions of the radial coordinate from the source.
The parameter $\mathcal{R}$, in particular, determines the effective temperature in the comoving frame and the corresponding peak energy of the spectrum, while $n_{cbm}$ determines the temporal behavior of the light curve.
It is found that the CBM is typically formed of ``clumps'' of width $\sim 10^{15-16}$ cm and average density contrast $10^{-1} \lesssim \langle \delta n/n \rangle \lesssim 10$ centered on the value of 4 $particles/cm^3$, see Fig. \ref{fig:rad}, and clumps of masses $M_{clump} \approx 10^{22-24}$ g. Particularly important is the determination of the average value of $n_{cbm}$. Values of the order of $0.1$-$10$ particles/cm$^3$ have been found for GRBs exploding inside star forming region galaxies, while values of the order of $10^{-3}$ particles/cm$^3$ have been found for GRBs exploding in galactic halos \citep{Bernardini2007,Caito2009,deBarros2011}.
The presence of such a clumpy medium, already predicted in pioneering works of Fermi in the theoretical study of interstellar matter in our galaxy \citep{Fermi1949,Fermi1954}, is by now well-established both from the GRB observations and by additional astrophysical observations, see e.g. the circum-burst medium observed in novae \citep{Shara1997}, or by theoretical considerations on supergiant, massive stars, clumpy wind \citep{Ducci2009}.
Interesting are the considerations by Arnett and Meakin \citep{Arnett2011}, who have shown how realistic 2D simulations of the late evolution of a core collapse show processes of violent emission of clouds: there the 2D simulations differ from the one in 1D, which show a much more regular and wind behavior around the collapsing core. Consequently, attention should be given also to instabilities prior to the latest phases of the evolution of the core, possibly giving origin to the cloud pattern observed in the CBM of GRB phenomenon (D.Arnett private communication).  

The determination of the $\mathcal{R}$ and $n_{CBM}$ parameters depends essentially on the reproduction of the shape of the extended-afterglow and of the respective spectral emission, in a fixed energy range.
Clearly, the simulation of a source within the fireshell model is much more complex than simply fitting the $N(E)$ spectrum with phenomenological analytic formulas for a finite temporal range of the data.
It is a consistent picture, which has to find the best value for the parameters of the source, the P-GRB \citep{Ruffini2001}, its spectrum, its temporal structure, as well as its energetics. 
For each spike in the light curve are computed the parameters of the corresponding CBM clumps, taking into account all the thousands of convolutions of comoving spectra over each EQTS leading to the observed spectrum \citep{Bianco2005a,Bianco2005b}. 
It is clear that, since the EQTS encompass emission processes occurring at different comoving times weighted by their Lorentz and Doppler factors, the ``fitting'' of a single spike is not only a function of the properties of the specific CBM clump but of the entire previous history of the source. 
Any mistake at any step of the simulation process affects the entire evolution that follows and, conversely,
at any step a fit must be made consistently with all the previous history: due to the non-linearity of the system and to the EQTS, any change in the simulation produces observable effects up to a much later time. 
This brings to an extremely complex procedure by trial and error in the data simulation, in which the variation of the parameters defining the source are further and further narrowed down, reaching very quickly the uniqueness. 
Of course, we cannot expect the latest parts of the simulation to be very accurate, since some of the basic hypothesis on the equations of motion, and possible fragmentation of the shell, can affect the procedure. 

In particular, the theoretical photon number spectrum to be compared with the observational data is obtained by an averaging procedure of instantaneous spectra. 
In turn, each instantaneous spectrum is linked to the simulation of the observed multiband light curves in the chosen time interval. 
Therefore, both the simulation of the spectrum and of the observed multiband light curves have to be performed together and simultaneously optimized.

\section{Spectral analysis of GRB 090618}\label{sec:4}

We proceed now to the detailed spectral analysis of GRB 090618. 
We divide the emission in six time intervals, shown in Table \ref{tab:no1}, each one identifying a significant feature in the emission process.
We then fit for each time interval the spectra by a Band model as well as by a blackbody with an extra power-law component, following \citet{Ryde2004}. 
In particular, we are interested in the estimation of the temperature $kT$ and the observed energy flux $\phi_{obs}$ of the blackbody component.
The specific intensity of emission of a thermal spectrum at energy $E$ in energy range $d E$ into solid angle $\Delta \Omega$ is 
\begin{equation}\label{eq:no2a}
I(E)dE = \frac{2}{h^3 c^2} \frac{E^3}{\exp(E/kT) - 1} \Delta \Omega dE.
\end{equation}
The source of radius $R$ is seen within a solid angle $\Delta \Omega = \pi R^2/D^2$, and its full luminosity is $L = 4 \pi R^2 \sigma T^4$. What we are fitting however is the background-subtracted photon spectra $A(E)$, which is obtained by dividing the specific intensity $I(E)$ by the energy $E$:
\begin{eqnarray}
A(E)dE \equiv \frac{I(E)}{E}dE &=& \frac{k^4 L}{2\sigma (kT)^4 D^2 h^3 c^2}\frac{E^2 d E}{\exp(E/kT)-1} \nonumber\\
&=& \frac{15 \phi_{obs}}{\pi^4 (kT)^4}\frac{E^2 d E}{\exp(E/kT)-1},
\label{eq:no2b}
\end{eqnarray}
where $h$, $k$ and $\sigma$ are respectively the Planck, the Boltzmann and the Stefan-Boltzmann constants, $c$ is the speed of light and $\phi_{obs} = L/(4 \pi D^2)$ is the observed energy flux of the blackbody emitter.
The great advantage of Eq. (\ref{eq:no2b}) is that it is written in terms of the observables $\phi_{obs}$ and $T$, so from a spectral fitting procedure we can obtain the values of these quantities for each time interval considered.
In order to determine these parameters, we must perform an integration of the actual photon spectrum $A(E)$ over the instrumental response $R(i,E)$ of the detector which observe the source, where $i$ denotes the different instrument energy channels. The result is a predicted count spectrum 
\begin{equation}
C_{p}(i) = \int_{E_{min}(i)}^{E_{max}(i)} A(E) R(i,E) dE,
\label{eq_int}
\end{equation}
where $E_{min}(i)$ and $E_{max}(i)$ are the boundaries of the $i$-th energy channel of the instrument. Eq. (\ref{eq_int}) must be compared with the observed data by a fit statistic.

\begin{figure}
\includegraphics[height=6cm,width=8cm]{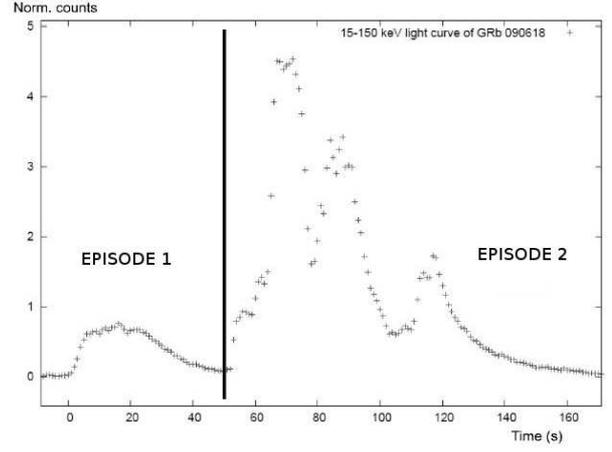}
\caption{The two episode nature of GRB 090618.}\label{fig:cospar}
\end{figure}

\begin{figure*}
\begin{tabular}{c|c}
\includegraphics[height=6cm,width=8cm]{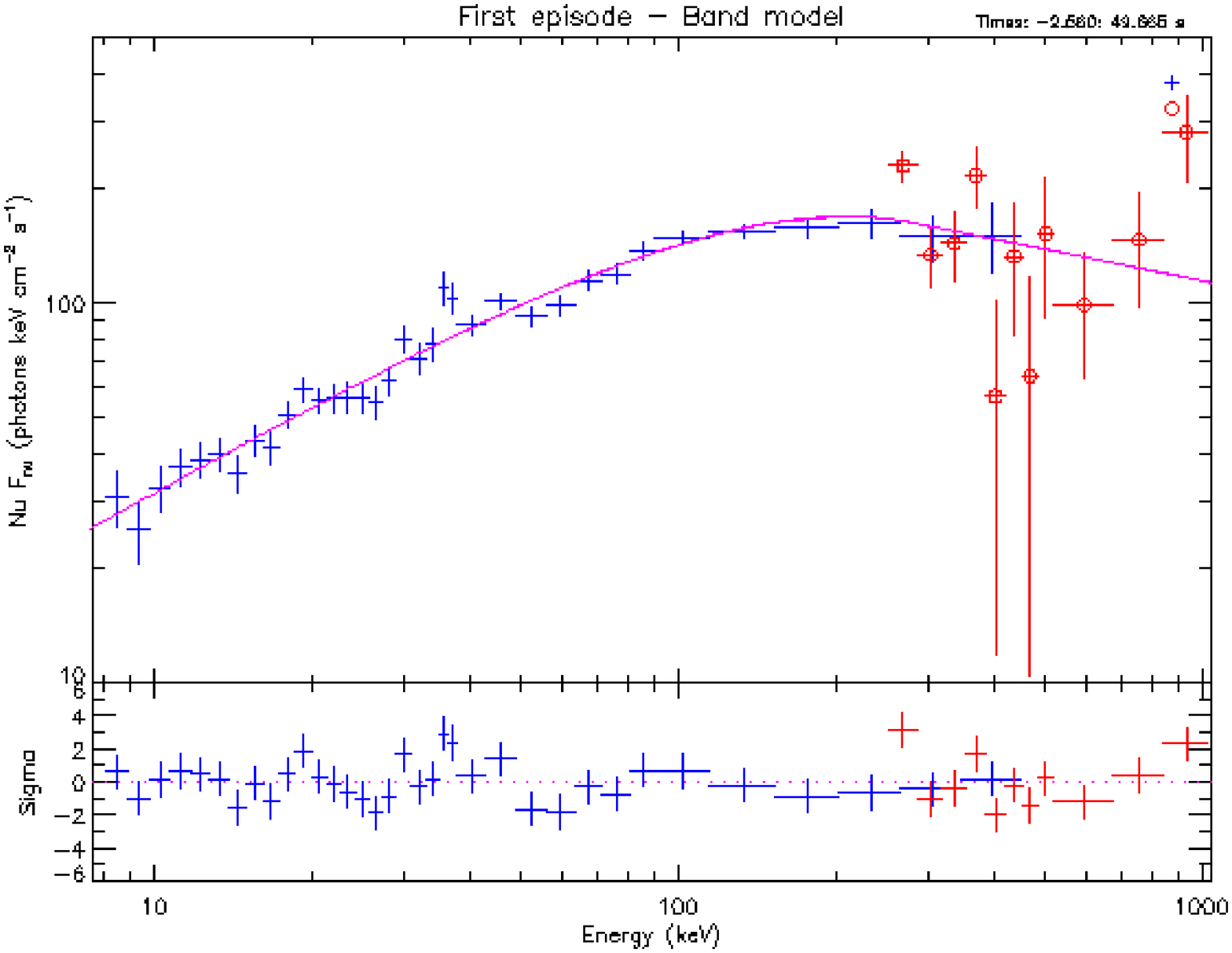}
\includegraphics[height=6cm,width=8cm]{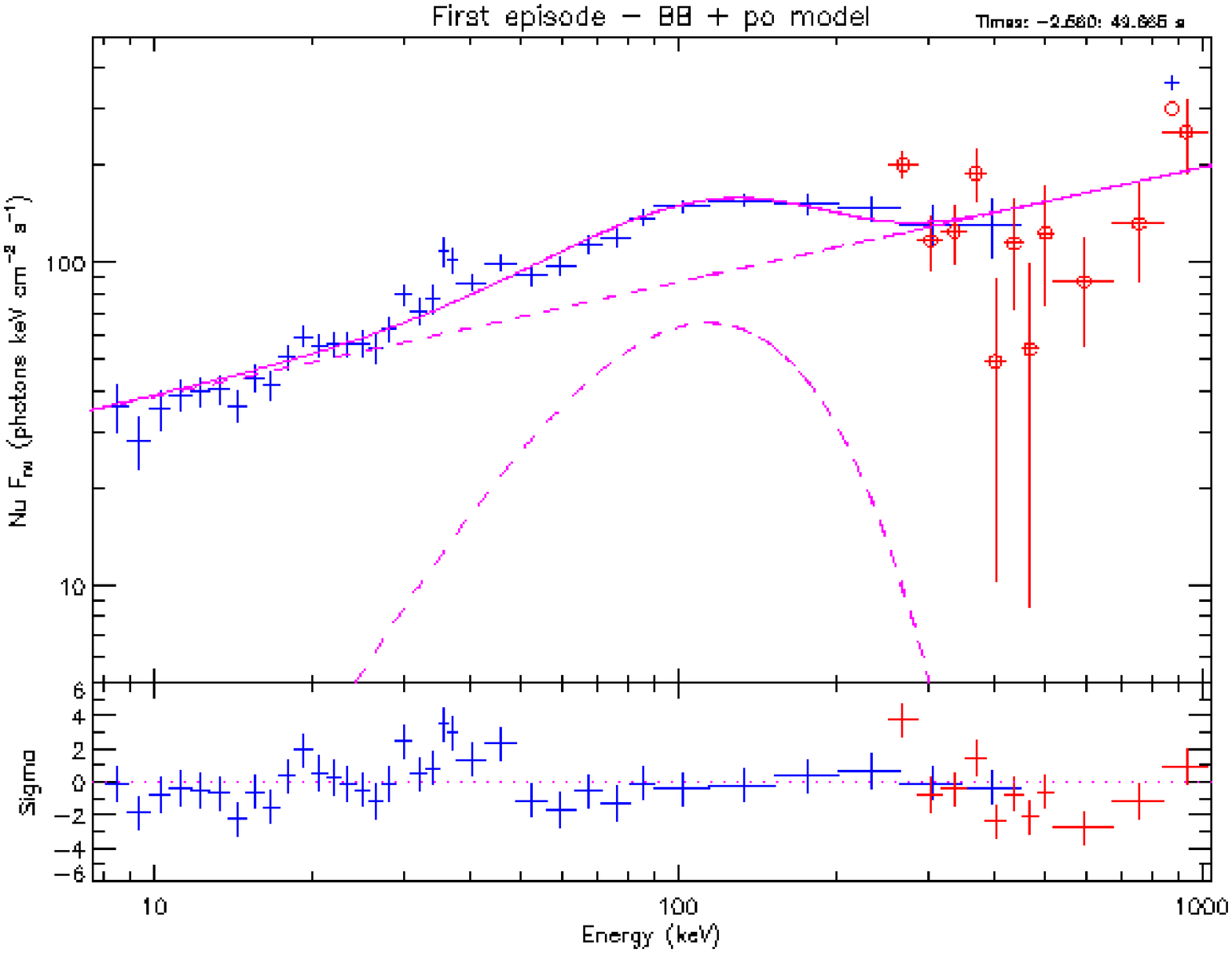}\\
\end{tabular}
\caption{Time-integrated spectra for the first episode (from 0 to 50 s) of GRB 090618 fitted with the Band, $\tilde{\chi}^2$ = 1.12 (left) and blackbody + power-law (right) models, $\tilde{\chi}^2$ = 1.28. In the following we will consider the case of a blackbody + power-law model and infer some physical consequences. The corresponding considerations in the case of the Band model are currently being considered and will be published elsewhere. }\label{fig:big1}
\end{figure*}

\begin{figure*}
\begin{tabular}{c|c}
\includegraphics[height=6cm,width=8cm]{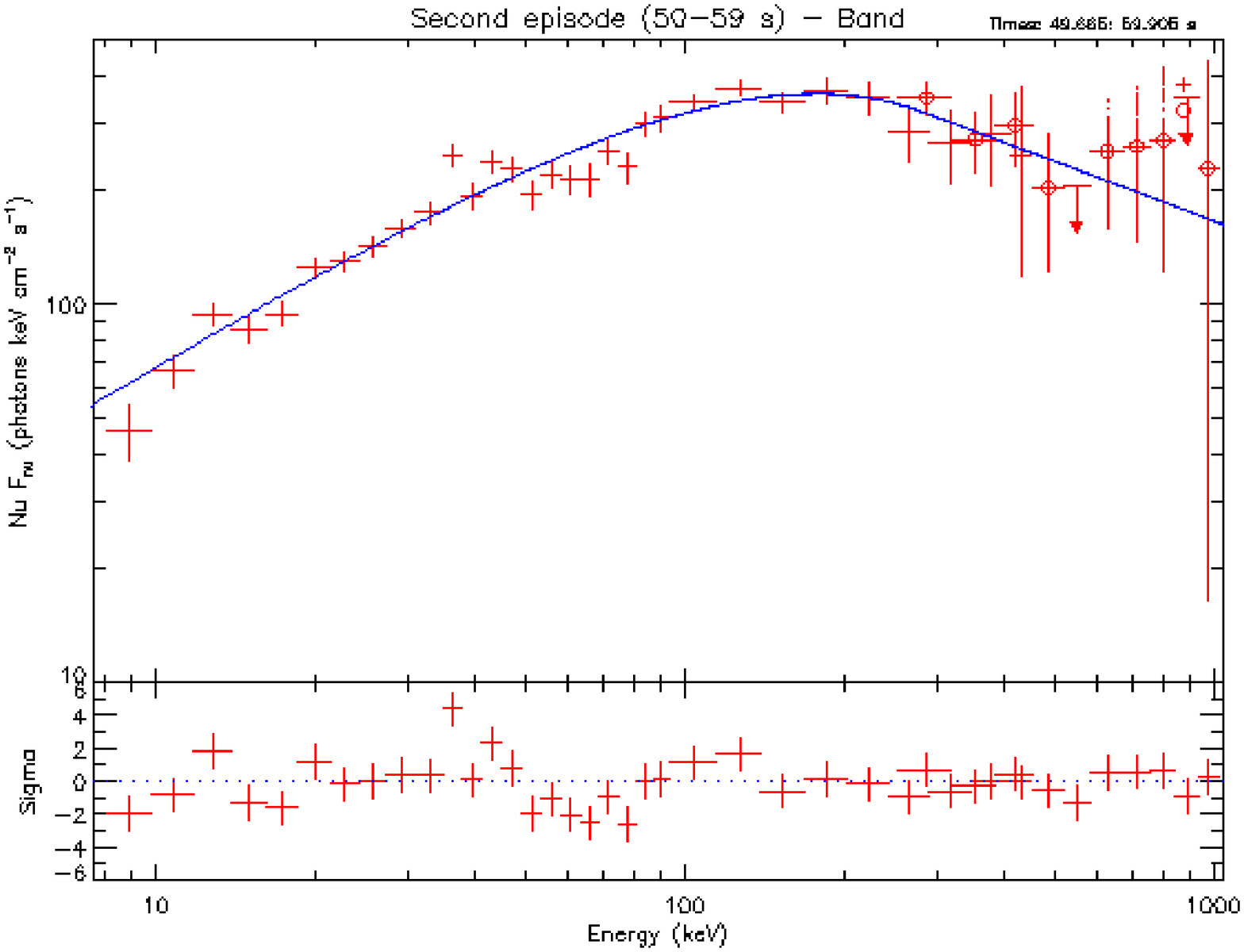}
\includegraphics[height=6cm,width=8cm]{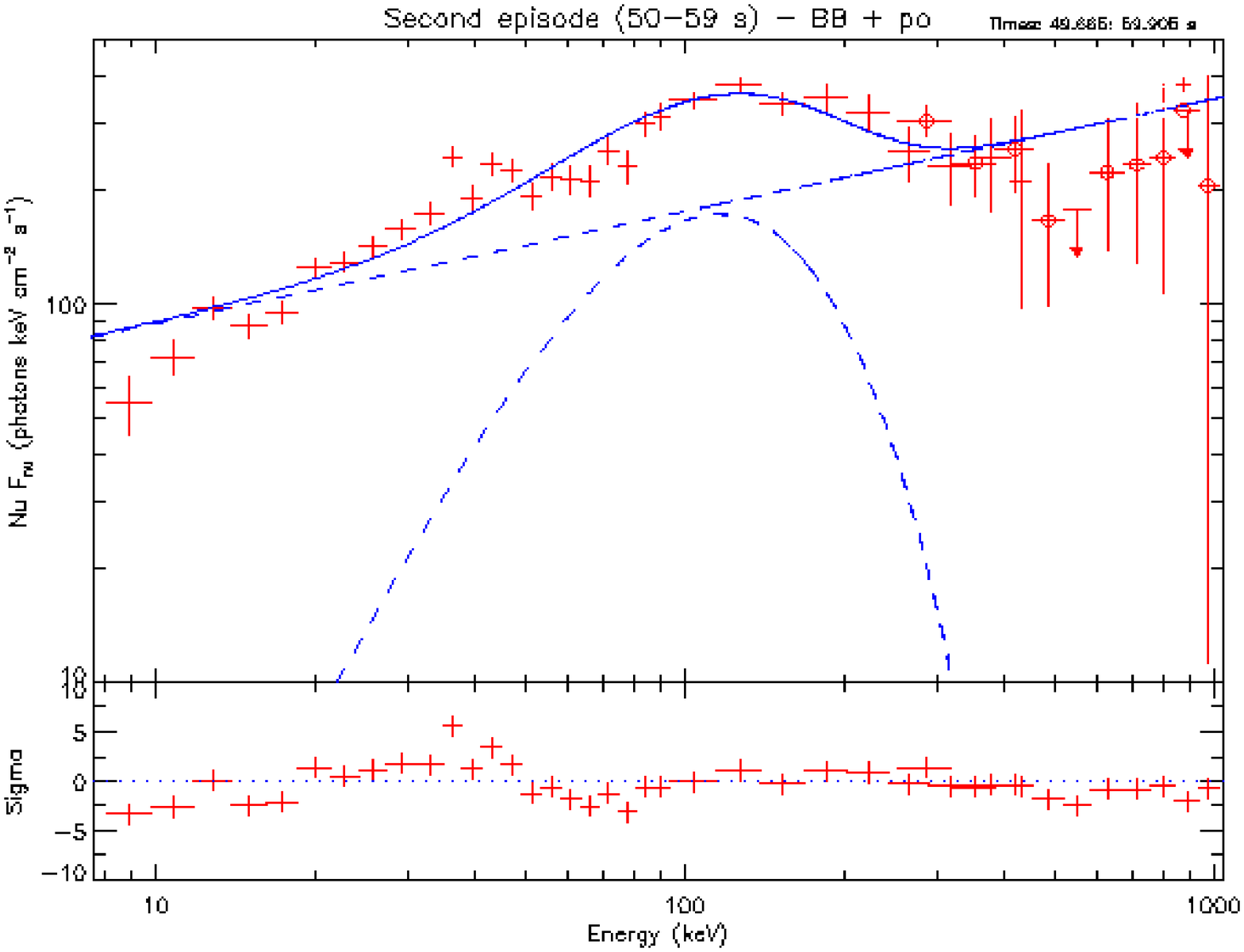}\\
\end{tabular}
\caption{Time-integrated spectra for the first 9 s of the second episode (from 50 to 59 s after the trigger time) of GRB 090618 fitted with the Band, $\tilde{\chi}^2$ = 1.23 (left) and blackbody + power-law (right) models, $\tilde{\chi}^2$ = 1.52. }\label{fig:big2}
\end{figure*}

The main parameters obtained from the fitting procedure are shown in Table \ref{tab:no2b}.
We divide the entire GRB in two main episodes, as advanced in \citet{TEXAS}: one lasting the first 50 s and the other from 50 to 151 s after the GRB trigger time, see Fig. \ref{fig:cospar}. 
It is easy to see that the first 50 s of emission, corresponding to the first episode, are well fitted by a Band model as well as a black-body with an extra power-law model, Fig. \ref{fig:big1}. 
The same happens for the first 9 s of the second episode (from 50 to 59 s after the trigger time), Fig. \ref{fig:big2}. 
For the subsequent three intervals corresponding to the main peaks in the light curve, the black-body plus a power-law model does not provide a satisfactory fit. 
Only the Band model fits the spectrum with good accuracy, with the exception of the first main spike (compare the values of $\chi^2$ in the table). 
We find also that the last peak can be fitted by a simple power-law model with a photon index $\gamma$ = 2.20 $\pm$ 0.03, better than by a Band model.

The result of this analysis points to a different emission mechanism in the first 50 s of GRB 090618 and in the following 9 s. A sequence of very large pulses follow, which spectral energy distribution is not attributable either to a blackbody or a blackbody and an extra power-law component.
The evidence for the transition is well represented by the test of the data fitting, whose indicator is given by the changing of the $\tilde{\chi^2}$ ($N_{dof} = 169$) for the blackbody plus a power-law model for the different time intervals, see table \ref{tab:no2b}.
Although the Band spectral model is an empirical model without a clear physical origin, we do check its validity in all of the time-detailed spectra with the sole exception of the first main pulse of the second episode.
The $\chi^2$ corresponding to the Band model for such a main pulse, although better than the one corresponding to the blackbody and power-law case, is unsatisfactory. 
We are now going to a direct application of the fireshell model in order to make more stringent the above conclusions and reach a better understanding of the source. 

\begin{table*}
\centering
\caption{Time-resolved spectral analysis (8 keV - 10 MeV) of the second episode in GRB 090618.} 
\label{tab:no2b} 
\begin{tabular}{l l c c c c c c c}
\hline\hline
 & Time Interval (s) & $\alpha$ & $\beta$ & $E_0(keV)$ & $\tilde{\chi}^2_{BAND}$ & $kT(keV)$ & $\gamma$ & $\tilde{\chi}^2_{BB+po} $\\ 
\hline 
A & 0 - 50 & -0.74 $\pm$ 0.10 & -2.32 $\pm$ 0.16 & 118.99 $\pm$ 21.71  & 1.12 & 32.07 $\pm$ 1.85  & 1.75 $\pm$ 0.04 & 1.21 \\
B & 50 - 59 & -1.07 $\pm$ 0.06 & -3.18 $\pm$ 0.97 & 195.01 $\pm$ 30.94  & 1.23 & 31.22 $\pm$ 1.49  & 1.78 $\pm$ 0.03 & 1.52 \\
C & 59 - 69 & -0.99 $\pm$ 0.02 & -2.60 $\pm$ 0.09 & 321.74 $\pm$ 14.60 & 2.09 & 47.29 $\pm$ 0.68  & 1.67 $\pm$ 0.08   & 7.05\\
D & 69 - 78 & -1.04 $\pm$ 0.03 & -2.42 $\pm$ 0.06 & 161.53 $\pm$ 11.64    &  1.55 & 29.29 $\pm$ 0.57   & 1.78 $\pm$ 0.01 & 3.05\\
E & 78 - 105 &  -1.06 $\pm$ 0.03 & -2.62 $\pm$ 0.09 & 124.51 $\pm$ 7.93    &  1.20 & 24.42 $\pm$ 0.43    &  1.86 $\pm$ 0.01  & 2.28\\     
F & 105 - 151 & -2.63 $\pm$ -1    & -2.06 $\pm$ 0.02 & unconstrained    &   1.74 & 16.24 $\pm$ 0.84  & 2.23 $\pm$ 0.05   & 1.15\\
\hline
\end{tabular}
\end{table*}

\section{Analysis of GRB 090618 in the fireshell scenario: from a single GRB to a multi-component GRB}\label{sec:5}

\subsection{Attempt for a single GRB scenario: the role of the first episode}

We first approach the analysis of GRB 090618 by assuming that we are in presence of a single GRB and attempt to identify its components in a canonical GRB scenario, based on the fireshell model.
We first attempt the identification of the P-GRB emission.
We have already seen that the integrated first 50 s can be well-fitted with a black-body at a temperature $kT$ = 32.07 $\pm$ 1.85 keV and an extra power-law component with the photon index $\gamma$ = -1.75 $\pm$ 0.04, see panel A in Fig. \ref{fig:big2} and Table \ref{tab:no2b}. 
Being the presence of a blackbody component the distinctive feature of the P-GRB, we have first attempted an interpretation of GRB 090618 as a single GRB with the first 50 s as the P-GRB \cite{COSPAR}.
We have first proceeded to evaluate if the energetics of the emission in the first 50 s can be interpreted as due to a P-GRB.
The energy emitted by the sole blackbody is $E_{BB}$ = 8.35$^{+0.27}_{-0.36}$ $\times$ 10$^{51}$ ergs.
Recalling that the isotropic energy of the entire GRB 090618 is $E_{iso}$ = (2.90 $\pm$ 0.02) $\times$ 10$^{53}$ ergs, we have that the blackbody component would be $\sim$ 2.9 $\%$ of the total energy emitted in the burst.
This would imply, see lower panel in fig. \ref{fig:no4}, a baryon load $B \sim 10^{-3}$ with a corresponding Lorentz $\Gamma$ factor of $\sim$ 800 and a temperature of $\sim$ 52 keV.
This value is in disagreement with the observed temperature $kT_{obs}$ = 32.07 keV.

One may attempt to reconcile the value of the theoretically predicted GRB temperature with the observed one by increasing $E_{tot}^{e^+e^-}$. 
This would lead to an $E_{tot}^{e^+e^-}$ = 4 $\times$ 10$^{54}$ ergs and a corresponding baryon load of $B \approx 10^{-4}$.
This would imply three major discrepancies:
a) there would be an unjustified complementary unobserved energy;
b) in view of the value of the baryon load, and the corresponding Lorentz $\Gamma$ factor, the duration of the extended-afterglow emission would be more than an order of magnitude smaller than the observed 100 s \citep{Bianco2008};
c) the duration of this first 50 s is much longer than the one typically expected for all P-GRBs identified in other GRBs \citep{Ruffini2007b}, which is at maximum of the order of $\sim$ 10 s.
We have therefore considered hopeless this approach and proceeded to a different one looking for multiple components. 

\subsection{The multi-component scenario: the second episode as an independent GRB}

\subsubsection{The identification of the P-GRB of the second episode}

\begin{figure*}
\begin{tabular}{c|c}
\includegraphics[height=6cm,width=8cm]{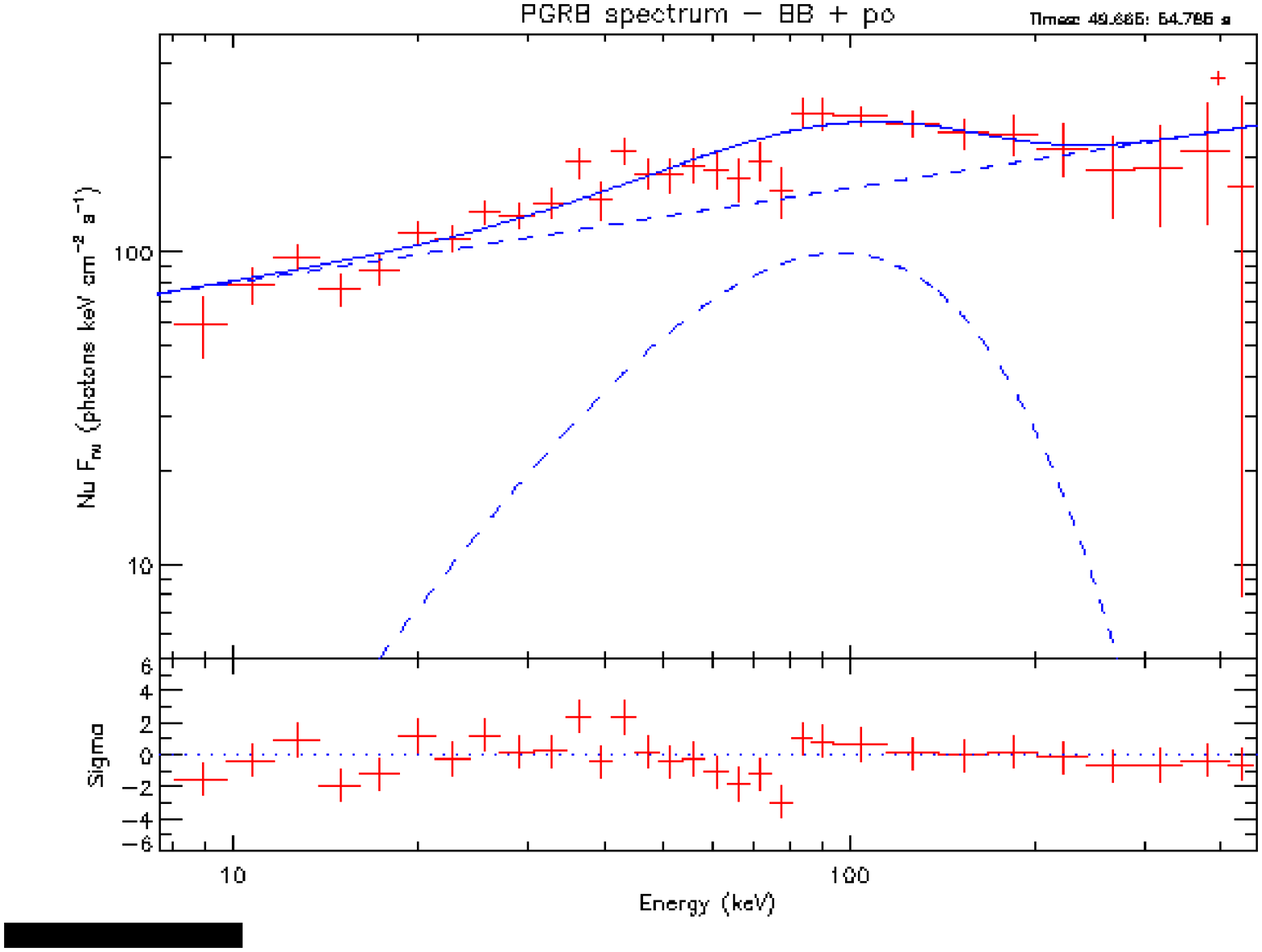}
\includegraphics[height=6cm,width=8cm]{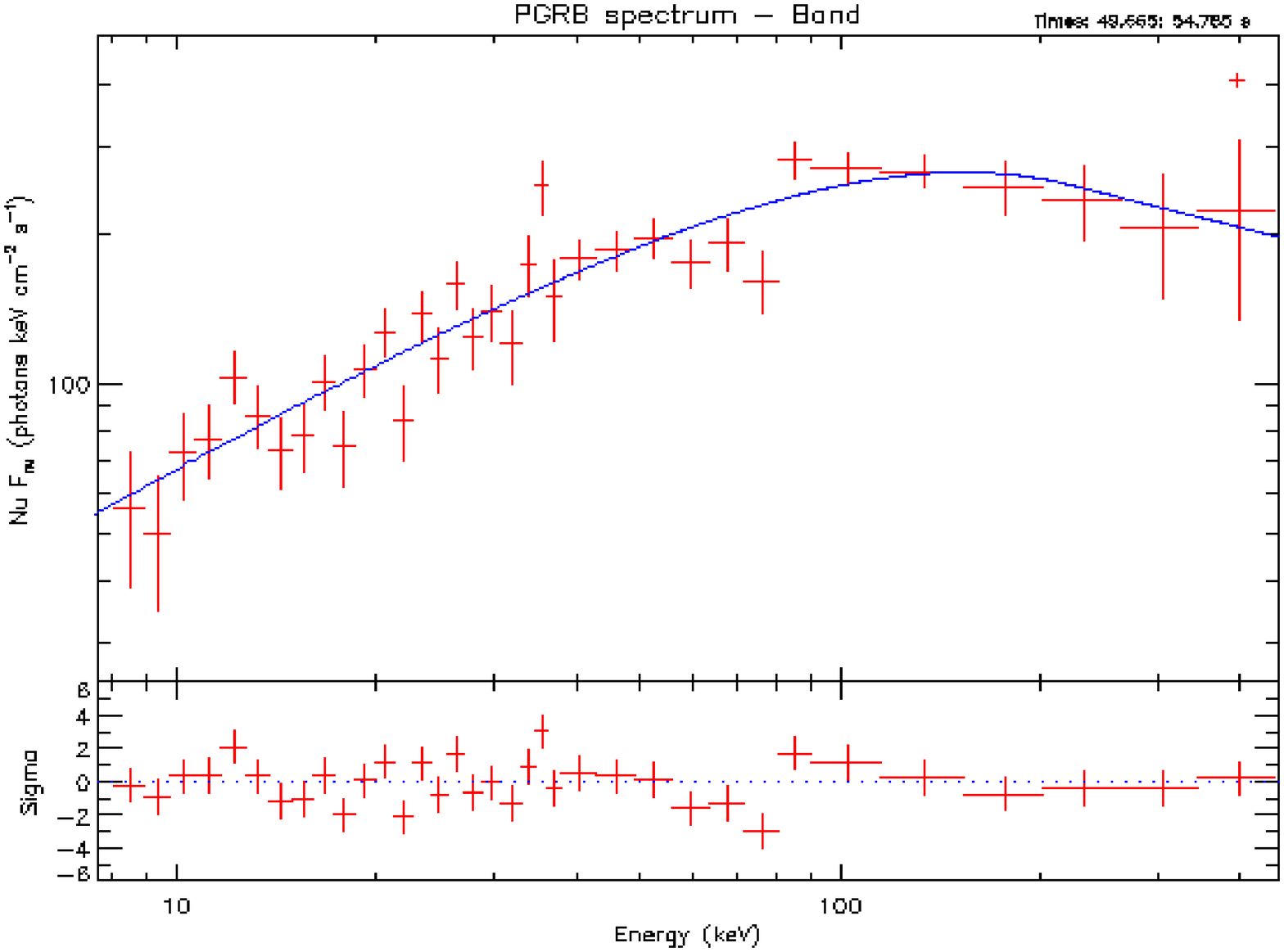}\\
\end{tabular}
\caption{On the left panel it is shown the time-integrated power spectra (8-440 keV) for the P-GRB emission episode (from 50 to 54 s after the trigger time) of GRB 090618 fitted with the blackbody + power-law models, $\tilde{\chi}^2$ = 1.15, while on the right it is shown the fit with a Band model, $\tilde{\chi}^2$ = 1.25.}
\label{fig:pgrb}
\end{figure*}

We now proceed to the analysis of the data between 50 and 150 s after the trigger time, as a canonical GRB in the fireshell scenario, namely the second episode, see Fig. \ref{fig:cospar}, \citep{TEXAS}.
We proceed to identify the P-GRB within the emission between 50 and 59 s, since we find a blackbody signature in this early second-episode emission.
Considerations based on the time variability of the thermal component bring us to consider the first 4 s of such time interval as due to the P-GRB emission. 
The corresponding spectrum (8-440 keV) is well fitted ($\tilde{\chi}^2 = 1.15$) with a blackbody of a temperature $kT = 29.22 \pm 2.21$ keV (norm = 3.51 $\pm$ 0.49), and an extra power-law component with photon index $\gamma$ = 1.85 $\pm$ 0.06, (norm = 46.25 $\pm$ 10.21), see Fig. \ref{fig:pgrb}. 
The fit with the Band model is also acceptable ($\tilde{\chi}^2 = 1.25$). The fit gives a low energy power-law index $\alpha=-1.22 \pm 0.08$, a high energy index $\beta=-2.32 \pm 0.21$ and a break energy $E_0 = 193.2 \pm 50.8$, see Fig. \ref{fig:pgrb}.
In view of the theoretical understanding of the thermal component in the P-GRB, see Section 3.2, we shall focus in the following on the blackbody + power-law spectral model.

The isotropic energy of the second episode is $E_{iso}$ = (2.49 $\pm$ 0.02) $\times$ 10$^{53}$ ergs.
The simulation within the fireshell scenario is done assuming $E_{tot}^{e^+e^-} \equiv E_{iso}$. 
From the upper panel in Fig. \ref{fig:no4} and the observed temperature, we can then derive the corresponding value of the baryon load.
The observed temperature of the blackbody component is $kT = 29.22 \pm 2.21$, so that we can determine a value of the baryon load of $B = 1.98 \pm 0.15 \times$ 10$^{-3}$, and deduce the energy of the P-GRB as a fraction of the total $E_{tot}^{e^+e^-}$. We so obtain a value of the P-GRB energy of 4.33$^{+0.25}_{-0.28}$ $\times$ 10$^{51}$ erg.

Now, from the second panel in Fig. \ref{fig:no4} we can derive the radius of the transparency condition, to occur at $r_{tr}$ = 1.46 $\times$ 10$^{14}$ cm.
From the third panel we derive the bulk Lorentz factor of $\Gamma_{th}$ = 495.
We compare this value with the energy measured in the sole blackbody component of $E_{BB}$ = 9.24$^{+0.50}_{-0.58}$ $\times$ 10$^{50}$ erg, and with the energy in the blackbody plus the power-law component of $E_{BB+po}$ = 5.43$^{+0.07}_{-0.11}$ $\times$ 10$^{51}$ erg, and verify that the theoretical value is in between these observed energies.
We have found this result quite satisfactory: it represents the first attempt to relate the GRB properties to the details of the black hole responsible for the overall GRB energetics. The above theoretical estimates have been based on a non rotating black hole of 10 M$_{\sun}$, a total energy of $E_{tot}^{e^+e^-}$ = 2.49 $\times$ 10$^{53}$ erg and a mean temperature of the initial plasma of $e^+ e^-$ of 2.4 MeV, derived from the expression of the dyadosphere radius, Eq. \ref{eq:rh}. Any refinement of the direct comparison between theory and observations will have to address a variety of fundamental issues such as, for example: 1) the possible effect of rotation of the black hole, leading to a more complex dyadotorus structure; 2) a more detailed analysis of the transparency condition of the $e^+e^-$ plasma, simply derived from the condition $\tau$ = $\int_R dr (n_{e^{\pm}} + n_{e^-}^b) \sigma_T = 0.67$ \citep{RSWX2}; 3) an analysis of the  general relativistic, electrodynamical, strong interactions descriptions of the gravitational collapse core leading to a black hole formation, \citep{Cherubini,Ruffini2003,RSWX2}.

\begin{figure}
\includegraphics[width=8cm, height=6cm]{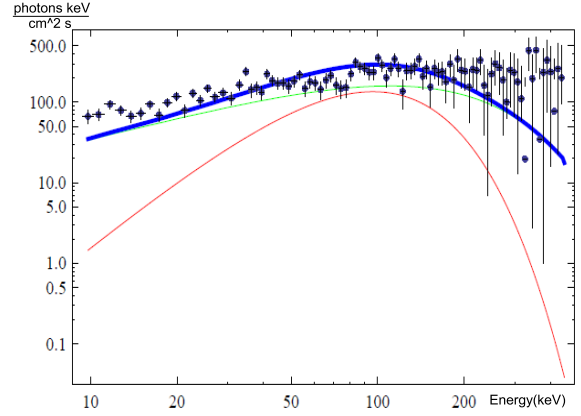}
\caption{The fireshell simulation, green line, and the sole blackbody emission, red line, of the time integrated (t0+50, t0+54 s) spectrum of the P-GRB emission. The sum of the two components, the blue line, is the total simulated emission in the first 4 s of the second episode.}
\label{fig:pgrb2}
\end{figure}

\subsubsection{The analysis of the extended-afterglow of the second episode}

\begin{figure}
\includegraphics[width=8cm, height=6cm]{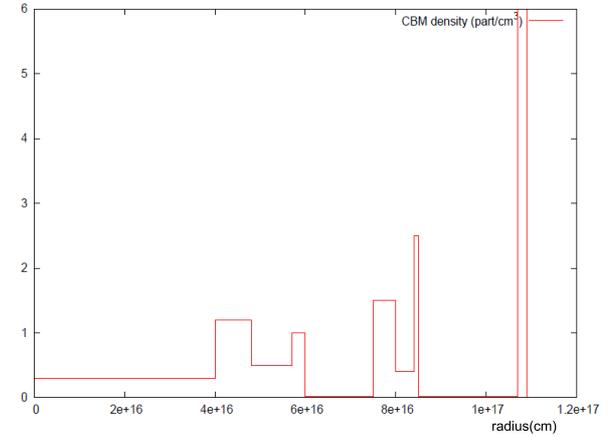}
\caption{Radial CBM density distribution in the case of GRB 090618. The characteristic masses of each cloud are of the order of $\sim$ 10$^{22-24}$ g and 10$^{16}$ cm in radii.}
\label{fig:rad}
\end{figure}

The extended-afterglow starts at the above given radius of the transparency, with an initial value of the Lorentz $\Gamma$ factor of $\Gamma_0$ = 495.
In order to simulate the extended-afterglow emission, we need to determine the radial distribution of the CBM around the burst site, which we assume for simplicity to be spherically symmetric, we infer characteristic size of $\Delta R = 10^{15-16}$ cm.
We have already recalled how the simulate of the spectra and of the observed multi band light curves have to be performed together and jointly optimized, leading to the determination of the fundamental parameters characterizing the CBM medium \citep{Ruffini2007b}.
This radial distribution is shown in Fig. \ref{fig:rad}, and is characterized by a mean value of $<n>$ = 0.6 part/cm$^3$ and an average density contrast with a $< \delta n/n >$ $\approx$ 2, see Fig. \ref{fig:rad} and Table \ref{tab:dens}.
The data up to 8.5 $\times$ 10$^{16}$ cm are simulated with a value for the filling factor $\mathcal{R} = 3 \times 10^{-9}$, while the data from this value on with $\mathcal{R} = 9 \times 10^{-9}$.
From the radial distribution of the CBM density, and considering the $1/\Gamma$ effect on the fireshell visible area, we found that the CBM clumps originating the spikes in the extended-afterglow emission have masses of the order of $10^{22-24}$ g.
The value of the $\alpha$ parameter has been found to be -1.8 along the total duration of the GRB. 

\begin{center}
\begin{figure}
\begin{tabular}{c}
\includegraphics[height=8cm,width=6cm,angle=270]{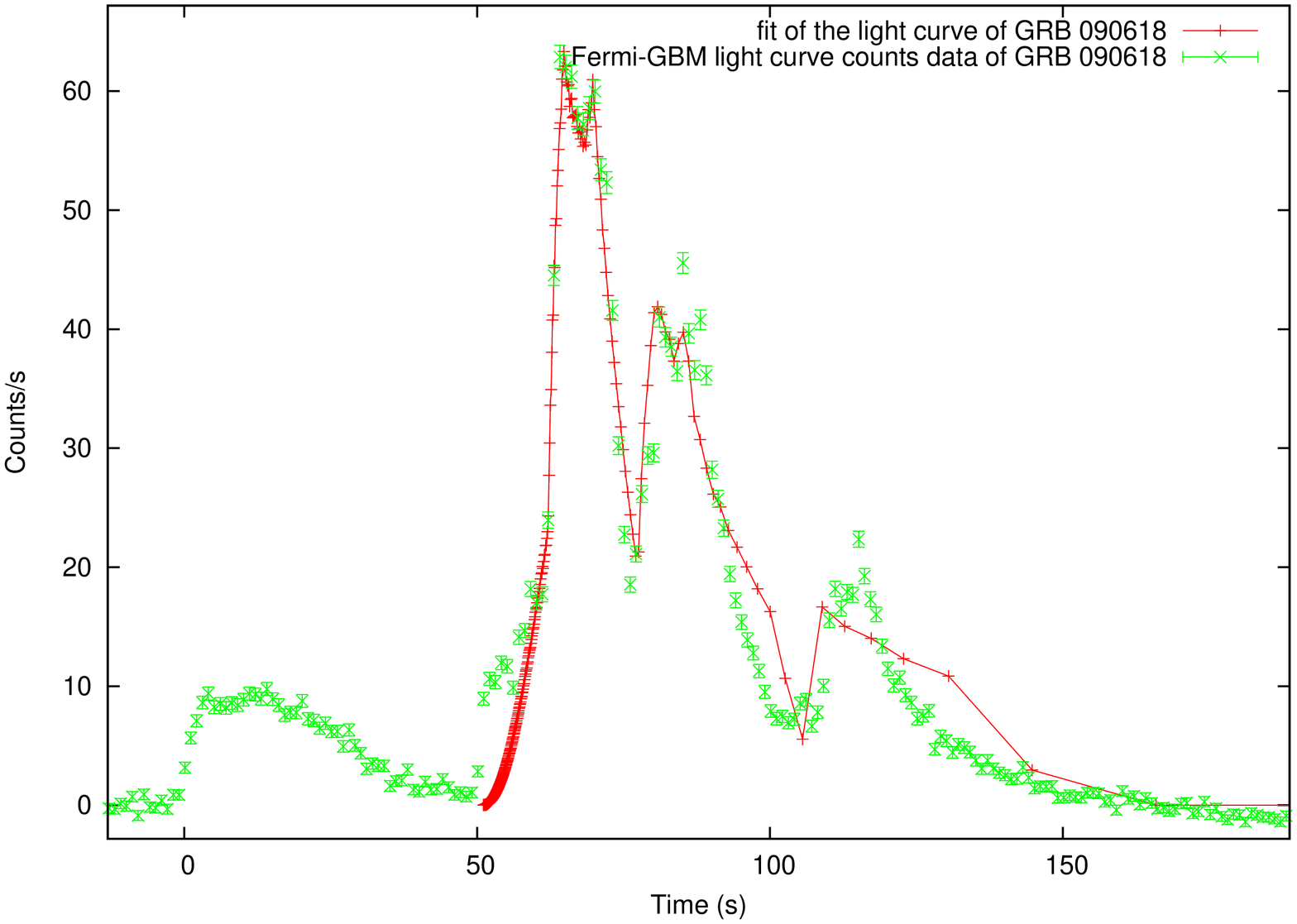}\\
\includegraphics[height=8cm,width=6cm,angle=270]{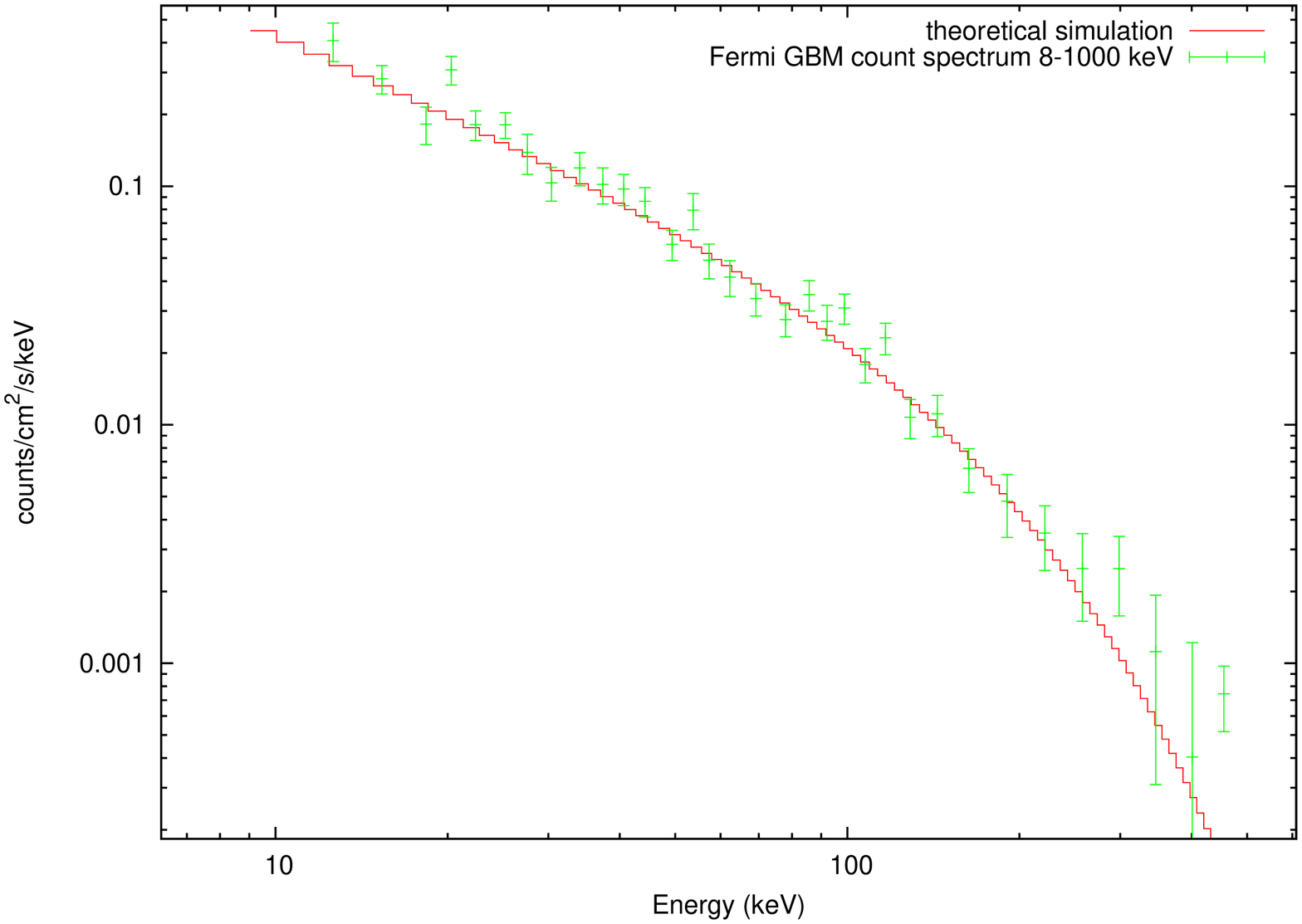}\\
\end{tabular}
\caption{Simulated light curve and time integrated (t0+58, t0+150 s) spectrum (8-440 keV) of the extended-afterglow of GRB 090618.}
\label{fig:firesh}
\end{figure}
\end{center}

In Figs. \ref{fig:firesh} we show the simulated light curve (8-1000 keV) of GRB and the corresponding spectrum, using the spectral model described in \citep{Bianco2004}, \citep{Patricelli2011}.
 
\begin{center}
\begin{figure}
\begin{tabular}{c}
\includegraphics[height=6cm,width=8cm]{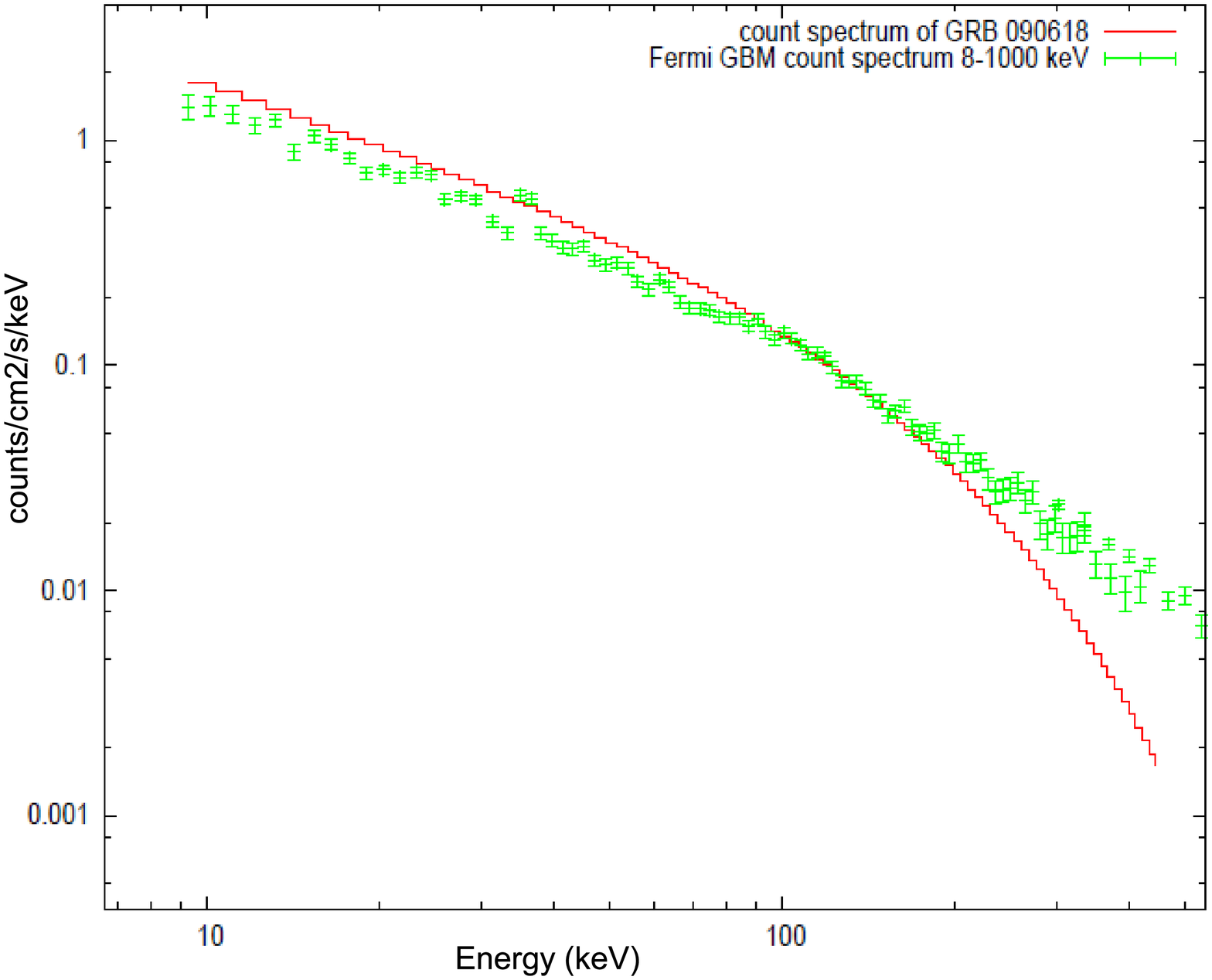}\\
\includegraphics[height=6cm,width=8cm]{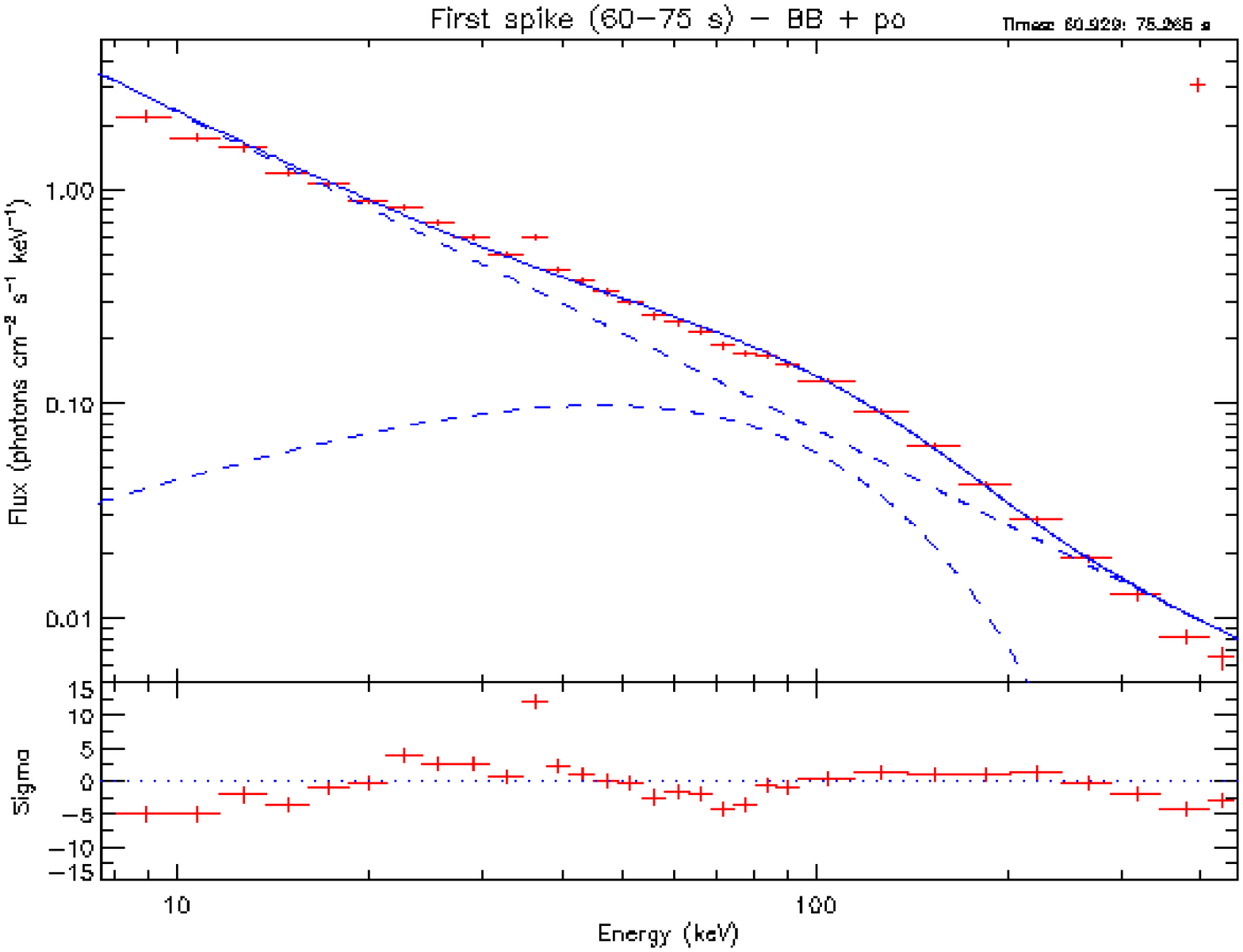}\\
\includegraphics[height=6cm,width=8cm]{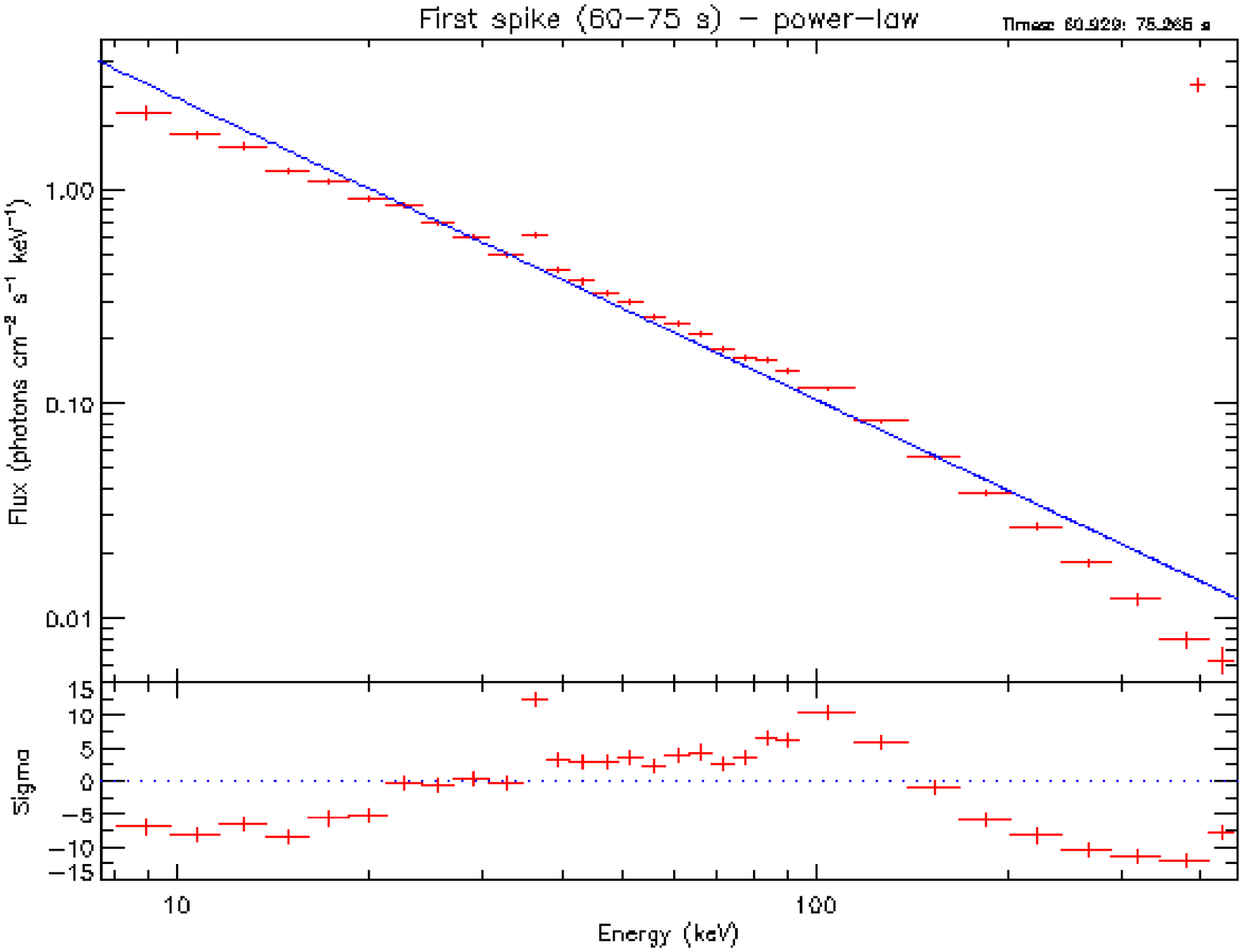}\\
\end{tabular}
\caption{Simulated time integrated (t0+58, t0+66 s) count spectrum (8-440 keV) of the extended-afterglow of GRB 090618 (upper panel), count spectrum (8 keV - 10 MeV) of the main pulse emission (t0+58, t0+66) and best fit with a blackbody + power-law model (middle panel) and a simple power-law model (lower panel).}
\label{fig:comp}
\end{figure}
\end{center}

We focus our attention, in particular, on the structure of the first spikes. 
The comparison between the spectra of the first main spike (t0+59, t0+66 s) of the extended-afterglow of GRB 090618, obtained with three different assumptions is shown in Fig. \ref{fig:comp}: in the upper panel we show the fireshell simulation of the integrated spectrum (t0+59, t0+66 s) of the first main spike, in the middle panel we show the best fit with a blackbody and a power-law component model and in the lower panel the best fit using a simple power-law spectral model.

We can see that the fit with the last two models is not satisfactory: the corresponding $\tilde{\chi}^2$ is 7 for the blackbody + power-law and $\sim$ 15 for the simple power-law.
We cannot give the $\tilde{\chi}^2$ of the fireshell simulation, since it is not represented by an explicit analytic fitting function, but it originates by a sequence of complex high non-linear procedure, summarized in Sec. \ref{sec:3}.
It is clear by a direct scrutiny that it correctly reproduces the low energy emission, thanks in particular to the role of the $\alpha$ parameter, which was described previously.
At higher energies, the theoretically predicted spectrum is affected by the cut-off induced by the thermal spectrum.
The temporal variability of the first two spikes are well simulated. 

We are not able to accurately reproduce the last spikes of the light curve, since the equations of motion of the accelerated baryons become very complicated after the first interactions of the fireshell with the CBM \citep{Ruffini2007b}.
This happens for different reasons.
First, a possible fragmentation of the fireshell can occur \citep{Ruffini2007b}.
Moreover, at larger distances from the progenitor the fireshell visible area becomes larger than the transverse dimension of a typical blob of matter, consequently a modification of the code for a three-dimensional description of the interstellar medium will be needed.
This is unlike the early phases in the prompt emission, which is the main topic we address at the moment, where a spherically simmetric approximation applies. 
The fireshell visible area is smaller than the typical size of the CBM clouds in the early phases of the prompt radiation, \citep{Izzo2010}.

The second episode, lasting from 50 to 151 s, agrees with a canonical GRB in the fireshell scenario.
Particularly relevant is the problematic of the P-GRB.
It interfaces with the fundamental physics issues, related to the physics of the gravitational collapse and the black hole formation.
There is an interface between the reaching of transparency of the P-GRB and the early part of the extended-afterglow.
This connection has already been introduced in literature \citep{Peer2010}.
We have studied this interface in the fireshell by analyzing the thermal emission at the transparency with the early interaction of the baryons with the CBM matter, see Fig. \ref{fig:pgrb2}.

We turn now to reach a better understanding of the meaning of the first episode, between 0 and 50 s of the GRB emission.
To this end we examine the two episodes in respect to: 1) the Amati relation, 2) the hardness variation and 3) the observed time lag.

\begin{table}
\centering
\caption{Final results of the simulation of GRB 090618 in the fireshell scenario} 
\label{tab:ris} 
\begin{tabular}{l c}
\hline\hline
Parameter & Value \\ 
\hline 
$E_{tot}^{e^+e^-}$ &  2.49 $\pm$ 0.02 $\times$ 10$^{53}$ ergs\\
$B$ &  1.98 $\pm$ 0.15 $\times$ 10$^{-3}$\\
$\Gamma_0$ & 495 $\pm$ 40\\
$kT_{th}$ & 29.22 $\pm$ 2.21 keV\\
$E_{P-GRB,th}$ & 4.33 $\pm$ 0.28 $\times$ 10$^{51}$ ergs\\     
$<n>$ & $0.6 \,  part/cm^3$\\
$<\delta n/n>$ & $2 \, part/cm^3$\\
\hline
\end{tabular}
\end{table}

\begin{table}
\centering
\caption{Physical properties of the three clouds surrounding the burst site: the Distance from the burst site (2$^{nd}$ column, the radius $r$ of the cloud, 3$^{rd}$ column, the particle density $\rho$, 4$^{th}$ column and the mass $M$ in the last column  } 
\label{tab:dens} 
\begin{tabular}{l c c c c}
\hline\hline
Cloud & Distance (cm) & r (cm) & $\rho$ (\#/cm$^3$) & M (g)\\ 
\hline 
First & 4.0 $\times$ 10$^{16}$ & 1 $\times$ 10$^{16}$ & 1 & 2.5 $\times$ 10$^{24}$ \\
Second & 7.4 $\times$ 10$^{16}$ & 5 $\times$ 10$^{15}$ & 1 & 3.1 $\times$ 10$^{23}$ \\
Third & 1.1 $\times$ 10$^{17}$ & 2 $\times$ 10$^{15}$ & 4 &  2.0 $\times$ 10$^{22}$ \\
\hline
\end{tabular}
\end{table}

\section{The Amati relation, the HR and the time lag of the two episodes}\label{sec:6}

\subsection{The first episode as an independent GRB?}

\begin{figure}
\includegraphics[width=8cm, height=6cm]{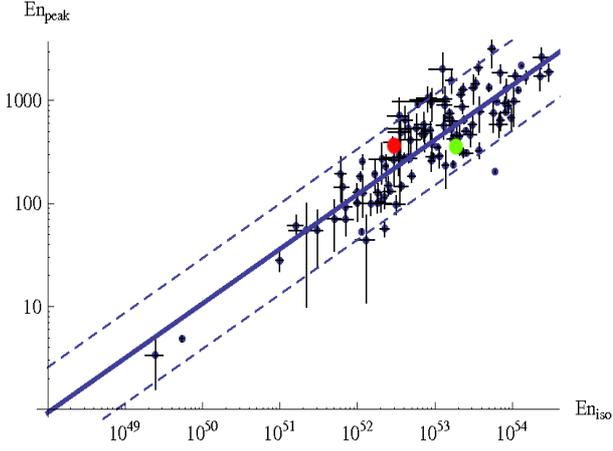}
\caption{Position of the first and second component of GRB 090618 in the $E_{p,i}$ - $E_{iso}$ plane respect the best fit of the Amati relation, as derived following the procedure described in \citep{CapozzielloIzzo}. The red circle corresponds to the first emission while the green circle corresponds to the second one.}
\label{fig:amati}
\end{figure}

We first check if the two episodes fulfill separately the Amati relation, \citep{Amati2002}.
By using the Band spectrum we verify that the first episode presents an intrinsic peak energy value of $E_{p,1st}$ = 223.01 $\pm$ 24.15 keV, while the second episode presents an $E_{p,2nd}$ = 224.57 $\pm$ 17.4 keV.
The isotropic energies emitted in each single episode are $E_{iso,1st}$ = 4.09 $\pm$ 0.07 $\times$ 10$^{52}$ ergs and $E_{iso,2nd}$ = 2.49 $\pm$ 0.02 $\times$ 10$^{53}$ ergs, so we have that both episodes satisfy the Amati relation, see fig. \ref{fig:amati}.
The fulfillment of the Amati relation of episode 2 was expected, being the second episode a canonical GRB.
What we find surprising is the fulfillment of the Amati relation of the first episode.

We first examine the episode 1 as a single GRB.
We notice a sharp rise in the luminosity in the first 6 s of emission.
We therefore attempted a first interpretation by assuming the first 6 s  as the P-GRB component of this independent GRB, as opposed to the remaining 44 s as the extended-afterglow of this GRB.
A value of the fit gives $E_{tot}^{e^+e^-} = 3.87 \times 10^{52}$ ergs and $B= 1.5 \times 10^{-4}$.
This would imply a very high value for the Lorentz factor at the transparency of $\sim$ 5000.
In turn, this value would imply \citep{RSWX2} a spectrum of the P-GRB peaking around  $\sim$ 300 keV, which is in contrast with the observed temperature of 58 keV.
Alternatively, we have attempted a second simulation by assuming all the observed
data be part of the extended-afterglow of a GRB, with a P-GRB below the detector threshold.
Assuming in this case $E_{iso}$ = $E_{tot}^{e^+e^-}$, $B$ = 10$^{-2}$, and assuming for the P-GRB a duration smaller than 10 s, as confirmed from the observations of all existing P-GRBs \citep{Ruffini2007b}, we should obtain an energy of the P-GRB  greater than 10$^{-8}$ ergs/cm$^2$/s, which should have been easily detectable from Fermi and Swift.
Also this second possibility is therefore not viable.
We can then conclude generally that in no way we can interpret this episode either as a P-GRB 
of the second episode, as proved in paragraph 3.2 or, as proved here, as a separate GRB.
We then conclude that the fulfillment of the Amati relation does not imply for the source to be necessarily a GRB.

\subsection{The HR variation and the time lag of the two episodes}

We finally address a further difference between the two episodes, related to the Hardness-Ratio behavior (HR) and their observed time-lag.
The first evidence of an evolution of the GRBs power-law slope indexes with time was observed in the BATSE GRB photon spectra \citep{Crider1997}. 
In the context of the fireshell scenario, as recalled earlier, the spectral evolution comes out naturally from the evolution of the comoving temperature, the decrease of the bulk Lorentz $\Gamma$ factor and from the curvature effect \citep{Bianco2004}, with theoretically predicted values, in excellent agreement with observations in past GRBs.

\begin{figure}
\includegraphics[width=8cm, height=6cm]{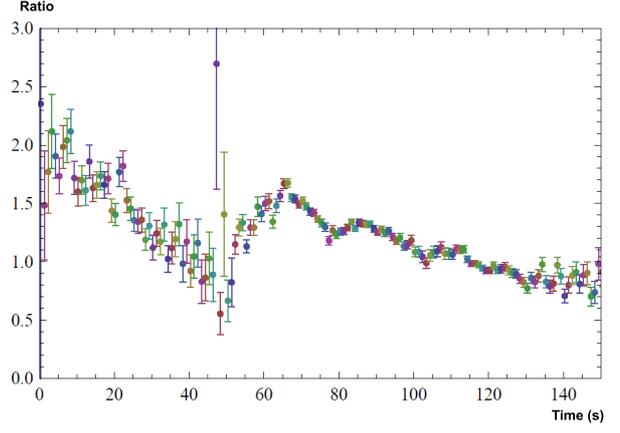}
\includegraphics[width=8cm, height=6cm]{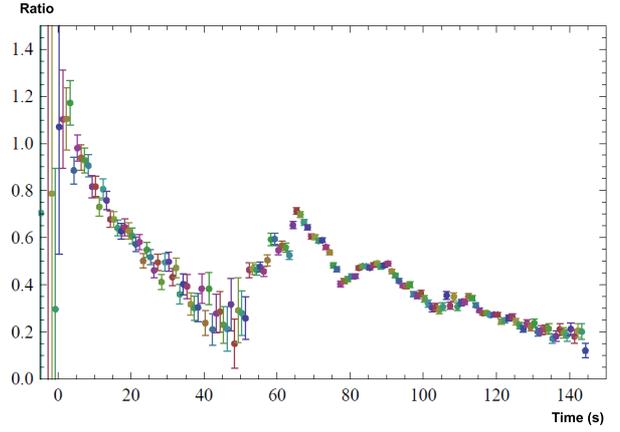}
\caption{Hardness-Ratio ratios for the Swift BAT data in two different energy channels: HR1 = cts(25-50 keV)/cts(15-25 keV), HR2 = cts(50-150 keV)/cts(15-50 keV). }
\label{fig:no5}
\end{figure}

In order to build the HR ratio, we considered the data from three different instruments: Swift-BAT, Fermi-GBM and the CORONAS-PHOTON-RT-2.
The plots obtained with these instruments confirm the existence of a peculiar trend of the hardness behavior: in the first 50 s it is evident a monotonic hard-to-soft behavior, as due to the blackbody evolution of the first episode. For the second episode, the following 50 to 151 s of the emission, there is a soft-to-hard trend in the first 4 s of emission, and a hard-to-soft behavior modulated by the spiky emission in the following 100 s. 
For the HR ratio we considered the ratio of the count rate detected from a higher energy channel to that of a lower energy channel: HR = ctg(HE)/ctg(LE).
In particular, we considered the count rate subtracted for the background, even when this choice provides bad HR data in time region dominated by the background, where the count rate can be zero or negative.
For the Swift data, we consider the HR ratio for two different energy subranges: the HR1 ratio shows the ratio of the (50-150 keV) over the (15-50 keV) emission while the HR2 ratio shows the ratio of the (25-50 keV) over the (15-25 keV) emission, see Fig. \ref{fig:no5}. 

\begin{figure}
\includegraphics[width=8cm, height=6cm]{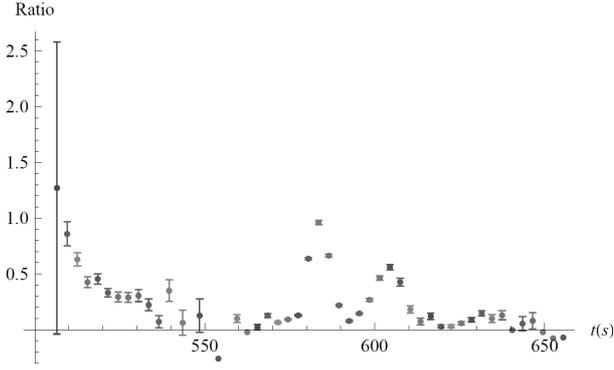}
\caption{Hardness-Ratio ratio for the Fermi data. We considered the cts observed in the (260 keV - 40 MeV) energy range over the (8 - 260 keV) energy range. The time reported on the x-axis is in terms of the Mission Elapsed Time (MET). The presence of some negative data points is due to the presence of noise, in other terms the non-presence of GRB emission, in the background-subtracted BGO count light curve.}
\label{fig:no5b}
\end{figure}

A similar trend is found for the Fermi-GBM NaI and RT-2 instruments, see Fig.\ref{fig:no5b}.
In particular, the HR from Fermi observations was done considering the counts observed by the b0 BGO detector in the range (260 keV - 40 MeV) and the ones observed by the n4 NaI detector in the range (8 - 260 keV).
In Fig. \ref{fig:no5b} it is shown the HR ratio for the Fermi observations, where we rebinned the counts in time intervals of 3 seconds.
From this analysis we see that the HR ratio peaks at the beginning of each pulse, also for the second episode pulses, but each peak of the second episode pulses is softer than the previous one, suggesting that these pulses are consequential in the second episode and are in general agreement with the advance of a fireshell in the CBM. 
Since RT-2 data clearly show both the episodes up to 1 MeV it complements the results obtained by Swift (up to 200 keV) and FERMI (up to 440 keV) in the
high and the most interesting energy range. 
Hardness ratio plot of (250-1000 keV)/(8-250 keV) indicates that first phase of both episodes are the hardest.

Finally, the evident asymmetry of the first episode, supported by the observations of a large time lag in the high and low energy channels, see fig. \ref{fig:1}, suggests a different process at work.
There is a very significant softening of the first episode, as reported in \citet{Rao2011}, where it is observed a large time lag between the 15-25 keV energy range and the 100-150 keV one: the high energy photons peak $\sim$ 7 seconds before the photons detected in the 15-25 keV energy range.
This large time lag is not observed in the second episode, where the lags are of the order of $\sim$ 1 s.

Motivated by these results, we proceed to a most accurate time-resolved spectral analysis of the first episode to identify its physical and astrophysical origin.

\section{A different emission process in the first episode}\label{sec:7}

\subsection{The time resolved spectra and temperature variation}

One of the most significant outcome of the multi-year work of Felix Ryde and his collaborators, (see e.g. \citet{Ryde2010} and references therein), has been the identification and the detailed analysis of the thermal plus power-law features observed in a time limited intervals in selected BATSE GRBs.
Similar features have been also observed recently in the data acquired by the Fermi satellite \citep{Ryde2010,Guiriec2011}.
We propose to divide these observations in two broad families.
The first family presents a thermal plus power-law(s) feature, with a temperature changing in time following precise power-law behavior.
The second family is also characterized by a thermal plus power-law component, but with the blackbody emission generally varying without specific power-law behavior and on shorter time scales.
It is our goal to study these features within the fireshell scenario, in order to possibly identify the underlying physical processes.
We have already identified in Sec. 4 that the emission of the thermal plus power-law component characterizes the P-GRB emission.
We have also emphasized that the P-GRB emission is the most relativistic regime occurring in GRBs, uniquely linked to the process of the black hole formation, see Sec. 5.
This process appears to belong to the second family above considered. 
Our aim here is to see if the first episode of GRB 090618 can lead to the identification of the above first family of events: the ones with temperature changing with time following a power-law behavior on time scales from 1 to 50 s.
We have already pointed out in the previous section that the hardness-ratio evolution and the large time lag observed for the first episode \citep{Rao2011} points to a distinct origin for the first 50 s of emission, corresponding to the first episode.

We have made a detailed time-resolved analysis of the first episode, considering different time bin durations in order to have a good statistic in the spectra and to take into account the sub-structures in the light curve.
We have then used two different spectral models to fit the observed data, a classical Band spectrum \citep{Band1993}, and a blackbody with a power-law component.

\begin{figure}
\includegraphics[width=8cm, height=6cm]{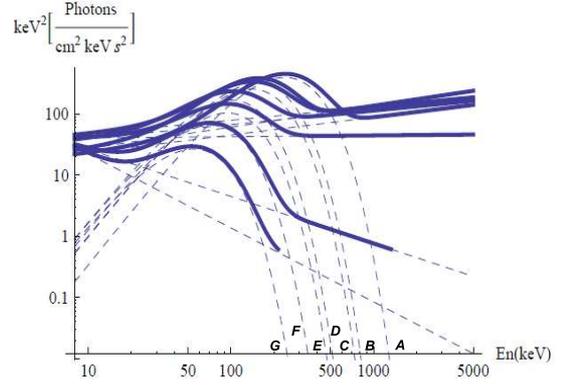}
\caption{Evolution of the BB+powerlaw spectral model in the $\nu\,F(\nu)$ spectrum of the first emission of GRB 090618. It is evident the cooling of the black-body and of the associated non-thermal component with the time. In this picture we have preferred to plot just the fitting functions, in order to prevent some confusion.}
\label{fig:no6}
\end{figure}

\begin{figure}
\begin{tabular}{c}
\includegraphics[width=8cm, height=6.5cm]{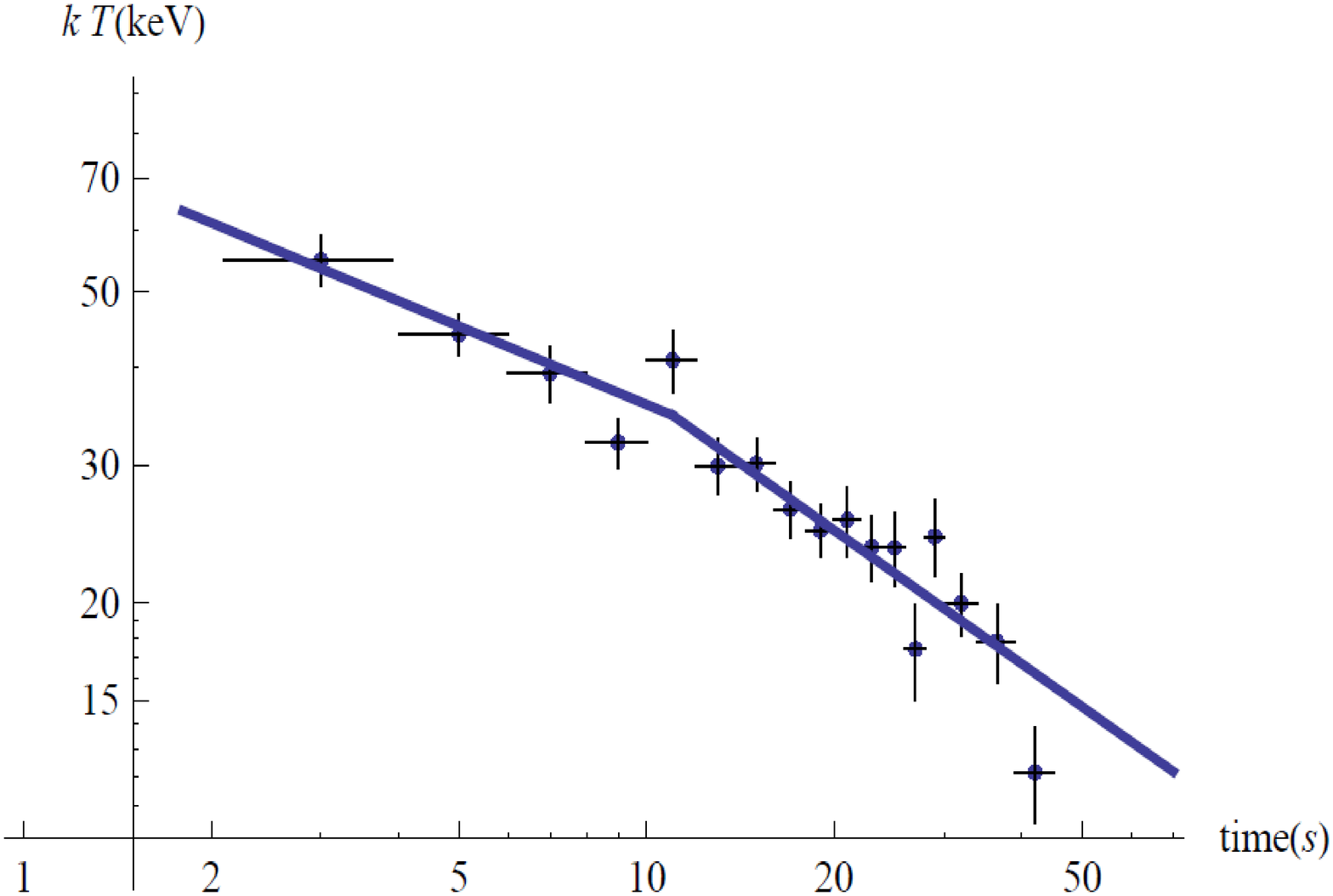}\\
\includegraphics[width=8cm, height=6.5cm]{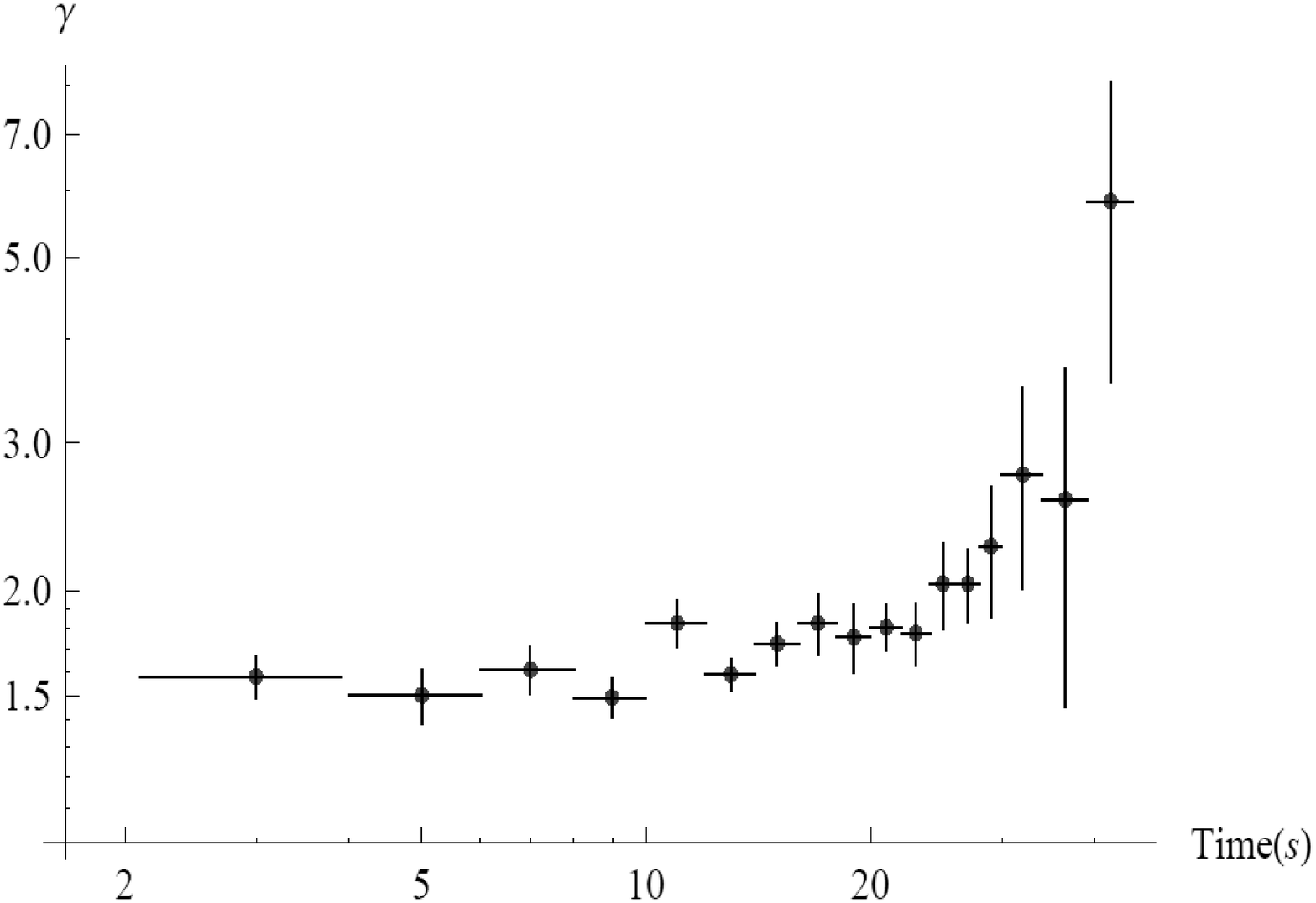}\\
\end{tabular}
\caption{Evolution of the $kT$ observed temperature of the black-body component and the corresponding evolution of the photon index of the power-law. The blue line in the upper panel corresponds to the fit of the time evolution of the temperature with a broken power-law function. It is evident a break time $t_b$ around 11 s after the trigger time, as obtained from the fitting procedure.}
\label{fig:no17}
\end{figure}

In order to have more accurate constraints on the spectral parameters, we made a joint fit considering the observations from both the n4 NaI and the b0 BGO detectors, covering in this way a wider energy range, from 8 keV to 40 MeV.
To avoid some bias due to low photon statistic, we considered an energy upper limit of the value of 10 MeV.
We report in the last three columns of the Table \ref{tab:no} also the spectral analysis performed in the energy range of the BATSE LAD instrument (20-1900 keV), as analyzed in \citet{Ryde2009}, just as a comparison tool with the results described in that paper.
Our analysis has been summarized in Figs. \ref{fig:no6}, \ref{fig:no17} and in Table \ref{tab:no}, where we report the residual ratio diagram as well as the reduced-$\chi^2$ values for the spectral models considered. 

\begin{table*}
\centering
\caption{Time-resolved spectral analysis of the first episode in GRB 090618. We have considered seven time intervals, as described in the text, and we used two spectral models, whose best-fit parameters are shown here. The last three columns, marked with a LAD subscript, report the same analysis but in the energy range $20 - 1900$ keV, which is the same energy range of the BATSE-LAD detector as used in the work of \citet{Ryde2009}.} 
\label{tab:no} 
{\tiny
\begin{tabular}{l c c c c c c c c c c}
\hline\hline
Time  & $\alpha$ & $\beta$ & $E_0$ (keV) & $\tilde{\chi}^2_{BAND}$ & $kT$ (keV) & $\gamma$ & $\tilde{\chi}^2_{BB+po}$ & $kT_{LAD}$ (keV) & $\gamma_{LAD}$ & $\tilde{\chi}^2_{BB+po,LAD}$\\ 
\hline 
A:0 - 5 & -0.45 $\pm$ 0.11 & -2.89 $\pm$ 0.78 & 208.9 $\pm$ 36.13 & 0.93 & 59.86 $\pm$ 2.72 & 1.62 $\pm$ 0.07 & 1.07 & 52.52 $\pm$ 23.63 & 1.42 $\pm$ 0.06 & 0.93 \\
B:5 - 10 & -0.16 $\pm$ 0.17 & -2.34 $\pm$ 0.18 & 89.84 $\pm$ 17.69 & 1.14 & 37.57 $\pm$ 1.76 & 1.56 $\pm$ 0.05 & 1.36 & 37.39 $\pm$ 2.46 & 1.55 $\pm$ 0.06 & 1.27\\
C:10 - 17 & -0.74 $\pm$ 0.08 & -3.36 $\pm$ 1.34 & 149.7 $\pm$ 21.1 & 0.98 & 34.90 $\pm$ 1.63 & 1.72 $\pm$ 0.05 & 1.20 & 36.89 $\pm$ 2.40 & 1.75 $\pm$ 0.06 & 1.10\\
D:17 - 23 & -0.51 $\pm$ 0.17 & -2.56 $\pm$ 0.26 & 75.57 $\pm$ 16.35 & 1.11 & 25.47 $\pm$ 1.38 & 1.75 $\pm$ 0.06 & 1.19 & 25.70 $\pm$ 1.76 & 1.75 $\pm$ 0.08 & 1.19\\
E:23 - 31 & -0.93 $\pm$ 0.13 & unconstr. & 104.7 $\pm$ 21.29 & 1.08 & 23.75 $\pm$ 1.68 & 1.93 $\pm$ 0.10 & 1.13 & 24.45 $\pm$ 2.24 & 1.95 $\pm$ 0.12 & 1.31\\           
F:31 - 39 & -1.27 $\pm$ 0.28 & -3.20 $\pm$ 1.00 & 113.28 $\pm$ 64.7 & 1.17 & 18.44 $\pm$ 1.46 & 2.77 $\pm$ 0.83 & 1.10 & 18.69 $\pm$ 1.89 & 4.69 $\pm$ 4.2 & 1.08\\
G:39 - 49 & -3.62 $\pm$ 1.00 & -2.19 $\pm$ 0.17 & 57.48 $\pm$ 50.0 & 1.15 & 14.03 $\pm$ 2.35 & 3.20 $\pm$ 1.38 & 1.10 & 14.71 $\pm$ 3.52 & 3.06 $\pm$ 3.50 & 1.09\\
\hline
\end{tabular}
}
\end{table*}

We conclude that both the Band and the proposed blackbody + power-law spectral models fit very well the observed data.
Particularly interesting is the clear evolution in the time-resolved spectra, corresponding to the blackbody and power-law component, see Fig. \ref{fig:no6}.
In particular the $kT$ parameter of the blackbody presents a strong decay, with a temporal behavior well described by a double broken power-law function, see upper panel in Fig. \ref{fig:no17}.
From a fitting procedure we obtain the best fit (R$^2$-statistic = 0.992) for the two decay indexes for the temperature variation are $a_{kT}$ = -0.33 $\pm$ 0.07 and $b_{kT}$ = -0.57 $\pm$ 0.11.
In \citet{Ryde2009} an average value for these parameters on a set of 49 GRBs is given: $\left\langle a_{kT} \right\rangle$ = -0.07 $\pm$ 0.19 and $\left\langle b_{kT} \right\rangle$ = -0.68 $\pm$ 0.24.
We note however that in the sample considered in \citet{Ryde2009} only few bursts shows a break time around 10 s, as it is in our case, see Fig. \ref{fig:no17}.
There are two of these bursts, whose analysis presents many similarities with our presented source GRB 090618: GRB 930214 and GRB 990102.
These bursts are characterized by a simple FRED pulse, whose total duration is $\sim$ 40 s, quite close to the one corresponding to the first episode of GRB 090618.
The break time $t_b$ in these two burst are respectively at 12.9 and 8.1 s, while the decay indexes are $a_{kT}$ = -0.25 $\pm$ 0.02 and $b_{kT}$ = -0.78 $\pm$ 0.04 for GRB 930214 and $a_{kT}$ = -0.36 $\pm$ 0.03 and $b_{kT}$ = -0.64 $\pm$ 0.04 for GRB 990102, see Table 1 in \citet{Ryde2009}, in very good agreement with the values observed for the first episode of GRB 090618.
We conclude that the values we observe in GRB 090618 are very close to the values of these two bursts.
We shall return to compare and contrast our results with the other sources considered in \citep{Ryde2009}, as well as GRB 970828 \citep{Peer2007} in a forthcoming publication.

The results presented in Figs. \ref{fig:no6},\ref{fig:no17}, as well as in Table \ref{tab:no}, point to a rapid cooling of the thermal emission with time of the first episode.
The evolution of the corresponding power-law spectral component, also, appears to be strictly related to the change of the temperature $kT$.
The power-law $\gamma$ index falls, or softens, with the temperature, see Fig. \ref{fig:no6}.
An interesting feature appears to occur at the transition of the two power-law describing the observed decrease of the temperature. 
The large time lag observed in the first episode and reported in section 6.1 has a clear explanation in the power-law behavior of the temperature and corresponding evolution of the photon index $\gamma$, Figs. \ref{fig:no6},\ref{fig:no17}.

\subsection{The radius of the emitting region}

We turn now to estimate an additional crucial parameter for the identification of the nature of the blackbody component: the radius of the emitter $r_{em}$.
We have proved that the first episode is not an independent GRB, not a part of a GRB.
We can therefore provide the estimate of the radius of the emitter from non-relativistic considerations, just corrected for the cosmological redshift $z$.
We have, in fact, that the temperature of the emitter $T_{em} = T_{obs} (1+z)$, and that the luminosity of the emitter, due to the blackbody emission, is

\begin{equation}\label{eq:8.1}
L = 4 \pi r_{em}^2 \sigma T_{em}^4 = 4 \pi r_{em}^2 \sigma T_{obs}^4 (1+z)^4,
\end{equation}

where $r_{em}$ is the radius of the emitter and $\sigma$ is the Stefan constant. From the luminosity distance definition, we also have that the observed flux $\phi_{obs}$ is given by:

\begin{equation}\label{eq:8.2}
\phi_{obs} = \frac{L}{4 \pi D^2} = \frac{r_{em}^2 \sigma T_{obs}^4 (1+z)^4}{D^2}.
\end{equation}

We then obtain

\begin{equation}\label{eq:radius}
r_{em} = \left(\frac{\phi_{obs}}{\sigma T_{ob}^4}\right)^{1/2} \frac{D}{(1+z)^2}.
\end{equation}

The above radius differs from the radius $r_{ph}$ given in Eq. (1) of \citet{Ryde2009} and clearly obtained by interpreting the early evolution of GRB 970828 as belonging to the photospheric emission of a GRB and assuming a relativistic expansion with a Lorentz gamma factor $\Gamma$:
\begin{equation}
r_{ph} = \hat{\mathcal{R}} D \left(\frac{\Gamma}{(1.06) (1+z)^2}\right),
\end{equation}
where $\hat{\mathcal{R}} = \left(\phi_{obs}/(\sigma T_{ob}^4)\right)^{1/2}$ and the prefactor 1.06 arises from the dependence of $r_{ph}$ on the angle of sight \citep{Peer2008}.
Typical values of $r_{ph}$ are at least two orders of magnitude larger than our radius $r_{em}$. 
We shall return on the analysis of GRB 970828 in a forthcoming paper.

\begin{figure}
\includegraphics[width=8cm, height=6cm]{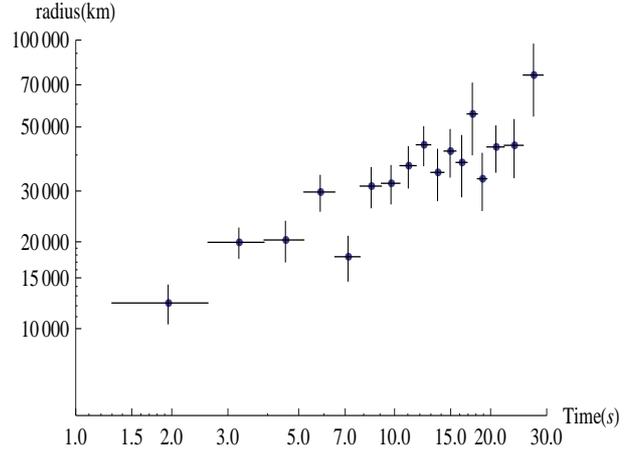}
\caption{Evolution of the radius of the first episode emitter, as given by Eq. (\ref{eq:radius}).}
\label{fig:no18}
\end{figure}

Assuming a standard cosmological model ($H_0 = 70$ km/s/Mpc, $\Omega_m = 0.27$ and $\Omega_{\Lambda} = 0.73$) for the estimate of the
luminosity distance $D$, and using the values for the observed flux $\phi_{obs}$ and the temperature $kT_{obs}$, we have given in Fig. \ref{fig:no18} the evolution of the radius of the surface emitting the blackbody $r_{em}$ as a function of time.

Assuming an exponential evolution with time $t^{\delta}$ of the radius in the comoving frame, we obtain from a fitting procedure the value $\delta = 0.59 \pm 0.11$, well compatible with $\delta =0.5$.
We also notice a steeper behavior for the variation of the radius with time corresponding to the first 10 s, which corresponds to the emission before the break of the double power-law behavior of the temperature.
We estimate an average velocity of $\bar{v} = 4067 \pm 918$ km/s, R$^2$ = 0.91, in these first 10 s of emission.
In episode 1 the observations lead to a core of an initial radius of $\sim$ 12000 km expanding in the early phase with a sharper initial velocity of $\sim$ 4000 km/s.
The effective Lorentz $\Gamma$ factor is very low, $\Gamma - 1 \sim$ 10$^{-5}$.

\section{Conclusions}\label{sec:8}

GRB 090618 is one of the closest ($z = 0.54$) and most energetic ($E_{iso}$ = 2.9 $\times$ 10$^{53}$ ergs ) GRBs up to date.
It has been observed simultaneously by the largest number of X and $\gamma$ ray telescopes: Fermi, Swift, AGILE, Konus-WIND, Suzaku-WAM and the CORONAS-PHOTON-RT2.
These circumstances have produced an unprecedented set of high quality data as well as the coverage of the instantaneous spectral properties and of the time variability in luminosity of selected bandwidth of the source, see e.g. Figs. \ref{fig:Chak},\ref{fig:1}.
In addition there is also the possibility of identifying an underlying supernova event from the optical observations in the light curve of well-defined bumps, as well as from the correspective change in colour after around 10 days from the main event \citep{Cano2010}.
Unfortunately a spectroscopic confirmation of the presence of such supernova is lacking.
We have restricted our attention in this paper to the sole X and $\gamma$ ray emission of the GRB, without addressing the possible supernova component.

By applying our analysis within the fireshell scenario, see section 4, we have supported that GRB 090618 is actually composed of two different episodes \citep{COSPAR}: episode 1, lasting from 0 to 50 s and episode 2 from 50 s to 151 s after the trigger time.
We have also illustrated the recent conclusions presented in \citet{TEXAS}, that episode 1 cannot be either a GRB nor a part of a GRB, see section 5.
By a time-resolved spectral analysis we have fitted the instantaneous spectra by a blackbody plus an extra power-law component.
The temperature of the blackbody appears to have a regular dependence with time, described by two power-law functions:
a first power-law with decay index a$_{kT}$ = -0.33 $\pm$ 0.07 and the second one with b$_{kT}$ = -0.57 $\pm$ 0.11, see Section 7 .
All these features follow precisely some of the results obtained by Felix Ryde and his collaborators \citep{Ryde2009}, where the authors analyzed selected temporal episodes in some GRBs observed by BATSE.

We have also examined with particular attention, see section 6, the radius $r_{em}$ of the blackbody emitter observed in the first episode, given by Eq. (\ref{eq:radius}).
We interpret the nature of this episode 1 as originating from what we have defined a proto-black hole, \citep{COSPAR}: the collapsing bare core leading to the black hole formation.
Within this interpretation, the radius $r_{em}$ depends only on the observed energy flux of the blackbody component $\phi_{obs}$, the temperature $kT$ as well as on the luminosity distance of the source $D$.
We obtained a radius of the emitting region smoothly varying between $\sim$ 12000 and 70000 km, see Fig. \ref{fig:no18}.
Other interpretations associating the origin of this early emission to the GRB main event \citep{Peer2007} lead to a different definition for the radius of the blackbody emitter, which results to be larger than our radius by at least two orders of magnitude. 
We are planning a systematic search for other systems presenting these particular features.

Episode 2 is identified as a canonical long GRB which originates from the black hole formation process and lasts in arrival time from 50 s to 151 s after the trigger time. The good quality of data allowed us to search for the P-GRB signature in the early emission of the episode 2.
From a detailed analysis we find that the first 4 s of episode 2 are in good agreement with the theoretically predicted P-GRB emission, see section 5.2. The observed spectrum integrated over these 4 s is well fitted by a blackbody with an extra power-law component, where this latter component is mainly due to the early emission of the extended-afterglow, see Fig. \ref{fig:pgrb}.
From the temperature observed in the P-GRB, $kT_{PGRB}$ = 29.22 $\pm$ 2.21, and the $E_{tot}^{e^+e^-}$ energy of the second episode, which we assumed equal to the isotropic equivalent energy of this episode, $E_{tot}^{e^+e^-}$ = 2.49 $\times$ 10$^{53}$ ergs, we obtained the value of the baryon load of the GRB, see also Fig. \ref{fig:no4}, $B = (1.98 \pm 0.15) \times 10^{-3}$, and a consequent Lorentz $\Gamma$ factor at the transparency of $\Gamma_\circ = 495 \pm 40$.
We have been able to simulate the temporal and the spectral emission of the second episode, as seen by the Fermi-GBM instrument (8 keV -- 10 MeV).
As we have shown in Fig. \ref{fig:comp}, our simulation succeeds in fitting the light curves as well as the spectral energy distribution emitted in the first main spike of the second episode.
The residual emission of the last spikes is reasonably fitted, taking into due account the difficulties in integrating the equations of motion, which after the first interactions of the fireshell with the CBM become hardly predictable.
The energetic of the simulation is fulfilled and we find that the emission is due to blobs of matter in the CBM with typical dimensions of $r_{bl} = 10^{16}$ cm and average density contrast $\delta n/n$ $\simeq$ 2 particles/cm$^3$ in an overall  average density of 1 particle/cm$^3$.
We need to find additional cases of such phenomena to augment our statistic and improve its comprehension.

Particularly relevant are the first two-dimensional
hydrodynamical simulations of the progenitor evolution of a $23
M_{\sun}$ star close to core-collapse, leading to a naked core, as
shown in the recent work of Arnett and Meakin \citep{Arnett2011}. In
that work, pronounced asymmetries and strong dynamical interactions
between burning shells are seen: the dynamical behavior proceeds to
large amplitudes, enlarging deviations from the spherical symmetry in
the burning shells. It is of clear interest to find a possible
connection between the proto black hole concept, introduced in this
work, with the Arnett and Meakin results: to compare the radius, the
temperature and the dynamics of the core we have found in the present
work with the naked core obtained by Arnett and Meakin from the
thermonuclear evolution of the progenitor star. Particularly relevant is the presence, during this phase of collapse, of strong waves, originated in the mixing of the different element' shells. Such waves should become compressional, as they propagate inward, but they should also dissipate in non-convective regions, causing heating and slow mixing in these regions of the star. Since the wave heating is faster than radiative diffusion (which is very slow), an expansion phase of the boundary layers will occur, while the iron (Fe) core will contract \citep{Arnett2011}.
There is also the interesting possibility that the CBM clouds observed in GRBs be
related to the vigorous dynamics in violent activity of matter ejected
in the evolution of the original massive star, well before the formation
of the naked core (Arnett D., private communication).  

It is appropriate to emphasize that these results have no relation with the study of precursors in GRBs done in the current literature \citep[see e.g.][and references therein]{Burlon2008}.
Episode 1 and episode 2 are not temporally separated by a quiescent time. The spectral feature of episode 1 and episode 2 are strikingly different and, moreover, the episode 1 is very energetic, which is quite unusual for a typical precursor event.
We finally conclude that for the first time we witness the process of formation of the black hole from the phases just preceding the gravitational collapse all the way up to the GRB emission.

There is now evidence that the Proto Black Hole formation has been observed also in other GRB sources. After the submission of this article a second example has been found in GRB 101023, then and a paper about this source was submitted on November 4th 2011 and then published on February 1st 2012 \citep{Penacchioni2012}. There, extremely novel considerations in the structure of the late phase of the emission in X-ray at times larger than 200 s have been presented in favour of a standard signature in these sources \citep[see also the considerations made in][]{Page2011}. The possible use of this new family of GRBs as distance indicators is being considered.

\begin{acknowledgements}
We thank David Arnett for most fruitful discussions, the participants of the Les Houches workshop ``From Nuclei to White Dwarfs and Neutron Stars'' held in April 2011 (Eds. A. Mezzacappa and R. Ruffini, World Scientific 2011, in press), as well as the members of the AlbaNova University High Energy Astrophysics group. 
We are thankful to an anonymous referee for her/his important remarks both on the content and the presentation of our work which have improved the presentation of our paper. 
LI is especially grateful to Marco Muccino for fruitful discussions about the work concerning this manuscript.
We are also greateful to the Swift and Fermi teams for their assistance. 
One of us, AVP, acknowledges the support for the fellowship awarded for the Erasmus Mundus IRAP PhD program. This work made use of data supplied by the UK Swift Science Data Centre at the University of Leicester. 
\end{acknowledgements}



\begin{thebibliography}{141}
\expandafter\ifx\csname natexlab\endcsname\relax\def\natexlab#1{#1}\fi

\bibitem[{{Aksenov} {et~al.}(2007){Aksenov}, {Ruffini}, \&
  {Vereshchagin}}]{2007PhRvL..99l5003A}
{Aksenov}, A.~G., {Ruffini}, R., \& {Vereshchagin}, G.~V. 2007, Physical Review
  Letters, 99, 125003

\bibitem[{{Amati} {et~al.}(2002){Amati}, {Frontera}, {Tavani}, {in't Zand},
  {Antonelli}, {Costa}, {Feroci}, {Guidorzi}, {Heise}, {Masetti}, {Montanari},
  {Nicastro}, {Palazzi}, {Pian}, {Piro}, \& {Soffitta}}]{Amati2002}
{Amati}, L., {Frontera}, F., {Tavani}, M., {et~al.} 2002, \aap, 390, 81

\bibitem[{{Arnaud}(1996)}]{XSPEC}
{Arnaud}, K.~A. 1996, in Astronomical Society of the Pacific Conference Series,
  Vol. 101, Astronomical Data Analysis Software and Systems V, ed.
  {G.~H.~Jacoby \& J.~Barnes}, 17

\bibitem[{{Arnett} \& {Meakin}(2011)}]{Arnett2011}
{Arnett}, W.~D. \& {Meakin}, C. 2011, \apj, 733, 78

\bibitem[{{Atwood} {et~al.}(2009){Atwood}, {Abdo}, {Ackermann}, {Althouse},
  {Anderson}, {Axelsson}, {Baldini}, {Ballet}, {Band}, {Barbiellini}, \&
  et~al.}]{Atwood2009}
{Atwood}, W.~B., {Abdo}, A.~A., {Ackermann}, M., {et~al.} 2009, \apj, 697, 1071

\bibitem[{{Band} {et~al.}(1993){Band}, {Matteson}, {Ford}, {Schaefer},
  {Palmer}, {Teegarden}, {Cline}, {Briggs}, {Paciesas}, {Pendleton}, {Fishman},
  {Kouveliotou}, {Meegan}, {Wilson}, \& {Lestrade}}]{Band1993}
{Band}, D., {Matteson}, J., {Ford}, L., {et~al.} 1993, \apj, 413, 281

\bibitem[{{Baumgartner} {et~al.}(2009){Baumgartner}, {Barthelmy}, {Cummings},
  {Fenimore}, {Gehrels}, {Krimm}, {Markwardt}, {Palmer}, {Sakamoto}, {Sato},
  {Schady}, {Stamatikos}, {Tueller}, \& {Ukwatta}}]{GCN9530}
{Baumgartner}, W.~H., {Barthelmy}, S.~D., {Cummings}, J.~R., {et~al.} 2009, GRB
  Coordinates Network, 9530, 1

\bibitem[{{Beardmore} \& {Schady}(2009)}]{GCN9528}
{Beardmore}, A.~P. \& {Schady}, P. 2009, GRB Coordinates Network, 9528, 1

\bibitem[{{Bernardini} {et~al.}(2007){Bernardini}, {Bianco}, {Caito},
  {Dainotti}, {Guida}, \& {Ruffini}}]{Bernardini2007}
{Bernardini}, M.~G., {Bianco}, C.~L., {Caito}, L., {et~al.} 2007, \aap, 474,
  L13

\bibitem[{{Bianco} {et~al.}(2008){Bianco}, {Bernardini}, {Caito}, {Dainotti},
  {Guida}, \& {Ruffini}}]{Bianco2008}
{Bianco}, C.~L., {Bernardini}, M.~G., {Caito}, L., {et~al.} 2008, in American
  Institute of Physics Conference Series, Vol. 1065, American Institute of
  Physics Conference Series, ed. {Y.-F.~Huang, Z.-G.~Dai, \& B.~Zhang}, 223

\bibitem[{{Bianco} \& {Ruffini}(2004)}]{Bianco2004}
{Bianco}, C.~L. \& {Ruffini}, R. 2004, \apjl, 605, L1

\bibitem[{{Bianco} \& {Ruffini}(2005{\natexlab{a}})}]{Bianco2005b}
{Bianco}, C.~L. \& {Ruffini}, R. 2005{\natexlab{a}}, \apjl, 633, L13

\bibitem[{{Bianco} \& {Ruffini}(2005{\natexlab{b}})}]{Bianco2005a}
{Bianco}, C.~L. \& {Ruffini}, R. 2005{\natexlab{b}}, \apjl, 620, L23

\bibitem[{{Blandford} \& {McKee}(1976)}]{BlandfordMcKee}
{Blandford}, R.~D. \& {McKee}, C.~F. 1976, Phys. Fluids, 19, 1130

\bibitem[{{Bloom} {et~al.}(2006){Bloom}, {Prochaska}, {Pooley}, {Blake},
  {Foley}, {Jha}, {Ramirez-Ruiz}, {Granot}, {Filippenko}, {Sigurdsson},
  {Barth}, {Chen}, {Cooper}, {Falco}, {Gal}, {Gerke}, {Gladders}, {Greene},
  {Hennanwi}, {Ho}, {Hurley}, {Koester}, {Li}, {Lubin}, {Newman}, {Perley},
  {Squires}, \& {Wood-Vasey}}]{Bloom2006}
{Bloom}, J.~S., {Prochaska}, J.~X., {Pooley}, D., {et~al.} 2006, \apj, 638, 354

\bibitem[{{Burlon} {et~al.}(2008){Burlon}, {Ghirlanda}, {Ghisellini},
  {Lazzati}, {Nava}, {Nardini}, \& {Celotti}}]{Burlon2008}
{Burlon}, D., {Ghirlanda}, G., {Ghisellini}, G., {et~al.} 2008, \apjl, 685, L19

\bibitem[{Burrows {et~al.}(2005)Burrows, Hill, Nousek, Kennea, Wells, Osborne,
  Abbey, Beardmore, Mukerjee, Short, Chincarini, Campana, Citterio, Moretti,
  Pagani, Tagliaferri, Giommi, Capalbi, Tamburelli, Angelini, Cusumano,
  Brauninger, Burkert, \& Hartner}]{Burrows2005}
Burrows, D., Hill, J., Nousek, J., {et~al.} 2005, Space Science Reviews, 120,
  165, 10.1007/s11214-005-5097-2

\bibitem[{{Caito} {et~al.}(2010){Caito}, {Amati}, {Bernardini}, {Bianco}, {de
  Barros}, {Izzo}, {Patricelli}, \& {Ruffini}}]{Caito2010}
{Caito}, L., {Amati}, L., {Bernardini}, M.~G., {et~al.} 2010, \aap, 521, A80+

\bibitem[{{Caito} {et~al.}(2009){Caito}, {Bernardini}, {Bianco}, {Dainotti},
  {Guida}, \& {Ruffini}}]{Caito2009}
{Caito}, L., {Bernardini}, M.~G., {Bianco}, C.~L., {et~al.} 2009, \aap, 498,
  501

\bibitem[{{Cano} {et~al.}(2011){Cano}, {Bersier}, {Guidorzi}, {Margutti},
  {Svensson}, {Kobayashi}, {Melandri}, {Wiersema}, {Pozanenko}, {van der
  Horst}, {Pooley}, {Fernandez-Soto}, {Castro-Tirado}, {Postigo}, {Im},
  {Kamble}, {Sahu}, {Alonso-Lorite}, {Anupama}, {Bibby}, {Burgdorf}, {Clay},
  {Curran}, {Fatkhullin}, {Fruchter}, {Garnavich}, {Gomboc}, {Gorosabel},
  {Graham}, {Gurugubelli}, {Haislip}, {Huang}, {Huxor}, {Ibrahimov}, {Jeon},
  {Jeon}, {Ivarsen}, {Kasen}, {Klunko}, {Kouveliotou}, {Lacluyze}, {Levan},
  {Loznikov}, {Mazzali}, {Moskvitin}, {Mottram}, {Mundell}, {Nugent},
  {Nysewander}, {O'Brien}, {Park}, {Peris}, {Pian}, {Reichart}, {Rhoads},
  {Rol}, {Rumyantsev}, {Scowcroft}, {Shakhovskoy}, {Small}, {Smith}, {Sokolov},
  {Starling}, {Steele}, {Strom}, {Tanvir}, {Tsapras}, {Urata}, {Vaduvescu},
  {Volnova}, {Volvach}, {Wijers}, {Woosley}, \& {Young}}]{Cano2010}
{Cano}, Z., {Bersier}, D., {Guidorzi}, C., {et~al.} 2011, \mnras, 174

\bibitem[{{Capozziello} \& {Izzo}(2010)}]{CapozzielloIzzo}
{Capozziello}, S. \& {Izzo}, L. 2010, \aap, 519, A73

\bibitem[{{Cavallo} \& {Rees}(1978)}]{CavalloRees}
{Cavallo}, G. \& {Rees}, M.~J. 1978, \mnras, 183, 359

\bibitem[{{Cenko} {et~al.}(2009){Cenko}, {Perley}, {Junkkarinen}, {Burbidge},
  {Diego}, \& {Miller}}]{GCN9518}
{Cenko}, S.~B., {Perley}, D.~A., {Junkkarinen}, V., {et~al.} 2009, GRB
  Coordinates Network, 9518, 1

\bibitem[{{Cherubini} {et~al.}(2009){Cherubini}, {Geralico}, {J.~A.~Rueda}, \&
  {Ruffini}}]{Cherubini}
{Cherubini}, C., {Geralico}, A., {J.~A.~Rueda}, H., \& {Ruffini}, R. 2009,
  \prd, 79, 124002

\bibitem[{{Costa} {et~al.}(1997){Costa}, {Frontera}, {Heise}, {Feroci}, {in't
  Zand}, {Fiore}, {Cinti}, {Dal Fiume}, {Nicastro}, {Orlandini}, {Palazzi},
  {Rapisarda\#}, {Zavattini}, {Jager}, {Parmar}, {Owens}, {Molendi},
  {Cusumano}, {Maccarone}, {Giarrusso}, {Coletta}, {Antonelli}, {Giommi},
  {Muller}, {Piro}, \& {Butler}}]{Costa1997}
{Costa}, E., {Frontera}, F., {Heise}, J., {et~al.} 1997, \nat, 387, 783

\bibitem[{{Crider} {et~al.}(1997){Crider}, {Liang}, {Smith}, {Preece},
  {Briggs}, {Pendleton}, {Paciesas}, {Band}, \& {Matteson}}]{Crider1997}
{Crider}, A., {Liang}, E.~P., {Smith}, I.~A., {et~al.} 1997, \apjl, 479, L39

\bibitem[{{Daigne} {et~al.}(2009){Daigne}, {Bosnjak}, \& {Dubus}}]{Daigne2009}
{Daigne}, F., {Bosnjak}, Z., \& {Dubus}, G. 2009, ArXiv e-prints

\bibitem[{{Daigne} \& {Mochkovitch}(2002)}]{Daigne2002}
{Daigne}, F. \& {Mochkovitch}, R. 2002, \mnras, 336, 1271

\bibitem[{{Damour} \& {Ruffini}(1975)}]{Damour}
{Damour}, T. \& {Ruffini}, R. 1975, Physical Review Letters, 35, 463

\bibitem[{{de Barros} {et~al.}(2011){de Barros}, {Amati}, {Bernardini},
  {Bianco}, {Caito}, {Izzo}, {Patricelli}, \& {Ruffini}}]{deBarros2011}
{de Barros}, G., {Amati}, L., {Bernardini}, M.~G., {et~al.} 2011, \aap, 529,
  A130

\bibitem[{{Dezalay} {et~al.}(1992){Dezalay}, {Barat}, {Talon}, {Syunyaev},
  {Terekhov}, \& {Kuznetsov}}]{Dezalay1992}
{Dezalay}, J.-P., {Barat}, C., {Talon}, R., {et~al.} 1992, in American
  Institute of Physics Conference Series, Vol. 265, American Institute of
  Physics Conference Series, ed. {W.~S.~Paciesas \& G.~J.~Fishman}, 304--309

\bibitem[{{Ducci} {et~al.}(2009){Ducci}, {Sidoli}, {Mereghetti}, {Paizis}, \&
  {Romano}}]{Ducci2009}
{Ducci}, L., {Sidoli}, L., {Mereghetti}, S., {Paizis}, A., \& {Romano}, P.
  2009, \mnras, 398, 2152

\bibitem[{{Eichler} \& {Levinson}(2000)}]{Eichler2000}
{Eichler}, D. \& {Levinson}, A. 2000, \apj, 529, 146

\bibitem[{{Evans} {et~al.}(2009){Evans}, {Beardmore}, {Page}, {Osborne},
  {O'Brien}, {Willingale}, {Starling}, {Burrows}, {Godet}, {Vetere}, {Racusin},
  {Goad}, {Wiersema}, {Angelini}, {Capalbi}, {Chincarini}, {Gehrels}, {Kennea},
  {Margutti}, {Morris}, {Mountford}, {Pagani}, {Perri}, {Romano}, \&
  {Tanvir}}]{Evans2}
{Evans}, P.~A., {Beardmore}, A.~P., {Page}, K.~L., {et~al.} 2009, \mnras, 397,
  1177

\bibitem[{{Evans} {et~al.}(2007){Evans}, {Beardmore}, {Page}, {Tyler},
  {Osborne}, {Goad}, {O'Brien}, {Vetere}, {Racusin}, {Morris}, {Burrows},
  {Capalbi}, {Perri}, {Gehrels}, \& {Romano}}]{Evans}
{Evans}, P.~A., {Beardmore}, A.~P., {Page}, K.~L., {et~al.} 2007, \aap, 469,
  379

\bibitem[{{Fermi}(1949)}]{Fermi1949}
{Fermi}, E. 1949, Physical Review, 75, 1169

\bibitem[{{Fermi}(1954)}]{Fermi1954}
{Fermi}, E. 1954, \apj, 119, 1

\bibitem[{{Fishman} {et~al.}(1994){Fishman}, {Meegan}, {Wilson}, {Brock},
  {Horack}, {Kouveliotou}, {Howard}, {Paciesas}, {Briggs}, {Pendleton},
  {Koshut}, {Mallozzi}, {Stollberg}, \& {Lestrade}}]{Fishman1994}
{Fishman}, G.~J., {Meegan}, C.~A., {Wilson}, R.~B., {et~al.} 1994, \apjs, 92,
  229

\bibitem[{{Fong} {et~al.}(2010){Fong}, {Berger}, \& {Fox}}]{Fong2010}
{Fong}, W., {Berger}, E., \& {Fox}, D.~B. 2010, \apj, 708, 9

\bibitem[{{Frontera} {et~al.}(2000){Frontera}, {Amati}, {Costa}, {Muller},
  {Pian}, {Piro}, {Soffitta}, {Tavani}, {Castro-Tirado}, {Dal Fiume}, {Feroci},
  {Heise}, {Masetti}, {Nicastro}, {Orlandini}, {Palazzi}, \&
  {Sari}}]{Frontera2000}
{Frontera}, F., {Amati}, L., {Costa}, E., {et~al.} 2000, \apjs, 127, 59

\bibitem[{{Gehrels} {et~al.}(2009){Gehrels}, {Ramirez-Ruiz}, \&
  {Fox}}]{Gehrels2009}
{Gehrels}, N., {Ramirez-Ruiz}, E., \& {Fox}, D.~B. 2009, \araa, 47, 567

\bibitem[{{Ghirlanda} {et~al.}(2002){Ghirlanda}, {Celotti}, \&
  {Ghisellini}}]{Ghirlanda2002}
{Ghirlanda}, G., {Celotti}, A., \& {Ghisellini}, G. 2002, \aap, 393, 409

\bibitem[{{Ghirlanda} {et~al.}(2003){Ghirlanda}, {Celotti}, \&
  {Ghisellini}}]{Ghirlanda2003}
{Ghirlanda}, G., {Celotti}, A., \& {Ghisellini}, G. 2003, \aap, 406, 879

\bibitem[{{Ghisellini} \& {Celotti}(1999)}]{Ghisellini1999}
{Ghisellini}, G. \& {Celotti}, A. 1999, \aaps, 138, 527

\bibitem[{{Giannios}(2006)}]{Giannios2006}
{Giannios}, D. 2006, \aap, 457, 763

\bibitem[{{Golenetskii} {et~al.}(2009){Golenetskii}, {Aptekar}, {Mazets},
  {Pal'Shin}, {Frederiks}, {Oleynik}, {Ulanov}, {Svinkin}, \&
  {Cline}}]{GCN9553}
{Golenetskii}, S., {Aptekar}, R., {Mazets}, E., {et~al.} 2009, GRB Coordinates
  Network, 9553, 1

\bibitem[{{Goodman}(1986)}]{Goodman1986}
{Goodman}, J. 1986, \apjl, 308, L47

\bibitem[{{Granot} {et~al.}(1999){Granot}, {Piran}, \& {Sari}}]{Granot1997}
{Granot}, J., {Piran}, T., \& {Sari}, R. 1999, \apj, 513, 679

\bibitem[{{Gruzinov} \& {Waxman}(1999)}]{1999ApJ...511..852G}
{Gruzinov}, A. \& {Waxman}, E. 1999, \apj, 511, 852

\bibitem[{{Guetta} {et~al.}(2011){Guetta}, {Pian}, \& {Waxman}}]{Guetta2011}
{Guetta}, D., {Pian}, E., \& {Waxman}, E. 2011, \aap, 525, A53

\bibitem[{{Guiriec} {et~al.}(2011){Guiriec}, {Connaughton}, {Briggs},
  {Burgess}, {Ryde}, {Daigne}, {M{\'e}sz{\'a}ros}, {Goldstein}, {McEnery},
  {Omodei}, {Bhat}, {Bissaldi}, {Camero-Arranz}, {Chaplin}, {Diehl}, {Fishman},
  {Foley}, {Gibby}, {Giles}, {Greiner}, {Gruber}, {von Kienlin}, {Kippen},
  {Kouveliotou}, {McBreen}, {Meegan}, {Paciesas}, {Preece}, {Rau}, {Tierney},
  {van der Horst}, \& {Wilson-Hodge}}]{Guiriec2011}
{Guiriec}, S., {Connaughton}, V., {Briggs}, M.~S., {et~al.} 2011, \apjl, 727,
  L33

\bibitem[{{Izzo} {et~al.}(2010){Izzo}, {Bernardini}, {Bianco}, {Caito},
  {Patricelli}, \& {Ruffini}}]{Izzo2010}
{Izzo}, L., {Bernardini}, M.~G., {Bianco}, C.~L., {et~al.} 2010, JKPS, 57

\bibitem[{{Kaneko} {et~al.}(2006){Kaneko}, {Preece}, {Briggs}, {Paciesas},
  {Meegan}, \& {Band}}]{Kaneko2006}
{Kaneko}, Y., {Preece}, R.~D., {Briggs}, M.~S., {et~al.} 2006, \apjs, 166, 298

\bibitem[{{Klebesadel}(1992)}]{Klebesadel1992}
{Klebesadel}, R.~W. 1992, {The durations of gamma-ray bursts}, ed. {Ho, C.,
  Epstein, R.~I., \& Fenimore, E.~E.}, 161--168

\bibitem[{{Klebesadel} {et~al.}(1973){Klebesadel}, {Strong}, \&
  {Olson}}]{Klebesadel1973}
{Klebesadel}, R.~W., {Strong}, I.~B., \& {Olson}, R.~A. 1973, \apjl, 182, L85

\bibitem[{{Kono} {et~al.}(2009){Kono}, {Daikyuji}, {Sonoda}, {Ohmori},
  {Hayashi}, {Noda}, {Nishioka}, {Yamauchi}, {Ohno}, {Suzuki}, {Kokubun},
  {Takahashi}, {Yamaoka}, {Sugita}, {Nakagawa}, {Tamagawa}, {Hong}, {Vasquez},
  {Hanabata}, {Uehara}, {Fukazawa}, {Iwakiri}, {Tashiro}, {Terada}, {Endo},
  {Onda}, {Sugasahara}, {Urata}, {Enoto}, {Nakazawa}, \& {Makishima}}]{GCN9568}
{Kono}, K., {Daikyuji}, A., {Sonoda}, E., {et~al.} 2009, GRB Coordinates
  Network, 9568, 1

\bibitem[{{Kotov} {et~al.}(2008){Kotov}, {Kochemasov}, {Kuzin}, {Kuznetsov},
  {Sylwester}, \& {Yurov}}]{Kotov2008}
{Kotov}, Y., {Kochemasov}, A., {Kuzin}, S., {et~al.} 2008, in COSPAR, Plenary
  Meeting, Vol.~37, 37th COSPAR Scientific Assembly, 1596

\bibitem[{{Kouveliotou} {et~al.}(1993){Kouveliotou}, {Meegan}, {Fishman},
  {Bhat}, {Briggs}, {Koshut}, {Paciesas}, \& {Pendleton}}]{Koveliotou1993}
{Kouveliotou}, C., {Meegan}, C.~A., {Fishman}, G.~J., {et~al.} 1993, \apjl,
  413, L101

\bibitem[{{Kumar} \& {McMahon}(2008{\natexlab{a}})}]{Kumar2008}
{Kumar}, P. \& {McMahon}, E. 2008{\natexlab{a}}, \mnras, 384, 33

\bibitem[{{Kumar} \& {McMahon}(2008{\natexlab{b}})}]{KumarMcMahon2008}
{Kumar}, P. \& {McMahon}, E. 2008{\natexlab{b}}, \mnras, 384, 33

\bibitem[{{Lazzati} \& {Begelman}(2010)}]{Lazzati2010}
{Lazzati}, D. \& {Begelman}, M.~C. 2010, \apj, 725, 1137

\bibitem[{{Longo} {et~al.}(2009){Longo}, {Moretti}, {Barbiellini}, {Vallazza},
  {Trifoglio}, {Bulgarelli}, {Gianotti}, {Fuschino}, {Marisaldi}, {Labanti},
  {Galli}, {Di Cocco}, {Cutini}, {Pittori}, {Tavani}, {Striani}, {Pucella},
  {D'Ammando}, {Vittorini}, {Argan}, {Trois}, {Piano}, {Sabatini}, {Del},
  {Feroci}, {Evangelista}, {Donnarumma}, {Pacciani}, {Soffitta}, {Costa},
  {Lazzarotto}, {Lapshov}, {Rapisarda}, {Giuliani}, {Chen}, {Mereghetti},
  {Perotti}, {Caraveo}, {Pellizzoni}, {Pilia}, {Vercellone}, {Picozza},
  {Morselli}, {Prest}, {Lipari}, {Zanello}, {Rappoldi}, {Cattaneo}, {Giommi},
  {Santolamazza}, {Verrecchia}, \& {Salotti}}]{GCN9524}
{Longo}, F., {Moretti}, E., {Barbiellini}, G., {et~al.} 2009, GRB Coordinates
  Network, 9524, 1

\bibitem[{{McBreen}(2009)}]{GCN9535}
{McBreen}, S. 2009, GRB Coordinates Network, 9535, 1

\bibitem[{{Medvedev}(2000)}]{Medvedev2000}
{Medvedev}, M.~V. 2000, \apj, 540, 704

\bibitem[{{Medvedev} \& {Loeb}(1999)}]{MedvedevLoeb}
{Medvedev}, M.~V. \& {Loeb}, A. 1999, \apj, 526, 697

\bibitem[{{Medvedev} \& {Spitkovsky}(2009)}]{MedvedevSpitkovsky2009}
{Medvedev}, M.~V. \& {Spitkovsky}, A. 2009, \apj, 700, 956

\bibitem[{{Meegan} {et~al.}(2009){Meegan}, {Lichti}, {Bhat}, {Bissaldi},
  {Briggs}, {Connaughton}, {Diehl}, {Fishman}, {Greiner}, {Hoover}, {van der
  Horst}, {von Kienlin}, {Kippen}, {Kouveliotou}, {McBreen}, {Paciesas},
  {Preece}, {Steinle}, {Wallace}, {Wilson}, \& {Wilson-Hodge}}]{Meegan2009}
{Meegan}, C., {Lichti}, G., {Bhat}, P.~N., {et~al.} 2009, \apj, 702, 791

\bibitem[{{Meegan}(1997)}]{Meegan1997}
{Meegan}, C.~A. 1997, NASA STI/Recon Technical Report N, 1, 70758

\bibitem[{{Meegan} {et~al.}(1992){Meegan}, {Fishman}, {Wilson}, {Horack},
  {Brock}, {Paciesas}, {Pendleton}, \& {Kouveliotou}}]{Meegan1992}
{Meegan}, C.~A., {Fishman}, G.~J., {Wilson}, R.~B., {et~al.} 1992, \nat, 355,
  143

\bibitem[{{M{\'e}sz{\'a}ros}(2002)}]{Meszaros2002}
{M{\'e}sz{\'a}ros}, P. 2002, \araa, 40, 137

\bibitem[{Meszaros(2006)}]{Meszaros2006}
Meszaros, P. 2006, Reports on Progress in Physics, 69, 2259

\bibitem[{{Meszaros} {et~al.}(1993){Meszaros}, {Laguna}, \&
  {Rees}}]{1993ApJ...415..181M}
{Meszaros}, P., {Laguna}, P., \& {Rees}, M.~J. 1993, \apj, 415, 181

\bibitem[{{Meszaros} \& {Rees}(1993)}]{MeszarosRees1993}
{Meszaros}, P. \& {Rees}, M.~J. 1993, \apj, 405, 278

\bibitem[{{M{\'e}sz{\'a}ros} \& {Rees}(2000)}]{MeszarosRees2000}
{M{\'e}sz{\'a}ros}, P. \& {Rees}, M.~J. 2000, \apj, 530, 292

\bibitem[{{Molinari} {et~al.}(2007){Molinari}, {Vergani}, {Malesani}, {Covino},
  {D'Avanzo}, {Chincarini}, {Zerbi}, {Antonelli}, {Conconi}, {Testa}, {Tosti},
  {Vitali}, {D'Alessio}, {Malaspina}, {Nicastro}, {Palazzi}, {Guetta},
  {Campana}, {Goldoni}, {Masetti}, {Meurs}, {Monfardini}, {Norci}, {Pian},
  {Piranomonte}, {Rizzuto}, {Stefanon}, {Stella}, {Tagliaferri}, {Ward},
  {Ihle}, {Gonzalez}, {Pizarro}, {Sinclaire}, \& {Valenzuela}}]{Molinari2007}
{Molinari}, E., {Vergani}, S.~D., {Malesani}, D., {et~al.} 2007, \aap, 469, L13

\bibitem[{{Nandi} {et~al.}(2009){Nandi}, {Rao}, {Chakrabarti}, {Malkar},
  {Sreekumar}, {Debnath}, {Hingar}, {Kotoch}, {Kotovk}, \&
  {Arkhangelskiy}}]{Nandi2009}
{Nandi}, A., {Rao}, A.~R., {Chakrabarti}, S.~K., {et~al.} 2009, ArXiv e-prints

\bibitem[{{Nousek} {et~al.}(2006){Nousek}, {Kouveliotou}, {Grupe}, {Page},
  {Granot}, {Ramirez-Ruiz}, {Patel}, {Burrows}, {Mangano}, {Barthelmy},
  {Beardmore}, {Campana}, {Capalbi}, {Chincarini}, {Cusumano}, {Falcone},
  {Gehrels}, {Giommi}, {Goad}, {Godet}, {Hurkett}, {Kennea}, {Moretti},
  {O'Brien}, {Osborne}, {Romano}, {Tagliaferri}, \& {Wells}}]{Nousek2006}
{Nousek}, J.~A., {Kouveliotou}, C., {Grupe}, D., {et~al.} 2006, \apj, 642, 389

\bibitem[{{Paciesas} {et~al.}(1999){Paciesas}, {Meegan}, {Pendleton}, {Briggs},
  {Kouveliotou}, {Koshut}, {Lestrade}, {McCollough}, {Brainerd}, {Hakkila},
  {Henze}, {Preece}, {Connaughton}, {Kippen}, {Mallozzi}, {Fishman},
  {Richardson}, \& {Sahi}}]{Paciesas1999}
{Paciesas}, W.~S., {Meegan}, C.~A., {Pendleton}, G.~N., {et~al.} 1999, \apjs,
  122, 465

\bibitem[{{Paczynski}(1986)}]{Paczynski1986}
{Paczynski}, B. 1986, \apjl, 308, L43

\bibitem[{{Page} {et~al.}(2011){Page}, {Starling}, {Fitzpatrick}, {Pandey},
  {Osborne}, {Schady}, {McBreen}, {Campana}, {Ukwatta}, {Pagani}, {Beardmore},
  \& {Evans}}]{Page2011}
{Page}, K.~L., {Starling}, R.~L.~C., {Fitzpatrick}, G., {et~al.} 2011, \mnras,
  416, 2078

\bibitem[{{Panaitescu} \& {Meszaros}(1998)}]{Panaitescu1998}
{Panaitescu}, A. \& {Meszaros}, P. 1998, \apjl, 493, L31

\bibitem[{{Panaitescu} \& {M{\'e}sz{\'a}ros}(2000)}]{Panaitescu2000}
{Panaitescu}, A. \& {M{\'e}sz{\'a}ros}, P. 2000, \apjl, 544, L17

\bibitem[{{Patricelli} {et~al.}(2011){Patricelli}, {Bernardini}, {Bianco},
  {Caito}, {Izzo}, {Ruffini}, \& {Vereshchagin}}]{Patricelli2011}
{Patricelli}, B., {Bernardini}, M.~G., {Bianco}, C.~L., {et~al.} 2011,
  International Journal of Modern Physics D, 20, 1983

\bibitem[{{Pe'er}(2008)}]{Peer2008}
{Pe'er}, A. 2008, \apj, 682, 463

\bibitem[{{Pe'er} {et~al.}(2007){Pe'er}, {Ryde}, {Wijers}, {M{\'e}sz{\'a}ros},
  \& {Rees}}]{Peer2007}
{Pe'er}, A., {Ryde}, F., {Wijers}, R.~A.~M.~J., {M{\'e}sz{\'a}ros}, P., \&
  {Rees}, M.~J. 2007, \apjl, 664, L1

\bibitem[{{Pe'er} \& {Zhang}(2006)}]{Peer2006b}
{Pe'er}, A. \& {Zhang}, B. 2006, \apj, 653, 454

\bibitem[{{Pe'er} {et~al.}(2010){Pe'er}, {Zhang}, {Ryde}, {McGlynn}, {Zhang},
  {Preece}, \& {Kouveliotou}}]{Peer2010}
{Pe'er}, A., {Zhang}, B.-B., {Ryde}, F., {et~al.} 2010, ArXiv e-prints

\bibitem[{{Penacchioni} {et~al.}(2012){Penacchioni}, {Ruffini}, {Izzo},
  {Muccino}, {Bianco}, {Caito}, {Patricelli}, \& {Amati}}]{Penacchioni2012}
{Penacchioni}, A.~V., {Ruffini}, R., {Izzo}, L., {et~al.} 2012, \aap, 538, A58

\bibitem[{{Peng} {et~al.}(2011){Peng}, {Yin}, {Bi}, {Bao}, \& {Ma}}]{Peng2011}
{Peng}, Z.~Y., {Yin}, Y., {Bi}, X.~W., {Bao}, Y.~Y., \& {Ma}, L. 2011,
  Astronomische Nachrichten, 332, 92

\bibitem[{{Piran}(1999)}]{Piran1999}
{Piran}, T. 1999, \physrep, 314, 575

\bibitem[{Piran(2005)}]{Piran2005}
Piran, T. 2005, Rev. Mod. Phys., 76, 1143

\bibitem[{{Piran} {et~al.}(2009){Piran}, {Sari}, \& {Zou}}]{Piran2009}
{Piran}, T., {Sari}, R., \& {Zou}, Y. 2009, \mnras, 393, 1107

\bibitem[{{Piran} {et~al.}(1993){Piran}, {Shemi}, \&
  {Narayan}}]{PiranShemiNarayan}
{Piran}, T., {Shemi}, A., \& {Narayan}, R. 1993, \mnras, 263, 861

\bibitem[{{Preece} {et~al.}(2002){Preece}, {Briggs}, {Giblin}, {Mallozzi},
  {Pendleton}, {Paciesas}, \& {Band}}]{Preece2002}
{Preece}, R.~D., {Briggs}, M.~S., {Giblin}, T.~W., {et~al.} 2002, \apj, 581,
  1248

\bibitem[{{Qin}(2002)}]{Qin2002}
{Qin}, Y.-P. 2002, \aap, 396, 705

\bibitem[{{Ramirez-Ruiz} \& {Fenimore}(2000)}]{Ramirez2000}
{Ramirez-Ruiz}, E. \& {Fenimore}, E.~E. 2000, \apj, 539, 712

\bibitem[{{Rao} {et~al.}(2011){Rao}, {Malkar}, {Hingar}, {Agrawal},
  {Chakrabarti}, {Nandi}, {Debnath}, {Kotoch}, {Sarkar}, {Chidambaram},
  {Vinod}, {Sreekumar}, {Kotov}, {Buslov}, {Yurov}, {Tyshkevich},
  {Arkhangelskij}, {Zyatkov}, \& {Naik}}]{Rao2011}
{Rao}, A.~R., {Malkar}, J.~P., {Hingar}, M.~K., {et~al.} 2011, \apj, 728, 42

\bibitem[{{Rees} \& {Meszaros}(1992)}]{ReesMeszaros1992}
{Rees}, M.~J. \& {Meszaros}, P. 1992, \mnras, 258, 41

\bibitem[{{Rees} \& {Meszaros}(1994)}]{ReesMeszaros1994}
{Rees}, M.~J. \& {Meszaros}, P. 1994, \apjl, 430, L93

\bibitem[{{Rees} \& {Meszaros}(1998)}]{ReesMeszaros1998}
{Rees}, M.~J. \& {Meszaros}, P. 1998, \apjl, 496, L1

\bibitem[{Roming {et~al.}(2005)Roming, Kennedy, Mason, Nousek, Ahr, Bingham,
  Broos, Carter, Hancock, Huckle, Hunsberger, Kawakami, Killough, Koch,
  Mclelland, Smith, Smith, Soto, Boyd, Breeveld, Holland, Ivanushkina, Pryzby,
  Still, \& Stock}]{Roming2005}
Roming, P., Kennedy, T., Mason, K., {et~al.} 2005, Space Science Reviews, 120,
  95, 10.1007/s11214-005-5095-4

\bibitem[{{Ruffini}(1999)}]{RSWX2}
{Ruffini}, R. 1999, \aaps, 138, 513

\bibitem[{{Ruffini}(2011)}]{Ruffini2011}
{Ruffini}, R. 2011, ArXiv e-prints

\bibitem[{{Ruffini} {et~al.}(2007){Ruffini}, {Bernardini}, {Bianco}, {Caito},
  {Chardonnet}, {Dainotti}, {Fraschetti}, {Guida}, {Vereshchagin}, \&
  {Xue}}]{Ruffini2007b}
{Ruffini}, R., {Bernardini}, M.~G., {Bianco}, C.~L., {et~al.} 2007, in ESA
  Special Publication, Vol. 622, ESA Special Publication, 561

\bibitem[{{Ruffini} {et~al.}(2005){Ruffini}, {Bernardini}, {Bianco},
  {Chardonnet}, {Fraschetti}, {Gurzadyan}, {Vitagliano}, \&
  {Xue}}]{Ruffini2005}
{Ruffini}, R., {Bernardini}, M.~G., {Bianco}, C.~L., {et~al.} 2005, in American
  Institute of Physics Conference Series, Vol. 782, XIth Brazilian School of
  Cosmology and Gravitation, ed. {M.~Novello \& S.~E.~Perez Bergliaffa}, 42

\bibitem[{{Ruffini} {et~al.}(2003){Ruffini}, {Bianco}, {Chardonnet},
  {Fraschetti}, {Vitagliano}, \& {Xue}}]{Ruffini2003}
{Ruffini}, R., {Bianco}, C.~L., {Chardonnet}, P., {et~al.} 2003, in American
  Institute of Physics Conference Series, Vol. 668, Cosmology and Gravitation,
  ed. {M.~Novello \& S.~E.~Perez Bergliaffa}, 16

\bibitem[{{Ruffini} {et~al.}(2002){Ruffini}, {Bianco}, {Chardonnet},
  {Fraschetti}, \& {Xue}}]{Ruffini2002}
{Ruffini}, R., {Bianco}, C.~L., {Chardonnet}, P., {Fraschetti}, F., \& {Xue},
  S. 2002, \apjl, 581, L19

\bibitem[{{Ruffini} {et~al.}(2001{\natexlab{a}}){Ruffini}, {Bianco},
  {Fraschetti}, {Xue}, \& {Chardonnet}}]{Ruffini2001}
{Ruffini}, R., {Bianco}, C.~L., {Fraschetti}, F., {Xue}, S.-S., \&
  {Chardonnet}, P. 2001{\natexlab{a}}, \apjl, 555, L113

\bibitem[{{Ruffini} {et~al.}(2001{\natexlab{b}}){Ruffini}, {Bianco},
  {Fraschetti}, {Xue}, \& {Chardonnet}}]{Ruffini2001L107}
{Ruffini}, R., {Bianco}, C.~L., {Fraschetti}, F., {Xue}, S.-S., \&
  {Chardonnet}, P. 2001{\natexlab{b}}, \apjl, 555, L107

\bibitem[{{Ruffini} {et~al.}(2010{\natexlab{a}}){Ruffini}, {Chakrabarti}, \&
  {Izzo}}]{COSPAR}
{Ruffini}, R., {Chakrabarti}, S.~K., \& {Izzo}, L. 2010{\natexlab{a}},
  Submitted to Adv. Sp. Res.

\bibitem[{{Ruffini} {et~al.}(2011){Ruffini}, {Izzo}, {Penacchioni}, {Bianco},
  {Caito}, {Chakrabarti}, \& {Nandi}}]{TEXAS}
{Ruffini}, R., {Izzo}, L., {Penacchioni}, A.~V., {et~al.} 2011, PoS(Texas2010),
  101

\bibitem[{{Ruffini} {et~al.}(2000){Ruffini}, {Salmonson}, {Wilson}, \&
  {Xue}}]{RSWX}
{Ruffini}, R., {Salmonson}, J.~D., {Wilson}, J.~R., \& {Xue}, S. 2000, \aap,
  359, 855

\bibitem[{{Ruffini} {et~al.}(2010{\natexlab{b}}){Ruffini}, {Vereshchagin}, \&
  {Xue}}]{PhysRep}
{Ruffini}, R., {Vereshchagin}, G., \& {Xue}, S.-S. 2010{\natexlab{b}},
  \physrep, 487, 1

\bibitem[{{Ruffini}(2001)}]{Ruffini2001K}
{Ruffini}, R.~J. 2001, {Analogies, new paradigms and observational data as
  growing factors of Relativistic Astrophysics}

\bibitem[{{Ryde}(2004)}]{Ryde2004}
{Ryde}, F. 2004, \apj, 614, 827

\bibitem[{{Ryde} {et~al.}(2010){Ryde}, {Axelsson}, {Zhang}, {McGlynn}, {Pe'er},
  {Lundman}, {Larsson}, {Battelino}, {Zhang}, {Bissaldi}, {Bregeon}, {Briggs},
  {Chiang}, {de Palma}, {Guiriec}, {Larsson}, {Longo}, {McBreen}, {Omodei},
  {Petrosian}, {Preece}, \& {van der Horst}}]{Ryde2010}
{Ryde}, F., {Axelsson}, M., {Zhang}, B.~B., {et~al.} 2010, \apjl, 709, L172

\bibitem[{{Ryde} {et~al.}(2006){Ryde}, {Bj{\"o}rnsson}, {Kaneko},
  {M{\'e}sz{\'a}ros}, {Preece}, \& {Battelino}}]{Rydeetal2006}
{Ryde}, F., {Bj{\"o}rnsson}, C.-I., {Kaneko}, Y., {et~al.} 2006, \apj, 652,
  1400

\bibitem[{{Ryde} \& {Pe'er}(2009)}]{Ryde2009}
{Ryde}, F. \& {Pe'er}, A. 2009, \apj, 702, 1211

\bibitem[{{Rykoff} {et~al.}(2009){Rykoff}, {Aharonian}, {Akerlof}, {Ashley},
  {Barthelmy}, {Flewelling}, {Gehrels}, {G{\"o}{\v g}{\"u}{\c s}}, {G{\"u}ver},
  {Kizilo{\v g}lu}, {Krimm}, {McKay}, {{\"O}zel}, {Phillips}, {Quimby},
  {Rowell}, {Rujopakarn}, {Schaefer}, {Smith}, {Vestrand}, {Wheeler}, {Wren},
  {Yuan}, \& {Yost}}]{Rykoff2009}
{Rykoff}, E.~S., {Aharonian}, F., {Akerlof}, C.~W., {et~al.} 2009, \apj, 702,
  489

\bibitem[{{Sakamoto} {et~al.}(2009){Sakamoto}, {Ukwatta}, \&
  {Barthelmy}}]{GCN9534}
{Sakamoto}, T., {Ukwatta}, T.~N., \& {Barthelmy}, S.~D. 2009, GRB Coordinates
  Network, 9534, 1

\bibitem[{{Sari}(1997)}]{Sari1997}
{Sari}, R. 1997, \apjl, 489, L37

\bibitem[{{Sari}(1998)}]{Sari1998}
{Sari}, R. 1998, \apjl, 494, L49

\bibitem[{{Sari} \& {Piran}(1999)}]{SariPiran1999}
{Sari}, R. \& {Piran}, T. 1999, \apj, 520, 641

\bibitem[{{Sari} {et~al.}(1999){Sari}, {Piran}, \& {Halpern}}]{Sarietal1999}
{Sari}, R., {Piran}, T., \& {Halpern}, J.~P. 1999, \apjl, 519, L17

\bibitem[{{Schady}(2009)}]{GCN9527}
{Schady}, P. 2009, GRB Coordinates Network, 9527, 1

\bibitem[{{Schady} {et~al.}(2009){Schady}, {Baumgartner}, {Beardmore},
  {Campana}, {Curran}, {Guidorzi}, {Kennea}, {Mao}, {Margutti}, {Osborne},
  {Page}, {Romano}, {Siegel}, {Stratta}, \& {Ukwatta}}]{GCN9512}
{Schady}, P., {Baumgartner}, W.~H., {Beardmore}, A.~P., {et~al.} 2009, GRB
  Coordinates Network, 9512, 1

\bibitem[{{Schaefer}(2007)}]{Schaefer2007}
{Schaefer}, B.~E. 2007, \apj, 660, 16

\bibitem[{{Shara} {et~al.}(1997){Shara}, {Zurek}, {Williams}, {Prialnik},
  {Gilmozzi}, \& {Moffat}}]{Shara1997}
{Shara}, M.~M., {Zurek}, D.~R., {Williams}, R.~E., {et~al.} 1997, \aj, 114, 258

\bibitem[{{Shemi}(1994)}]{Shemi1994}
{Shemi}, A. 1994, \mnras, 269, 1112

\bibitem[{{Shemi} \& {Piran}(1990)}]{ShemiPiran1990}
{Shemi}, A. \& {Piran}, T. 1990, \apjl, 365, L55

\bibitem[{{Spitkovsky}(2008)}]{Spitkovsky2008}
{Spitkovsky}, A. 2008, \apjl, 673, L39

\bibitem[{{Stern} \& {Poutanen}(2004)}]{Stern2004}
{Stern}, B.~E. \& {Poutanen}, J. 2004, \mnras, 352, L35

\bibitem[{{Strong}(1975)}]{Strong1975}
{Strong}, I.~B. 1975, in Astrophysics and Space Science Library, Vol.~48,
  Neutron Stars, Black Holes and Binary X-ray Sources, ed. H.~{Gursky} \&
  R.~{Ruffini}, 47--58

\bibitem[{{Strong} \& {Klebesadel}(1974)}]{Strong1974a}
{Strong}, I.~B. \& {Klebesadel}, R.~W. 1974, \nat, 251, 396

\bibitem[{{Strong} {et~al.}(1974){Strong}, {Klebesadel}, \&
  {Olson}}]{Strong1974b}
{Strong}, I.~B., {Klebesadel}, R.~W., \& {Olson}, R.~A. 1974, \apjl, 188, L1

\bibitem[{{Tavani}(1996)}]{Tavani1996}
{Tavani}, M. 1996, \apj, 466, 768

\bibitem[{{Tavani}(1998)}]{Tavani1998}
{Tavani}, M. 1998, \apjl, 497, L21

\bibitem[{{van Paradijs} {et~al.}(1997){van Paradijs}, {Groot}, {Galama},
  {Kouveliotou}, {Strom}, {Telting}, {Rutten}, {Fishman}, {Meegan}, {Pettini},
  {Tanvir}, {Bloom}, {Pedersen}, {N{\o}rdgaard-Nielsen}, {Linden-V{\o}rnle},
  {Melnick}, {van der Steene}, {Bremer}, {Naber}, {Heise}, {in't Zand},
  {Costa}, {Feroci}, {Piro}, {Frontera}, {Zavattini}, {Nicastro}, {Palazzi},
  {Bennett}, {Hanlon}, \& {Parmar}}]{vanParadjis1997}
{van Paradijs}, J., {Groot}, P.~J., {Galama}, T., {et~al.} 1997, \nat, 386, 686

\bibitem[{{van Paradijs} {et~al.}(2000){van Paradijs}, {Kouveliotou}, \&
  {Wijers}}]{Van2000}
{van Paradijs}, J., {Kouveliotou}, C., \& {Wijers}, R.~A.~M.~J. 2000, \araa,
  38, 379

\bibitem[{{Waxman}(1997)}]{Waxman1997}
{Waxman}, E. 1997, \apjl, 491, L19

\bibitem[{{Zdziarski} {et~al.}(1991){Zdziarski}, {Svensson}, \&
  {Paczynski}}]{Zdziarski1991}
{Zdziarski}, A.~A., {Svensson}, R., \& {Paczynski}, B. 1991, \apj, 366, 343

\end{thebibliography}
\end{document}